\documentclass[reprint,
 amsmath,amssymb,
 aps,
longbibliography
]{revtex4-2}

\usepackage{braket}
\usepackage{multirow}
\usepackage{slashed} 
\usepackage{cancel}  
\usepackage{graphicx}
\usepackage{dcolumn}
\usepackage{bm}
\usepackage{hyperref}
\usepackage{color}
\usepackage{xcolor}

\usepackage{float}

\newcommand{\vv}{\mathbf{v}}
\newcommand{\rr}{\mathbf{r}}
\newcommand{\vva}{\mathbf{v}_1}
\newcommand{\vvb}{\mathbf{v}_2}
\newcommand{\vvab}{\mathbf{v}_{12}}
\newcommand{\cc}{\mathbf{c}}
\newcommand{\COMc}{\mathbf{C}}
\newcommand{\cca}{\mathbf{c}_1}
\newcommand{\ccb}{\mathbf{c}_2}
\newcommand{\ccab}{\mathbf{c}_{12}}
\newcommand{\dif}{\mathrm{d}}
\newcommand{\ssig}{\widehat{\boldsymbol{\sigma}}}
\newcommand{\kb}{k_B}
\newcommand{\KLD}{\mathcal{D}}
\newcommand{\thm}{\text{th}}
\newcommand{\eq}{\text{eq}}
\newcommand{\LE}{\text{LE}}
\newcommand{\kin}{\text{kin}}
\newcommand{\vs}{z}
\newcommand{\bb}{b}
\newcommand{\ts}{t}
\newcommand{\zs}{\zeta_0}

\definecolor{rojo}{rgb}{1,0,0}
\definecolor{naranja}{rgb}{1,0.65098,0}
\definecolor{amarillo}{rgb}{1,1,0}
\definecolor{verde}{rgb}{0,1,0}
\definecolor{cian}{rgb}{0,1,1}
\definecolor{magent}{rgb}{1,0,1}
\definecolor{marron}{rgb}{0.898039,0.596078,0.4}
\definecolor{gris}{rgb}{0.698039,0.729412,0.733333}

\newcommand{\mean}[1]{\left\langle #1 \right\rangle}

\begin{document}


\title{Thermal versus entropic Mpemba effect  in molecular gases with nonlinear drag}

\author{Alberto Meg\'ias}
 \email{albertom@unex.es}
 \affiliation{Departamento de F\'isica, Universidad de Extremadura, E-06006 Badajoz, Spain}

\author{Andr\'es Santos}
 \email{andres@unex.es}
 \affiliation{Departamento de F\'isica, Universidad de Extremadura, E-06006 Badajoz, Spain}
\affiliation{Instituto de Computaci\'on Cient\'ifica Avanzada (ICCAEx), Universidad de Extremadura, E-06006 Badajoz, Spain}

\author{Antonio Prados}
 \email{prados@us.es}
 \affiliation{F\'isica  Te\'orica,  Universidad  de  Sevilla,  Apartado  de  Correos  1065,  E-41080  Sevilla,  Spain}

\date{\today}

\begin{abstract}
Loosely speaking, the Mpemba effect appears when hotter systems cool sooner or, in a more abstract way, when systems further from equilibrium relax faster. In this paper, we investigate the Mpemba effect in a molecular gas with nonlinear drag, both analytically (by employing the tools of kinetic theory) and numerically (direct simulation Monte Carlo of the kinetic equation and event-driven molecular dynamics). The analysis is carried out via two alternative routes, recently considered in the literature: first, the kinetic or thermal route, in which the Mpemba effect is characterized by the crossing of the evolution curves of the kinetic temperature (average kinetic energy), and, second, the stochastic thermodynamics or entropic route, in which the Mpemba effect is characterized by the crossing of the distance to equilibrium in probability space. In general, a nonmutual correspondence between the thermal and entropic Mpemba effects is found, i.e., there may appear the thermal effect without its entropic counterpart or vice versa.  Furthermore, a nontrivial overshoot with respect to equilibrium of the thermal relaxation makes it necessary to revise the usual definition of the thermal Mpemba effect, which is shown to be better described in terms of the relaxation of the local equilibrium distribution.
Our theoretical framework, which involves an extended Sonine approximation in which not only the excess kurtosis but also the sixth cumulant is retained, gives an excellent account of the behavior observed in simulations.
\end{abstract}

\maketitle

\section{Introduction}
\label{sec1}

In recent years, memory effects have become a hot topic in nonequilibrium statistical physics research~\cite{KPZSN19}. Those phenomena usually imply counterintuitive effects that apparently contradict well-established standard physical laws. One of the most interesting is the Mpemba effect (ME): Given two samples of a fluid in a common thermal bath, the  initially hotter one may cool more rapidly than that initially cooler. The well-known Newton's law of cooling, according to which the temperature evolution is predetermined by its initial value, is thus violated in the presence of the ME. Original studies of the ME deal with water
\cite{MO69,K69,F71,D71,F74,G74,W77,O79,F79,K80,H81,WOB88,A95,K96,M96,J06,ERS08,K09,VM10,B11,VM12,BT12,ZHMZZZJS14,VK15,S15,BT15,R15,JG15,IC16,GLH19,BKC21},
and even today there is still a lack of consensus about its existence in this very complex system~\cite{BL16,BH20,ES21}.

In a more general context, the ME can be recast as ``the initially
further from equilibrium relaxes faster,'' with the separation from
equilibrium being defined in a suitable way, see below. With such an
interpretation, Mpemba-like effects have been investigated in a large
variety of many-body systems: molecular gases~\cite{SP20,PSP21},
mixtures~\cite{GKG21}, granular
gases~\cite{LVPS17,TLLVPS19,BPRR20,MLTVL21,GG21,BPR21,BPR22}, inertial
suspensions~\cite{THS21,T21}, spin glasses~\cite{Betal19}, carbon
nanotube resonators~\cite{GLCG11}, clathrate hydrates~\cite{AKKL16},
Markovian models~\cite{LR17,KRHV19,CKB21,BGM21,LLHRW22}, active
systems~\cite{SL21}, Ising models~\cite{GMLMS21,TYR21,VD21}, non-Markovian
mean-field systems~\cite{YH20,YH22}, or quantum systems~\cite{CLL21}.  Very recently, the ME has been analyzed in the framework of Landau's theory of phase transitions~\cite{HR22}. Also, it has been
experimentally observed in colloids~\cite{KB20,KCB22}.

There have been two main approaches to the ME: the kinetic-theory or ``thermal'' approach~\cite{SP20,PSP21,GKG21,LVPS17,TLLVPS19,BPRR20,MLTVL21,GG21,BPR21,THS21,T21} and the stochastic-process (or thermodynamics) or ``entropic'' approach \cite{LR17,KRHV19,BGM21,CKB21,SL21,CLL21,KB20,KCB22,LLHRW22}. In the thermal approach, kinetic theory makes it possible to define in a natural way an out-of-equilibrium time-dependent temperature $T(t)$ as basically the average kinetic energy, i.e.,
\begin{equation}\label{eq:noneq-temp-def-intro}
    T(t) = \frac{m}{d \kb}\langle v^2\rangle,
\end{equation}
where $d$ is the dimensionality of the system, $m$ is the mass of a particle, and $\kb$ is the Boltzmann constant.  This definition allows for a simple, and close in spirit to the original studies in water, characterization of the separation from equilibrium at temperature $T_\eq$: The initially hotter (colder) sample A (B) translates into that having the larger (smaller) initial value of the kinetic temperature, $T_{A}^{0}>T_{B}^{0}>T_\eq$. A \textit{thermal} Mpemba effect (TME) is observed  if the evolution curves for the temperature cross at a certain time $t_{\theta}$, $T_A(t_{\theta})=T_B(t_{\theta})$, and that of the initially hotter remains below the other one for longer times, $T_\eq<T_A(t)<T_B(t)$ for $t>t_{\theta}$.
Additionally,  a Mpemba effect may exist in the absence of a temperature crossing if $T_B(t)$ overshoots the equilibrium value at a certain time $t_O$, i.e., $T_B(t_O)=T_\eq$, then reaches  a minimum, and finally relaxes to equilibrium later than  sample A. This overshoot effect would be the analog of the supercooling phenomenon in water.

In the stochastic-process approach, the starting point is usually a Markov process $\bm{x}(t)$. The state of the system at time $t$ is determined by a probability distribution $P(\bm{x},t)$, which typically obeys a master equation, for discrete $\bm{x}$, or a Fokker--Planck equation, for continuous $\bm{x}$. The Kullback--Leibler divergence (KLD) or relative entropy~\cite{KL51} is defined as
\begin{equation}\label{eq:KLD-def-intro}
\KLD(t)\equiv \mean{\ln\frac{P}{P^\eq}}=\int \dif\bm{x}\, P(\bm{x},t) \ln \frac{P(\bm{x},t)}{P^\eq(\bm{x})},
\end{equation}
where $P^\eq(\bm{x})$ stands for the equilibrium probability distribution. The $H$-theorem~\cite{vK07} ensures that $\KLD(t)$ monotonically decreases to zero over a nonequilibrium process and thus $\KLD(t)$ can be interpreted as the distance to equilibrium from a physical standpoint~\cite{note_22_01_1}. Also, $\KLD$ can be understood as (the opposite of) the nonequilibrium entropy relative to the equilibrium state. The ME is translated as follows in this context: The further from (closer to) sample A (B) has the larger (smaller) initial value of $\KLD$, i.e., $\KLD_A^0>\KLD_B^0>0$. The \textit{entropic} Mpemba effect (EME) emerges when the evolution curves for $\KLD$ cross at a certain time $t_\KLD$, $\KLD_A(t_\KLD)=\KLD_B(t_\KLD)$, and $0<\KLD_A(t)<\KLD_B(t)$ for $t>t_\KLD$.

\begin{figure}
    \includegraphics[width=.9\columnwidth]{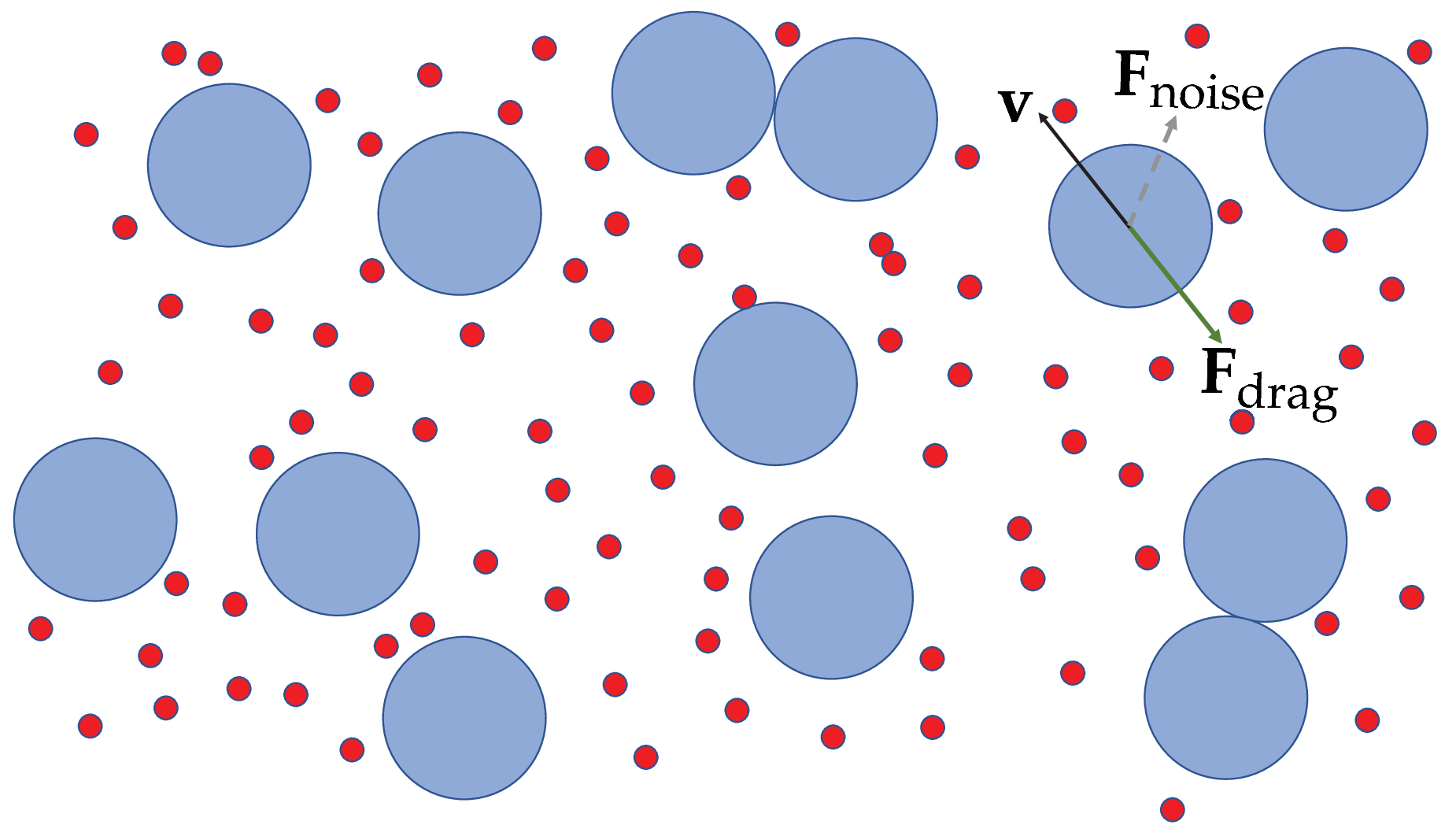}\\
    \caption{Illustration of the system considered in this paper.
A molecular gas of hard particles (represented by the large blue circles) is coupled to a thermal bath (made of particles represented by the small red circles) via a drag force $\mathbf{F}_{\text{drag}}=-m\zeta(v)\mathbf{v}$, where $\zeta(v)$ is a velocity-dependent drag coefficient, and a stochastic force $\mathbf{F}_{\text{noise}}=m\xi(v)\bm{\eta}$, where $\bm{\eta}$ is a Gaussian white-noise term.  In addition, the particles are subjected to binary elastic collisions.
      \label{fig:cartoon}}
\end{figure}

The two effects described above, TME and EME, are equivalent if a biunivocal correspondence between nonequilibrium temperature and (entropic) distance to equilibrium exists. Yet, this is not the case in general, as we will show. In fact, the main aim of this paper is to analyze the correspondence between the TME and the EME in a prototypical system, where the two approaches can be carried out analytically---at least in an approximate, systematic, way. Specifically, we consider a molecular gas of hard particles that is coupled to a thermal bath, with the resulting drag force being nonlinear in the velocity~\cite{SP20}.  In addition, there are binary elastic collisions between the particles. See Fig.~\ref{fig:cartoon} for an illustration of the system. The evolution equation of the velocity distribution function (VDF) is given by the Enskog--Fokker--Planck equation (EFPE)---the Enskog term accounts for binary collisions, whereas the Fokker--Planck term models the interaction with the thermal bath, see Sec.~\ref{sec:model} for details. To look into the system dynamics, we employ a hybrid approach that includes both a theoretical and a numerical analysis: kinetic-theory tools---via a Sonine approximation of the EFPE equation---for the former and direct simulation Monte Carlo (DSMC), together with event-driven molecular dynamics (EDMD), simulations for the latter.

Note that it is the nonlinearity of the drag force that the ME stems from. As a consequence, the time evolution of the kinetic temperature is coupled to other moments and the kinetic temperature of the nonlinear fluid shows algebraic nonexponential relaxation  and strong memory effects after a quench~\cite{PSP21}. Note also that elastic collisions do not change the average kinetic energy: Were the drag absent, the kinetic temperature would remain constant throughout the whole time evolution. Still, an initial nonequilibrium VDF would evolve toward the equilibrium Maxwellian---higher-order velocity cumulants would indeed be affected by collisions and tend to zero in the long-time limit.

The above characterizations of out-of-equilibrium temperature and distance to equilibrium, Eqs.~\eqref{eq:noneq-temp-def-intro} and \eqref{eq:KLD-def-intro},  are quite natural in the molecular fluid. Yet, a different choice may be more adequate in other systems. On the one hand, some kind of nonequilibrium temperature, e.g., in the spirit of the fictive or effective temperature for glassy systems~\cite{S90,C09,BB11}, may be introduced in systems where the kinetic temperature cannot be defined---for example, Ising models~\cite{BP94}. On the other hand, the $L_1$ and $L_2$ norms have been employed in  the literature to measure the distance of the VDF to equilibrium~\cite{LR17,BGM21,KB20}. Alternative choices for the observables characterizing the thermal relaxation and the distance to equilibrium may quantitatively affect the values of the crossing times $t_\theta$ and $t_\KLD$, and even the own existence of the TME and the EME.

With the above definitions, both the TME and the EME can be investigated. Some basic questions arise, though. Does the TME imply the EME, or vice versa?  When both of them are present, how close are the respective crossover times $t_\theta$ and $t_\KLD$? Is it possible to observe the ME if the kinetic temperature of at least one of the two samples overshoots its equilibrium value?
The theoretical framework developed in this paper,  which is supported by computer simulations,  answers these key questions.

The paper is organized as follows. Section~\ref{sec:model} puts forward our model system for a fluid with nonlinear drag. Also, the local equilibrium concept is introduced and its implications for the entropic distance are discussed. In Sec.~\ref{sec:evol-eqs}, we derive the evolution equations for the relevant physical quantities, within the Sonine approximation schemes developed in this paper. From this knowledge, the general phenomenology of TME and EME is predicted and described from heuristic arguments in Sec.~\ref{sec:TME_EME}. Afterwards, in Sec.~\ref{sec:OME} a singular case for TME, induced by the appearance of an overshoot effect, is investigated. Thus, Secs.~\ref{sec:model}--\ref{sec:OME} constitute the core of the theoretical framework developed in the paper. In addition, we present simulation results supporting the theoretical predictions in Sec.~\ref{sec:RESULTS}. Finally, conclusions are presented in Sec.~\ref{sec:Conclusions}, including a discussion on the definition of nonequilibrium temperature for a general system. Some technical parts are relegated to appendices.

\section{Model system and Local equilibrium}\label{sec:model}

Let us consider the following model for a fluid with nonlinear drag~\cite{F07,F14,HKLMSLW17,SP20,PSP21}: a $d$-dimensional fluid of elastic hard spheres of mass $m$ and diameter $\sigma$, with number density $n$, {subjected} to  a stochastic force composed by a white-noise term with nonlinear variance plus a nonlinear drag force. This scheme mimics a system of elastic spheres assumed to be suspended in a background fluid in equilibrium at temperature $T_\bb$, as depicted in Fig.~\ref{fig:cartoon}.

The (spatially uniform) EFPE for the one-body VDF $f(\vv,t)$ reads
\begin{equation}\label{eq:EFP}
    \partial_t f(\vv,t)-\frac{\partial}{\partial \vv}\cdot\left[\zeta(v)\vv+\frac{\xi^2(v)}{2} \frac{\partial}{\partial \vv}\right]f(\vv,t) = J[\vv|f,f],
\end{equation}
where
\begin{align}\label{eq:J-coll}
    J[\vva|f,f]=&\sigma^{d-1}g_c\int\dif\vvb\, \int_{+}\dif\ssig\,  \vvab\cdot\ssig\\ \nonumber
    &\times \left[f(\vva^\prime,t)f(\vvb^\prime,t)-f(\vva,t)f(\vvb,t)\right]
\end{align}
is the usual Boltzmann--Enskog collision operator with $\vvab\equiv
\vva-\vvb$, $g_c= \lim_{r\rightarrow \sigma^+} g(r)$ being the contact
value of the pair correlation function, and $\int_{+}\dif\ssig\equiv
\int\dif\ssig\, \Theta(\vvab\cdot\ssig)$. In addition, the drag
component of the stochastic force is  $-m\zeta(v)\vv$, while the
white-noise counterpart has a nonlinear variance $m^2\xi^2(v)$. The
functions $\zeta(v)$ and $\xi^2(v)$ are connected via the fluctuation-dissipation theorem as
\begin{equation}\label{eq:FDR}
    \xi^2(v) = \frac{2\kb T_\bb}{m}\zeta(v),
\end{equation}
where  $T_\bb$ is the temperature of the background fluid.
This ensures that the only stationary solution of the EFPE is the equilibrium Maxwellian,
\begin{equation}
f^\eq(\vv)=n \left(\frac{m}{2\pi \kb T_\bb}\right)^{d/2}e^{-mv^2/2\kb T_\bb}.
\end{equation}

A quadratic dependence of the drag coefficient naturally appears when the hard spheres and the background fluid particles have a comparable mass~\cite{F07,F14,HKLMSLW17,SP20,PSP21},
\begin{equation}\label{eq:4}
    \zeta(v) = \zeta_0\left(1+\gamma\frac{m v^2}{\kb T_\bb} \right).
\end{equation}
The coefficients $\zeta_0$ and $\gamma$ are both positive and measure the zero-velocity value of the drag coefficient and the degree of nonlinearity of the drag force, respectively. Note that, due to the nonlinearity of the drag force, the implementation of the Langevin equation associated with the free streaming of particles between collisions is far from trivial. This issue is discussed in Appendix~\ref{sec:Langevin}.

The two approaches to the ME can be implemented in the nonlinear fluid introduced above. Translating Eqs.~\eqref{eq:noneq-temp-def-intro} and \eqref{eq:KLD-def-intro} to our model system, we have that the nonequilibrium temperature $T(t)$ is given by
\begin{equation}\label{eq:temp-def}
    T(t) = \frac{m}{d \kb}\langle v^2\rangle = \frac{m}{nd\kb}\int\dif\vv\, v^2f(\vv,t),
\end{equation}
and the relative entropy is
\begin{equation}
\label{eq:KLD-kinetic}
    \KLD(t) =\mean{\ln\frac{f}{f^\eq}}=\frac{1}{n} \int \mathrm{d}\vv\, f(\vv,t)\ln\frac{f(\vv,t)}{f^\eq(\vv)},
\end{equation}
where $n\equiv \int\dif\vv\, f(\vv,t)$ is the number density.

On physical grounds, it is expected that the evolution of the gas toward equilibrium takes place along two stages~\cite{DvB77}. First, a rapid ``kinetic'' stage where the VDF approaches the so-called local equilibrium (LE) form,
\begin{equation}
\label{eq:f-LE}
    f^\LE(\vv;T(t))=n \left[\frac{m}{2\pi \kb T(t)}\right]^{d/2}e^{-mv^2/2\kb T(t)},
\end{equation}
i.e., $f^\LE$ has the Maxwellian shape but with the time-dependent temperature. Second, a slower ``hydrodynamic'' stage, where the VDF is close to $f^\LE$ and the evolution of the VDF takes place via the temperature.
\begin{figure}
    \includegraphics[width=3.25in]{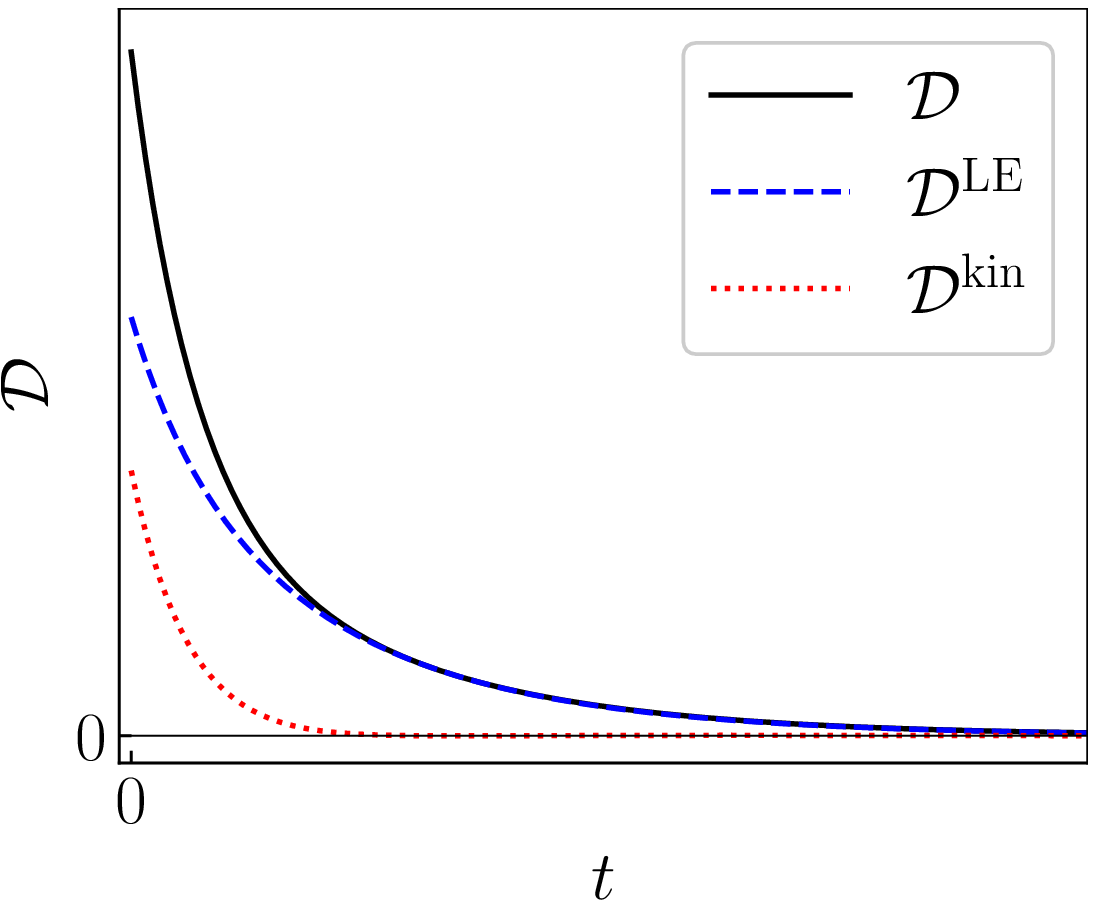}
    \caption{Sketch showing a typical evolution of the total KLD (solid line), its LE contribution (dashed line), and its kinetic contribution (dotted line).
      \label{fig:sketch}}
\end{figure}

The above discussion suggests the following decomposition for the relative entropy,
\begin{equation}
\label{eq:KLD_decomposition}
\KLD(t)=\KLD^{\kin}(t)+\KLD^\LE(T(t)),
\end{equation}
where
\begin{subequations}
\label{eq:KLD_decomposition_2}
\begin{equation}
\KLD^{\kin}(t)=\frac{1}{n} \int \mathrm{d}\vv\, f(\vv,t)\ln\frac{f(\vv,t)}{f^\LE(\vv;T(t))}
\end{equation}
and
\begin{align}
\label{eq:KLD^LE}
\KLD^\LE(T(t))
=&\frac{1}{n} \int \mathrm{d}\vv\, f^\LE(\vv;T(t))\ln\frac{f^\LE(\vv;T(t))}{f^\eq(\vv)}\nonumber\\
=&\frac{d}{2}\left[\theta(t)-1-\ln\theta(t)\right], \quad \theta(t)\equiv \frac{T(t)}{T_\bb}.
\end{align}
\end{subequations}
Both $\KLD^{\kin}$ and $\KLD^\LE$ are positive
definite~\cite{note_22_01_2}. To split $\KLD$ into the sum of $\KLD^{\kin}$ and $\KLD^\LE$, we have employed that the average of the kinetic energy with $f(\vv,t)$ is the same as with $f^\LE(\vv;T(t))$. A generalization of this idea makes it possible to define a nonequilibrium temperature and an analogous splitting of $\KLD$ in  quite a general class of systems, see Sec.~\ref{sec:Conclusions} for further details.

The first contribution to the total $\KLD$, $\KLD^{\kin}$, is
a measure of the departure of the true VDF from the LE one, and
depends explicitly on time through the whole VDF $f(\vv,t)$. In
contrast, the second contribution $\KLD^\LE$ measures the deviation of
the LE state from the asymptotic equilibrium state and  only depends
on time through the nonequilibrium temperature $T(t)$, namely on the
temperature ratio $\theta(t)$. More specifically, $\KLD^\LE$
monotonically increases as
$|\theta-1|$ increases in the domains $\theta>1$ and $\theta<1$ separately.
Figure \ref{fig:sketch} presents a sketch of the temporal evolution of $\KLD$ and its two contributions, $\KLD^\LE$ and $\KLD^\kin$ \cite{note_22_02_1}.

The TME and EME can be directly related if the crossing comes about in
the hydrodynamic regime, since therein $\KLD^{\kin}\approx 0$ and
$\KLD(t)\approx\KLD^\LE(\theta(t))$---which is a function of temperature
only, as explicitly stated by our notation. Therefore, the TME and EME
become equivalent during the hydrodynamic stage, and
$t_\theta\simeq t_\KLD$. However, we will show that in most situations
the ME occurs during the kinetic stage, the contribution $\KLD^{\kin}$
is then relevant, and the physical picture is much more complex.  In
fact, the two-stage relaxation picture may even break down under certain
conditions, as discussed in Ref.~\cite{KBD21}.

\section{Evolution equations}\label{sec:evol-eqs}

Multiplying both sides of Eq.~\eqref{eq:EFP} by $v^2$ and integrating over velocity one readily obtains the evolution equation for the time-dependent temperature,
\begin{equation}
\label{eq:Tdot}
\frac{\dot{T}}{\zeta_0} =- 2(T-T_\bb)\left[1+(d+2)\gamma\frac{T}{T_\bb} \right]-2(d+2)\gamma\frac{T^2}{T_\bb}a_2,
\end{equation}
where
\begin{equation}
\label{a2}
    a_2(t) \equiv \frac{d}{d+2}\frac{\langle v^4\rangle }{\langle v^2\rangle^2}-1
\end{equation}
is the excess kurtosis.

Using as unit of time the mean free time---average time between collisions---at equilibrium,
\begin{equation}
\tau_\bb=\frac{K_d}{g_cn\sigma^{d-1}\sqrt{2\kb T_\bb/m}},\quad K_d\equiv \frac{\sqrt{2}\Gamma(d/2)}{\pi^{\frac{d-1}{2}}},
\end{equation}
the dimensionless time $t^*$ and zero-velocity drag coefficient $\zeta_0^*$ can be defined as
\begin{equation}
t^*={t}/{\tau_\bb},\quad \zeta_0^*=\zeta_0\tau_\bb.
\end{equation}
The parameter $\zeta_0^*$ measures the relative relevance of the drag force (i.e., the interactions between the particles and the background fluid) and hard-sphere binary collisions. The limit $\zeta_0^*\to 0$ corresponds to negligible drag force, where the EFPE reduces to the Enskog equation. The limit $\zeta_0^*\to\infty$ corresponds to negligible collisions, where the EFPE reduces to the Fokker--Planck equation~\cite{note_22_01_3}. In this work, we typically consider the value $\zeta_0^*=1$, for which the drag force and binary collisions are comparable and act over the same timescale. Note that both the drag force and binary collisions drive by themselves the system to equilibrium, independently of the magnitude of the other interaction, with the entropic distance monotonically decreasing to zero \cite{JLPM21}.

In the remainder of the paper we employ dimensionless quantities. Dimensionless temperature is identified with the temperature ratio $\theta$ defined in Eq.~\eqref{eq:KLD^LE}. For simplicity, henceforth the stars on $t^*$ and $\zeta_0^*$ are dropped.  The evolution equation for the temperature, Eq.~\eqref{eq:Tdot}, thus reads
\begin{equation}
\label{eq:thetadot}
\frac{\dot{\theta}}{\zs} =- 2(\theta-1)\left[1+(d+2)\gamma\theta \right]-2(d+2)\gamma\theta^2a_2.
\end{equation}
Notice that  one gets Newton's cooling law $\dot{\theta}=-2\zeta_0(\theta-1)$ in the linear case $\gamma=0$.  However, if $\gamma\neq 0$, then the evolution of temperature is coupled to that of the fourth-degree moment $\langle v^4\rangle$ through $a_2$. Next, the evolution equation for  $\langle v^4\rangle$ stemming from the EFPE, Eq.~\eqref{eq:EFP}, is coupled to the sixth-degree moment $\langle v^6\rangle$ due to the nonlinear drag term and to all the moments $\langle v^\ell\rangle$ due to the collision term, and so on. Thus, the full evolution of $\theta(t)$ is coupled to the infinite hierarchy of moment equations, which are derived in Appendix~\ref{sec:evol-eqs-derivation} by introducing a Sonine expansion of the VDF. By retaining only the first two terms in the expansion, which involve the excess kurtosis (or fourth-order cumulant) $a_2$ and the sixth-order cumulant $a_3$,
\begin{equation}\label{eq:v6-a3-relation}
 a_3(t)=1+3a_2-\frac{d^2}{(d+2)(d+4)}\frac{\mean{v^6}}{\mean{v^2}^3},
\end{equation}
and neglecting nonlinear terms in the cumulants one gets

\begin{subequations}
\label{eq:cumul-evol-Sonine}
\begin{align}
    \frac{\dot{a_2}}{\zs}=& -8\gamma(\theta-1)+4\left[2\gamma-(d+8)\gamma\theta-\frac{1}{\theta}\right]a_2
  \nonumber\\
    &+4(d+4)\gamma\theta a_3-\frac{8(d-1)}{d(d+2)}\frac{\sqrt{\theta}}{\zs}\left(a_2-\frac{a_3}{4}\right), \label{eq:a2-evol-Sonine}
\end{align}
\begin{align}
    \frac{\dot{a_3}}{\zs}=&-24\gamma\left(2-3 \theta\right) a_2
    +6\left[4\gamma-(d+14)\gamma \theta-\frac{1}{\theta}\right] a_3\nonumber \\
&+\frac{\sqrt{\theta}}{\zs}\frac{3(d-1)}{d(d+2)(d+4)}\left[4a_2-(4d+19)a_3\right] .\label{eq:a3-evol-Sonine}
\end{align}
\end{subequations}
Equations \eqref{eq:thetadot} and \eqref{eq:cumul-evol-Sonine} make a closed set of three coupled differential equations, nonlinear in the temperature but linear in the cumulants.

In this paper, we consider two Sonine approximations. The roughest approximation consists of neglecting $a_3$, setting $a_3=0$ in Eq.~\eqref{eq:a2-evol-Sonine}, and dealing then with Eqs.~\eqref{eq:thetadot} and \eqref{eq:a2-evol-Sonine} for the pair $(\theta,a_2)$. Here, we term this approach the basic Sonine approximation (BSA)~\cite{note_22_01_4}.  A more sophisticated theory is obtained by keeping $a_3$ and dealing then with Eqs.~\eqref{eq:thetadot} and \eqref{eq:cumul-evol-Sonine}. We term this approach the extended Sonine approximation (ESA)~\cite{note_22_01_6}.

\section{Thermal versus entropic Mpemba effects}
\label{sec:TME_EME}
\subsection{Heuristic arguments}
\label{sec:TME_EME_A}

\begin{table*}
\caption{Summary of possible cases regarding the occurrence of the TME and the EME.\label{tab:cases}}
\begin{ruledtabular}
\begin{tabular}{lllll}
Case&Type of ME& Initial condition&If \ldots&then \ldots\\
\hline
 ET1&
Direct TME \& EME& $\theta_A^0>\theta_B^0>1,\quad  a_{2A}^0>a_{2B}^0,\quad  \KLD_A^0>\KLD_B^0$
& $|a_{2A}(\ts_\theta)|<|a_{2B}(\ts_\theta)|$&$\ts_\KLD<\ts_\theta$\\
TE1&Direct TME \& EME& $\theta_A^0>\theta_B^0>1,\quad  a_{2A}^0>a_{2B}^0,\quad  \KLD_A^0>\KLD_B^0$ &$|a_{2A}(\ts_\theta)|>|a_{2B}(\ts_\theta)|$&$\ts_\KLD>\ts_\theta$
\vspace{1mm}\\
ET2&
Inverse TME \& EME& $\theta_A^0<\theta_B^0<1,\quad  a_{2A}^0<a_{2B}^0,\quad  \KLD_A^0>\KLD_B^0$
& $|a_{2A}(\ts_\theta)|<|a_{2B}(\ts_\theta)|$&$\ts_\KLD<\ts_\theta$\\
TE2&Inverse TME \& EME& $\theta_A^0<\theta_B^0<1,\quad  a_{2A}^0<a_{2B}^0,\quad  \KLD_A^0>\KLD_B^0$ &$|a_{2A}(\ts_\theta)|>|a_{2B}(\ts_\theta)|$&$\ts_\KLD>\ts_\theta$
\vspace{1mm}\\
T1&Direct TME& $\theta_A^0>\theta_B^0>1,\quad  a_{2A}^0>a_{2B}^0$&$\KLD_B^0>\KLD_A^0$&No EME
\vspace{1mm}\\
T2&Inverse TME& $\theta_A^0<\theta_B^0<1,\quad  a_{2A}^0<a_{2B}^0$&$\KLD_B^0>\KLD_A^0$&No EME
\vspace{1mm}\\
E1&
EME& $\KLD_B^0>\KLD_A^0$&$\theta_A^0>\theta_B^0>1$&No Direct TME\\
E2&EME& $\KLD_B^0>\KLD_A^0$&$\theta_A^0<\theta_B^0<1$&No Inverse TME\\
\end{tabular}
\end{ruledtabular}
\end{table*}
Now we proceed to study the ME in the theoretical framework of the
Sonine approximations we have just introduced. To start with, let us
consider two samples (A and B) at the \emph{same} initial temperatures
$\theta_A(0)\equiv\theta_A^0$ and $\theta_B(0)\equiv\theta_B^0$, above
the equilibrium value, i.e., $\theta_A^0=\theta_B^0>1$. According to
Eq.~\eqref{eq:thetadot}, the initial slopes $\dot{\theta}_A(0)$ and
$\dot{\theta}_B(0)$ satisfy the inequality
$\dot{\theta}_A(0)<\dot{\theta}_B(0)$ if $a_{2A}(0)\equiv
a_{2A}^0>a_{2B}(0)\equiv a_{2B}^0$, in which case sample A is expected
to reach equilibrium before sample B. It must be brought to bear that
the latter statement is true if $\theta(t)-1$ keeps its initial sign
along the whole relaxation to equilibrium, a condition that is assumed
throughout this section. Exceptions to this fact, due to the overshoot of $\theta$ with respect to its equilibrium value, are discussed in Sec.~\ref{sec:OME}.

In order to analyze the TME described in Sec.~\ref{sec1}, let us take
now $\theta_A^0>\theta_B^0>1$. As discussed above,
$\dot{\theta}_A(0)<\dot{\theta}_B(0)$ if $a_{2A}^0>a_{2B}^0$. In that
way, it can be expected that, by a convenient choice of the
initial-condition values $(\theta_A^0,a_{2A}^0)$ and
$(\theta_B^0,a_{2B}^0)$, the evolution curves $\theta_A(\ts)$ and
$\theta_B(\ts)$ intersect at a certain crossover time
$\ts_\theta$. That is, $\theta_A(\ts_\theta)=\theta_B(\ts_\theta)$ and
$\dot{\theta}_A(\ts_\theta)<\dot{\theta}_B(\ts_\theta)$, which entails
$a_{2A}(\ts_\theta)>a_{2B}(\ts_\theta)$ and
$1<\theta_A(\ts)<\theta_B(\ts)$ for $\ts>\ts_\theta$. This is the
typical framework for the emergence of the (direct) TME in the kinetic
description~\cite{LVPS17,SP20,PSP21}.

The inverse TME is analogous, except that, instead of $\theta_A^0>\theta_B^0>1$, one now has  $\theta_A^0<\theta_B^0<1$. If now  $a_{2B}^0> a_{2A}^0$, then $\dot{\theta}_B(0)<\dot{\theta}_A(0)$, so that it is in principle possible that the evolution curve $\theta_A(\ts)$ intersects  $\theta_B(\ts)$ at a certain crossover time $\ts_\theta$.

Note that, without loss of generality, we denote by A the sample with an initial temperature farther from the equilibrium one, both in the direct and inverse TME.  Thus, the necessary (but, of course, not sufficient) conditions for the direct and inverse TME are $a_{2A}^0>a_{2B}^0$ and $a_{2B}^0>a_{2A}^0$, respectively.

In this work, we  analyze both the TME and the EME. In the latter, it is the evolution curves of the relative entropy $\KLD$ that intersect at a certain time $\ts_\KLD$, as described in Sec.~\ref{sec1} \cite{note_22_01_5}. In particular, we want to understand whether the TME implies the EME or not. Also, when both the TME and the EME are present, we would like to investigate the relation between the crossing times $\ts_\theta$ and $\ts_\KLD$.

Let us address the questions above by simple heuristic
arguments. First, we consider the case in which the further from
equilibrium sample in the kinetic approach (A) is also the further
from equilibrium in the entropic approach,
i.e., $\KLD_A^0>\KLD_B^0$. Therein, the existence of the TME implies
that of the EME, and vice versa, as shown below.
Note that $\KLD_A^0>\KLD_B^0$ if $\KLD^0$ increases
with $|\theta^0-1|$. This is indeed true for the  LE contribution
$\KLD^\LE$, but not necessarily so for the total KLD $\KLD$ if the kinetic contribution
$\KLD^{\kin}$ plays a relevant role.

For the direct TME, we have $\theta_A^0>\theta_B^0>1$ and
$\KLD_A^0>\KLD_B^0$.  If the TME exists, then one has
$\theta_B(\ts)>\theta_A(\ts)>1$ after the crossover. In particular,
this holds for sufficiently long times belonging to the hydrodynamic
stage, where both $\KLD^{\kin}_A$ and $\KLD^{\kin}_B$ are negligible,
and thus one has $\KLD_B(\ts)>\KLD_A(\ts)$ (EME) in the same
stage. Reciprocally, if the EME exists, then $\KLD_B(\ts)>\KLD_A(\ts)$
in the hydrodynamic stage after the crossover, implying
$\theta_B(\ts)>\theta_A(\ts)>1$ (TME) in the same regime. An analogous
reasoning applies to the inverse TME, i.e., $1>\theta_B^0>\theta_A^0$
and $\KLD_A^0>\KLD_B^0$.

Provided that the TME and EME are present, the argument above does not
tell us the relative positioning of the crossover times
$\ts_\theta$ and $\ts_\KLD$, i.e., whether $\ts_\theta>\ts_\KLD$ or
$\ts_\theta<\ts_\KLD$.
Let us start by considering that the direct TME takes place at
$\ts_\theta$. Therefore, we have that
$\KLD_A^\LE(\ts_\theta)=\KLD_B^\LE(\ts_\theta)$ and only the kinetic
part contributes to the KLD difference at $\ts_\theta$,
$\KLD_A(\ts_\theta)-\KLD_B(\ts_\theta)=\KLD^{\kin}_A(\ts_\theta)-\KLD^{\kin}_B(\ts_\theta)$.
This implies that $\KLD_A(\ts_\theta)<\KLD_B(\ts_\theta)$ (and hence
$\ts_\KLD<\ts_\theta$) if
$\KLD^{\kin}_A(\ts_\theta)<\KLD^{\kin}_B(\ts_\theta)$, while
$\KLD_A(\ts_\theta)>\KLD_B(\ts_\theta)$ (and hence
$\ts_\KLD>\ts_\theta$) otherwise.

For the sake of simplicity, and to go beyond the generic analysis of
the previous paragraph, let us assume that the values of the excess
kurtoses at the crossover time $\ts_\theta$ are small enough as to
approximate $\KLD^{\kin} \propto a_2^2$. The proportionality constant
may depend on the details of the VDF---see Eq.~\eqref{eq:deltaKLDgamma-approx} below for the specific
example of a gamma distribution.  Within this approximation, the first
case, $\ts_\KLD<\ts_\theta$, is expected if
$|a_{2A}(\ts_\theta)|<|a_{2B}(\ts_\theta)|$, while the second case,
$\ts_\KLD>\ts_\theta$, is expected if
$|a_{2A}(\ts_\theta)|>|a_{2B}(\ts_\theta)|$. Both scenarios are
possible, even recalling that $a_{2A}(\ts_\theta)>a_{2B}(\ts_\theta)$
is a necessary condition to have the TME, because the sign of the
excess kurtoses of samples A and B may be
different. For the case of the
inverse TME, the sign of
$\ts_\KLD-\ts_\theta$ coincides again with that of
$|a_{2A}(\ts_\theta)|-|a_{2B}(\ts_\theta)|$.

The different possibilities analyzed above for the case
$\KLD_A^0>\KLD_B^0$ are summarized in Table~\ref{tab:cases},
specifically as cases labeled ET1, TE1 (for the direct ME)  and ET2,
TE2 (for the inverse ME).

Now we move onto the situation in which the further from equilibrium
sample in the kinetic approach (A) is, however, the closer to equilibrium in the
entropic approach, $\KLD_A^0<\KLD_B^0$. On account of
Eq.~\eqref{eq:KLD^LE}, the condition $\KLD_B^0>\KLD_A^0$ requires
\begin{equation}
\label{eq:locus-KLD}
\KLD_B^{\kin,0}-\KLD_A^{\kin,0}>\frac{d}{2}\left(\theta_A^0-\theta_B^0-\ln\frac{\theta_A^0}{\theta_B^0}
\right)>0.  \end{equation}
Additional cases are possible, which are labeled as
T1, T2, E1, and E2 in Table~\ref{tab:cases}. The TME and the EME are
no longer biunivocally related. For example, in the T1
case, the direct TME is present but no genuine EME takes place:
$\theta_A^0>\theta_B^0>1$ and $\theta_B(\ts)>\theta_A(\ts)>1$ for
$\ts>\ts_\theta$, but $\KLD_B>\KLD_A$ both initially and for
asymptotically long times. Note, however, that this does not prevent
the difference $\KLD_B(\ts)-\KLD_A(\ts)$ from changing its sign an
even number of times during the transient relaxation.
\begin{figure}
    \includegraphics[width=3.25in]{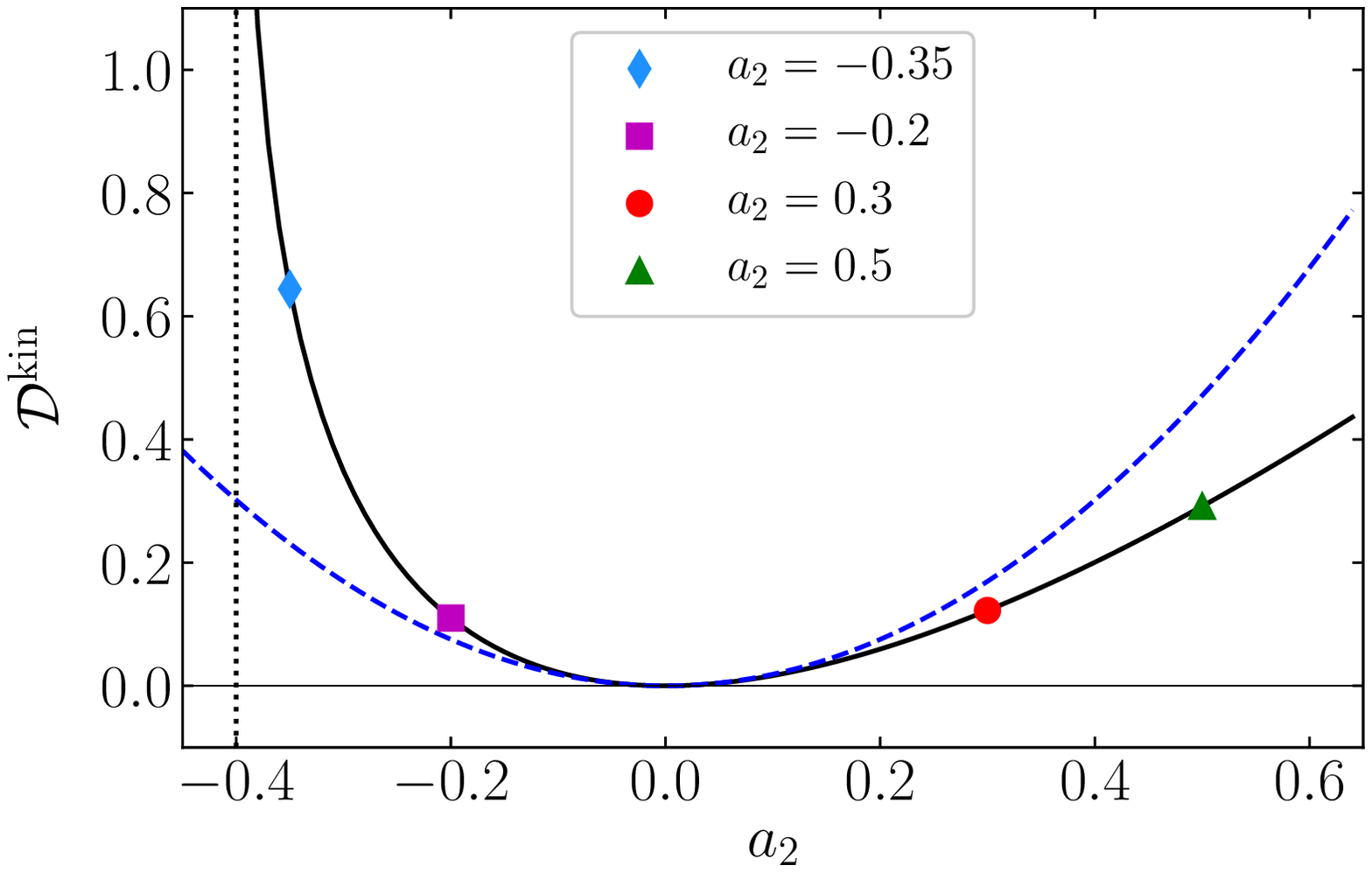}
    \caption{Dependence of $\KLD^{\kin}$  with $a_2$ for a
      gamma distribution. Specifically, we plot the exact expression
      (solid line), given by Eq.~\eqref{eq:deltaKLDgamma}, and the
      small $|a_2|$ approximation (dotted line), given by
      Eq.~\eqref{eq:deltaKLDgamma-approx}, for $d=3$.  Symbols
      correspond to the
      particular values at $a_2=-0.35$, $-0.20$, $0.30$, and $0.50$
      considered in Table \ref{tab:a2Aa2B}.
      \label{fig:delta-KLD}}
\end{figure}

To fix ideas and for further use, let us take the VDF corresponding to
a gamma distribution~\cite{HC78} for the probability density of the variable $x=c^2$. Using the condition $\langle c^2\rangle=\frac{d}{2}$, the reduced VDF associated with the gamma distribution reads
\begin{equation}
\label{eq:Gam-dist}
\phi(\mathbf{c})=\pi^{-d/2}\frac{\Gamma(\frac{d}{2})\vs^{d\vs/2}}{\Gamma(\frac{d\vs}{2})}c^{d(\vs-1)}e^{-\vs c^2},
\quad \vs\equiv \frac{1}{1+\frac{d+2}{2}a_2}.
\end{equation}
Note that this includes the LE distribution, Eq.~\eqref{eq:VDF-LE-reduced}, as the special case $a_2=0$.
Thus, the deviations of the distribution~\eqref{eq:Gam-dist} from LE
are monitored by the excess kurtosis $a_2$ only. In particular, the sixth cumulant is given by
\begin{equation}
\label{a3_Gamma}
a_3=\frac{4}{d+4}a_2\left(1-\frac{d+2}{2}a_2\right).
\end{equation}

The KLD of the gamma distribution with respect to the LE one is~\cite{webP10}
\begin{equation}
\label{eq:deltaKLDgamma}
    \KLD^{\kin} = \frac{d}{2}\left\{\ln\vs+(\vs-1)\left[\psi\left(\frac{d\vs}{2}\right)-1 \right]\right\}+\ln\frac{\Gamma\left(\frac{d}{2}\right)}{\Gamma\left(\frac{d\vs}{2}\right)},
\end{equation}
in which $\psi(x) = {\dif}\ln\Gamma(x)/{\dif x}$ is the digamma
function~\cite{AS72}. For small $|a_2|$, one has
\begin{equation}
\label{eq:deltaKLDgamma-approx}
\KLD^{\kin}\approx
\frac{d(d+2)^2}{16}\left[\frac{d}{2}\psi'\left(\frac{d}{2}\right)-1\right]a_2^2,
\end{equation}
where $\psi'(x)\equiv {\dif}\psi(x)/{\dif x}$.

The dependence of $\KLD^{\kin}$, as given by Eq.~\eqref{eq:deltaKLDgamma}, as a function of $a_2$ in the three-dimensional case ($d=3$) is shown in Fig.~\ref{fig:delta-KLD}. We observe that  $\KLD^{\kin}$ grows more rapidly for negative than for positive values of $a_2$, exhibiting a vertical asymptote at $a_2= -2/(d+2)$, which corresponds to $\vs\to\infty$.

\subsection{Linearized analysis}\label{sec:linear-analysis}

To provide a simple, but yet more quantitative, study, in the remainder of this section (and also in Sec.~\ref{sec:OME}) we adopt the linearization scheme put forward in Ref.~\cite{SP20}. The starting point is the BSA described by Eqs.~\eqref{eq:thetadot} and \eqref{eq:a2-evol-Sonine}, setting $a_3\to 0$ in the latter. Furthermore, the temperature ratio $\theta$ is linearized around a reference value $\theta_r$ close to $\theta^0\equiv \theta(0)$. The solution of the resulting set of two differential equations is~\cite{SP20}
\begin{subequations}
\label{eq:relax-explicit}
\begin{align}
\label{eq:temp-relax-explicit}
\theta(\ts)=&B_{1}+\left[A_{11}\left(\theta^{0}-B_{1}\right)-A_{12} \left(a_{2}^{0}-B_{2}\right)\right] e^{-\lambda_{-}\ts}\nonumber\\
  &-\left[(A_{11}-1)\left(\theta^{0}-B_{1}\right)-A_{12} \left(a_{2}^{0}-B_{2}\right)\right] e^{-\lambda_{+}\ts},
\end{align}
\begin{align}
\label{eq:a2-relax-explicit}
a_2(\ts)=&B_{2}+\left[A_{22}\left(a_{2}^{0}-B_{2}\right)-A_{21}\left(\theta^{0}-B_{1}\right) \right] e^{-\lambda_{-}\ts}\nonumber\\
  &
-\left[(A_{22}-1)\left(a_{2}^{0}-B_{2}\right)-A_{21}\left(\theta^{0}-B_{1}\right) \right] e^{-\lambda_{+}\ts},
\end{align}
\end{subequations}
in which $a_2^0\equiv a_2(0)$, and the expressions of the parameters $\lambda_\pm$, $B_i$, and $A_{ij}$ can be found in Appendix~\ref{appB}. We refer to Eqs.~\eqref{eq:relax-explicit} as the linearized basic Sonine approximation (LBSA). When using the LBSA to investigate the ME, we are assuming that $\theta(\ts)$ is close to $\theta_r$, which in turn is close to $\theta^0$. This entails that the LBSA is expected to be applicable to the kinetic stage only---i.e., when the ME comes about for short times.

The LBSA can be applied to the evolution of the two samples A and B with the convenient choice $\theta_r=\theta_B^0$~\cite{SP20}. It is then straightforward to find the crossover time $\ts_\theta$ as
\begin{equation}
\label{t_theta}
\ts_\theta=\frac{1}{\lambda_+-\lambda_-}\ln\left(1+\frac{A_{11}^{-1}}{R^0_{\max}/R^0-1}\right),
\end{equation}
where
\begin{equation}
\label{eq:t_theta2}
R^0\equiv\frac{\theta_A^0-\theta_B^0}{a_{2A}^0-a_{2B}^0},\quad R^0_{\max}\equiv\frac{A_{12}}{A_{11}}.
\end{equation}
Therefore, in the LBSA, the crossover time $\ts_\theta$ depends on the set of four initial values $\theta_A^0$, $a_{2A}^0$, $\theta_B^0$, and $a_{2B}^0$ only through the ratio $R^0$. Moreover, Eq.~\eqref{t_theta} is meaningful only if
\begin{equation}
\label{condition-TME}
0<R^0<R^0_{\max}.
\end{equation}
Otherwise, no TME---either direct or inverse---exists.

The determination of the crossover time $\ts_\KLD$ is much more
involved, even in the simple LBSA. It is obtained as the solution of a
transcendental equation and the solution depends on $\theta_A^0$,
$a_{2A}^0$, $\theta_B^0$, and $a_{2B}^0$. The locus separating the
region where $\ts_\KLD<\ts_\theta$ from the region where
$\ts_\KLD>\ts_\theta$ is approximately given by the condition
$|a_{2A}(\ts_\theta)|=|a_{2B}(\ts_\theta)|$; the sign of
$\ts_\KLD-\ts_\theta$ is the same as that of
$|a_{2A}(\ts_\theta)|-|a_{2B}(\ts_\theta)|$---as discussed in the
previous section.
See cases ET1, TE1, ET2, and TE2 in Table~\ref{tab:cases}. Furthermore, cases T1, T2, E1, and E2 are possible if the initial values of the KLD cross the locus $\KLD^0_A=\KLD^0_B$, as summarized in Table~\ref{tab:cases} and described in Sec.~\ref{sec:TME_EME_A}.
If the locus  $\KLD^0_A=\KLD^0_B$ happens to separate regions ET1 and T1 (or ET2 and T2), then one has $\ts_\KLD\to 0$ on the locus, so that $0<\ts_\KLD<\ts_\theta$ in region ET1 (or ET2) and formally $\ts_\KLD<0<\ts_\theta$ in region T1 (or T2).

\begin{table}
\caption{Four representative choices for the initial values $a_{2A}^0$ and $a_{2B}^0$ ($d=3$). The numerical values of $\KLD^\kin$, as given by Eq.~\eqref{eq:deltaKLDgamma} with $d=3$, are also included. The sixth column gives the cases (see Table \ref{tab:cases}) that, in principle, are associated with each pair $(a_{2A}^0,a_{2B}^0)$. However, some of them (enclosed in parentheses) are not actually observed (see Fig.~\ref{fig:phase_diag}). \label{tab:a2Aa2B}}
\begin{ruledtabular}
\begin{tabular}{cccccc}
Label&$a_{2A}^0$&$a_{2B}^0$&$\KLD^{\kin,0}_A$&$\KLD^{\kin,0}_B$&Cases\\
\hline
I&$0.50$&$-0.35$&$0.292$&$0.644$&ET1, (TE1), T1, E1, E2\\
II&$0.50$&$-0.20$&$0.292$&$0.110$&ET1, TE1\\
III&$-0.35$&$0.30$&$0.644$&$0.122$&ET2, TE2\\
IV&$-0.20$&$0.50$&$0.110$&$0.292$&(ET2), (TE2), T2, E1, E2\\
\end{tabular}
\end{ruledtabular}
\end{table}

\begin{figure*}
\includegraphics[width=7.in]{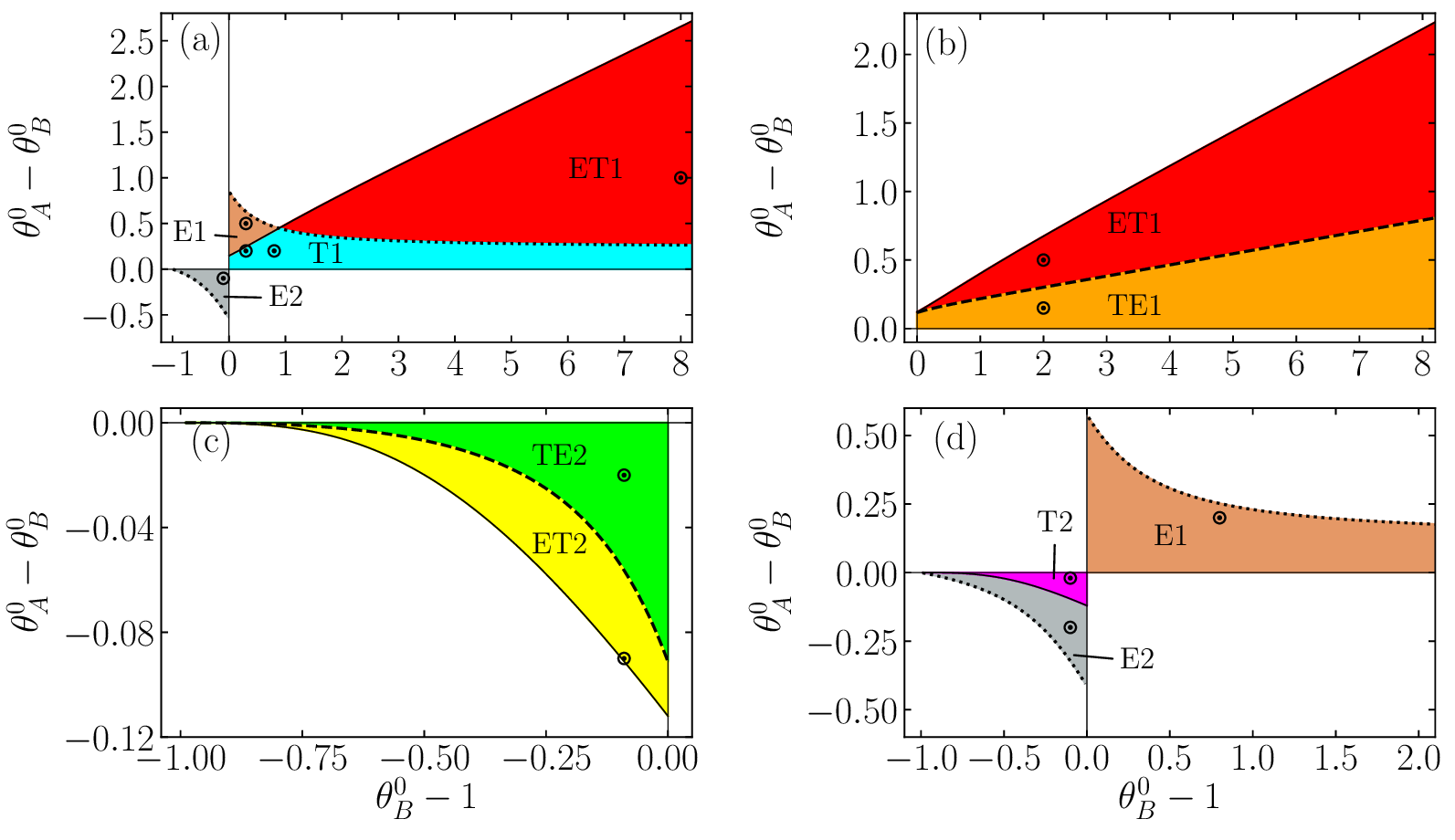}
\caption{Phase diagrams in the representation $\theta_A^0-\theta_B^0$
  vs $\theta_B^0-1$. Specifically, they are plotted for $\zs=1$, $\gamma=0.1$, $d=3$, and the four representative choices of $(a_{2A}^0,a_{2B}^0)$ displayed in Table \ref{tab:a2Aa2B}: (a) I, (b) II, (c) III, and (d) IV. The solid, dashed, and dotted lines represent the loci $R^0=R^0_{\max}$, $\ts_\theta=\ts_\KLD$, and $\KLD_A^0=\KLD^0_B$, respectively. The labels in each region correspond to the cases described in Table \ref{tab:cases} and the  circles represent the specific examples considered in Figs.~\ref{fig:I}--\ref{fig:IV}. Note that $\theta_A^0>\theta_B^0>1$ and $\theta_A^0<\theta_B^0<1$ refer to the direct TME and inverse TME, respectively.
}
\label{fig:phase_diag}
\end{figure*}
\begin{figure}
    \includegraphics[width=0.4\textwidth]{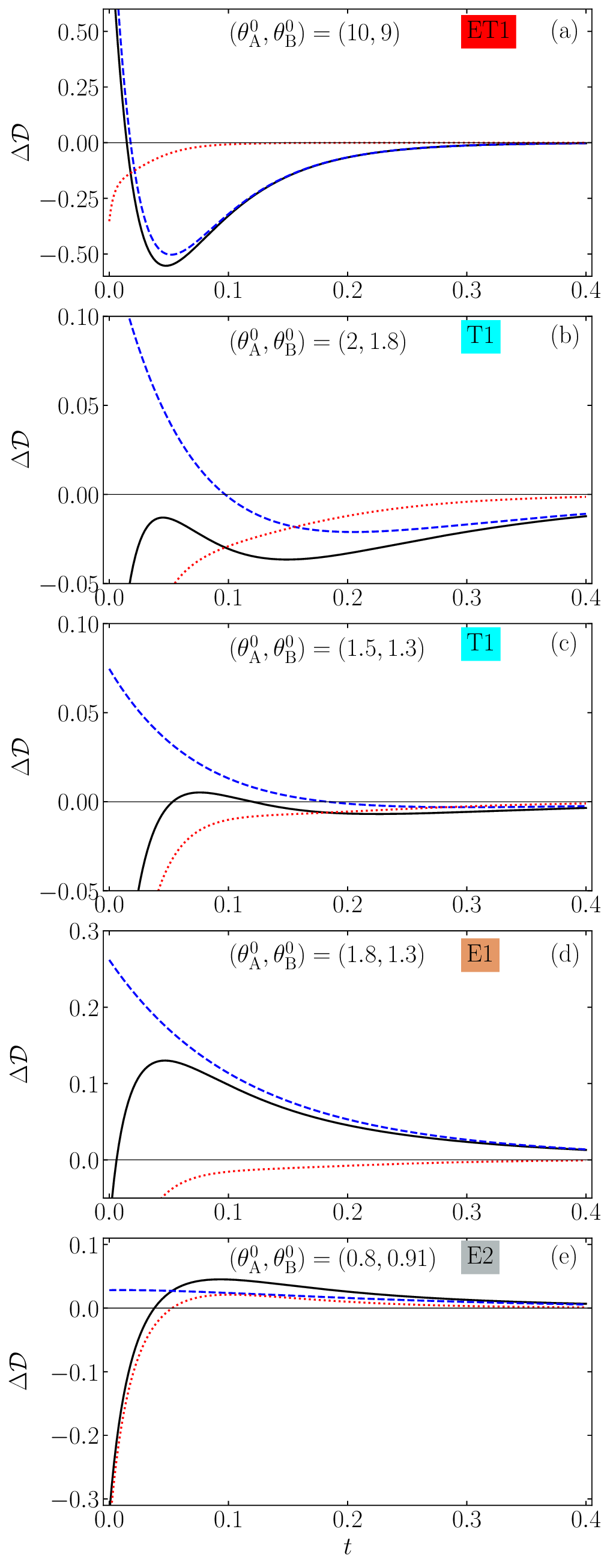}
       \caption{Time evolution of the difference of the relative entropies for the representative initial condition I  in Table~\ref{tab:a2Aa2B},  $(a_{2A}^0,a_{2B}^0)=(0.50,-0.35)$.  Parameter values are $\zs=1$, $\gamma=0.1$, $d=3$. Specifically, we represent $\KLD_A^{\text{kin}}-\KLD_B^{\text{kin}}$ (dotted lines), $\KLD_A^\LE-\KLD_B^\LE$ (dashed lines), and $\KLD_A-\KLD_B$ (solid lines) for different pairs of initial temperatures, namely: (a) $(\theta_B^0-1,\theta_A^0-\theta_B^0)=(8,1)$, (b) $(\theta_B^0-1,\theta_A^0-\theta_B^0)=(0.8,0.2)$,  (c) $(\theta_B^0-1,\theta_A^0-\theta_B^0)=(0.3,0.2)$, (d) $(\theta_B^0-1,\theta_A^0-\theta_B^0)=(0.3,0.5)$, and (e) $(\theta_B^0-1,\theta_A^0-\theta_B^0)=(-0.09,-0.11)$. Panels (a), (b), (c), (d) and (e) represent examples of cases ET1, T1, T1, E1, and E2, respectively [see Fig.~\ref{fig:phase_diag}(a)].
    \label{fig:I}}
\end{figure}
\begin{figure}
    \includegraphics[width=0.4\textwidth]{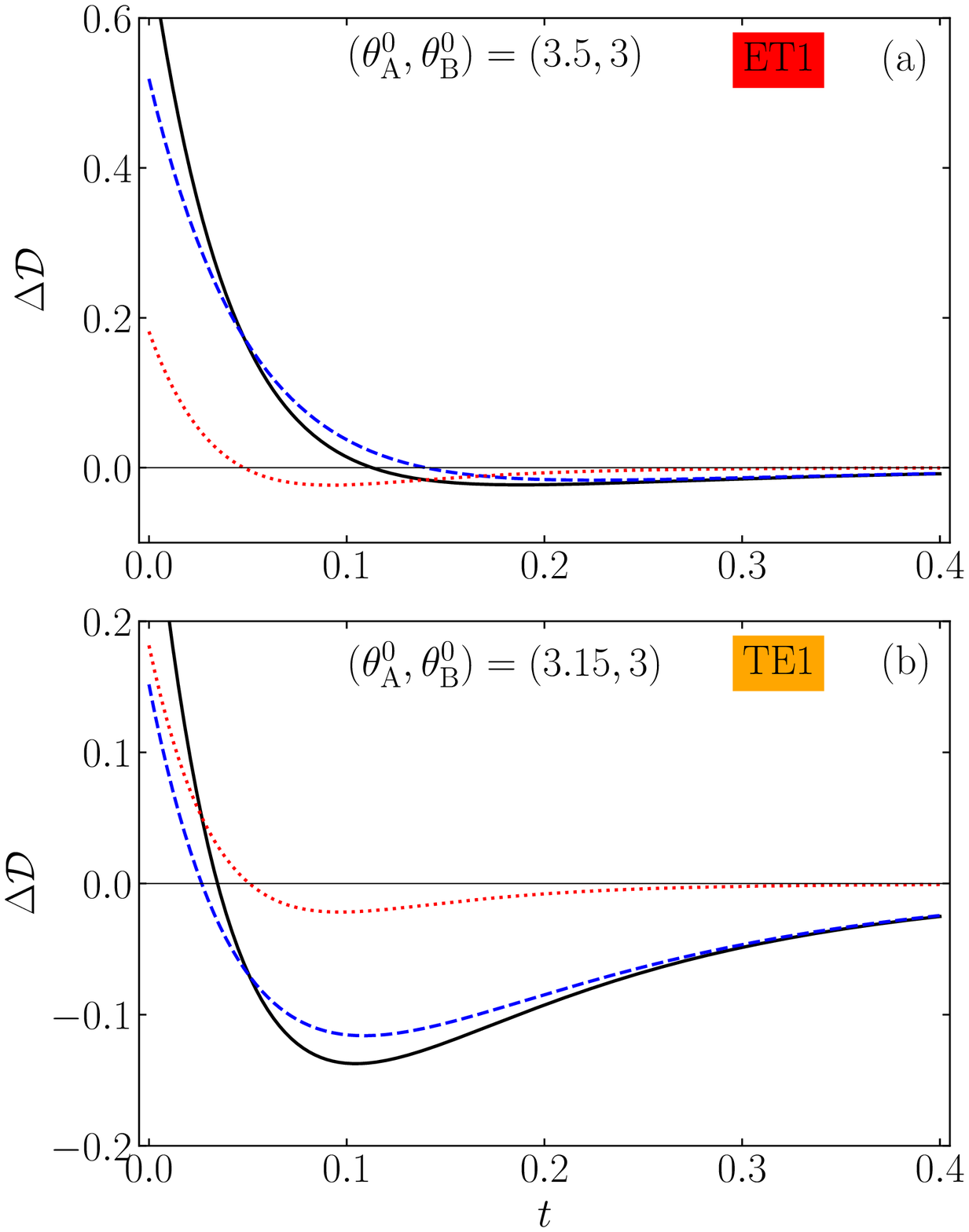}
        \caption{Same as in Fig.~\ref{fig:I}, but now for the
          representative initial condition II in Table~\ref{tab:a2Aa2B},  $(a_{2A}^0,a_{2B}^0)=(0.50,-0.20$. Here, initial conditions for the temperatures are: (a) $(\theta_B^0-1,\theta_A^0-\theta_B^0)=(2.0,0.5)$ and (b) $(\theta_B^0-1,\theta_A^0-\theta_B^0)=(2.00,0.15)$. Panels (a) and (b) represent examples of cases ET1 and TE1, respectively [see Fig.~\ref{fig:phase_diag}(b)].
    \label{fig:II}}
\end{figure}
\begin{figure}
    \includegraphics[width=0.4\textwidth]{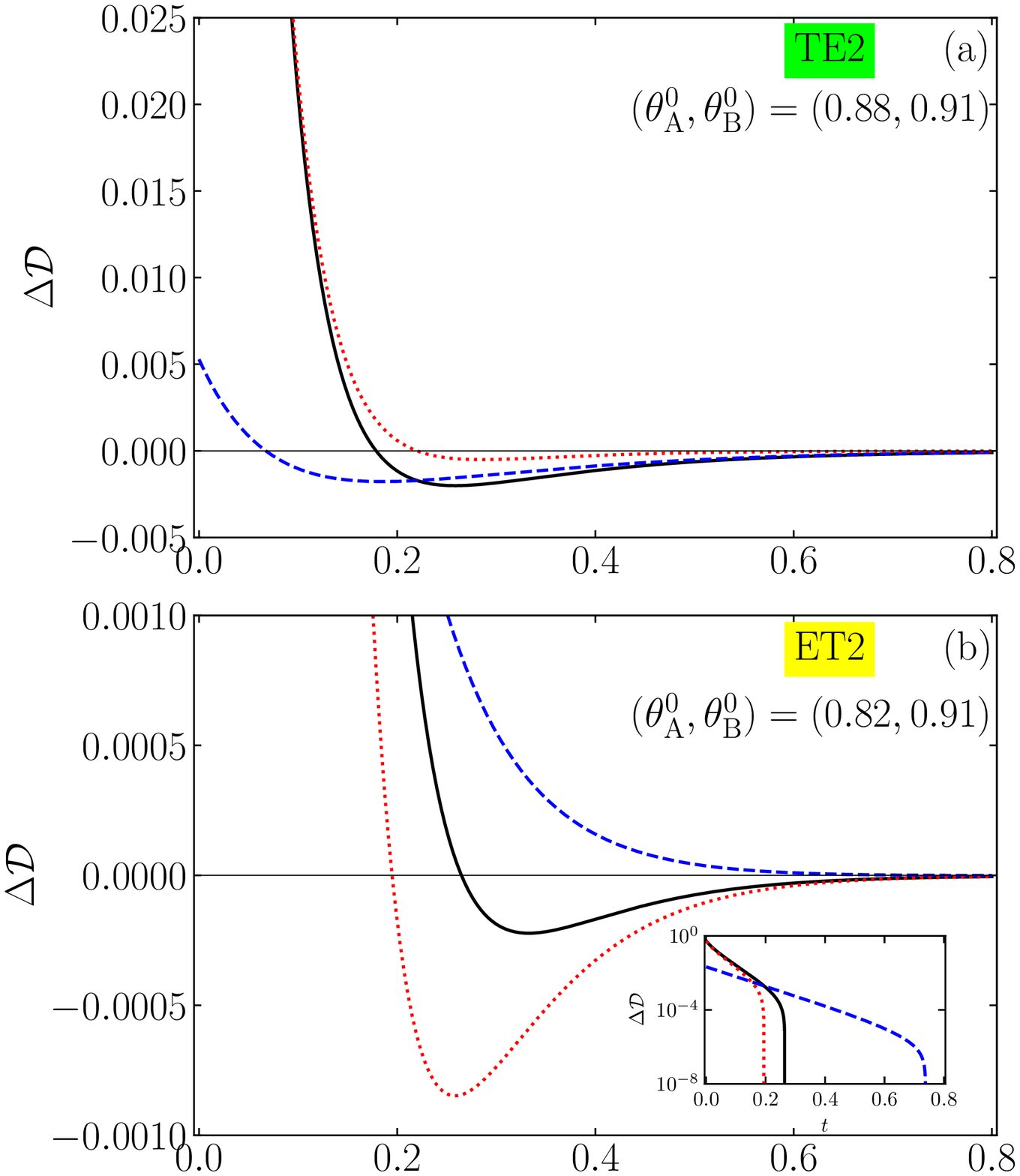}
    \caption{Same as in Fig.~\ref{fig:I}, but now for the representative initial condition III in Table~\ref{tab:a2Aa2B},  $(a_{2A}^0,a_{2B}^0)=(-0.35,0.30)$. In this case, the initial conditions for the temperatures chosen are: (a) $(\theta_B^0-1,\theta_A^0-\theta_B^0)=(-0.09,-0.03)$ and (b) $(\theta_B^0-1,\theta_A^0-\theta_B^0)=(-0.09,-0.09)$. Panels (a) and (b) represent examples of cases TE2 and ET2, respectively [see Fig.~\ref{fig:phase_diag}(c)]. {Note that $\Delta\KLD$ is plotted in logarithmic scale in the inset of panel (b) to favor the perception of the crossover times, at which $\Delta\KLD$ vanishes.}
    \label{fig:III}}
\end{figure}
\begin{figure}
    \includegraphics[width=0.4\textwidth]{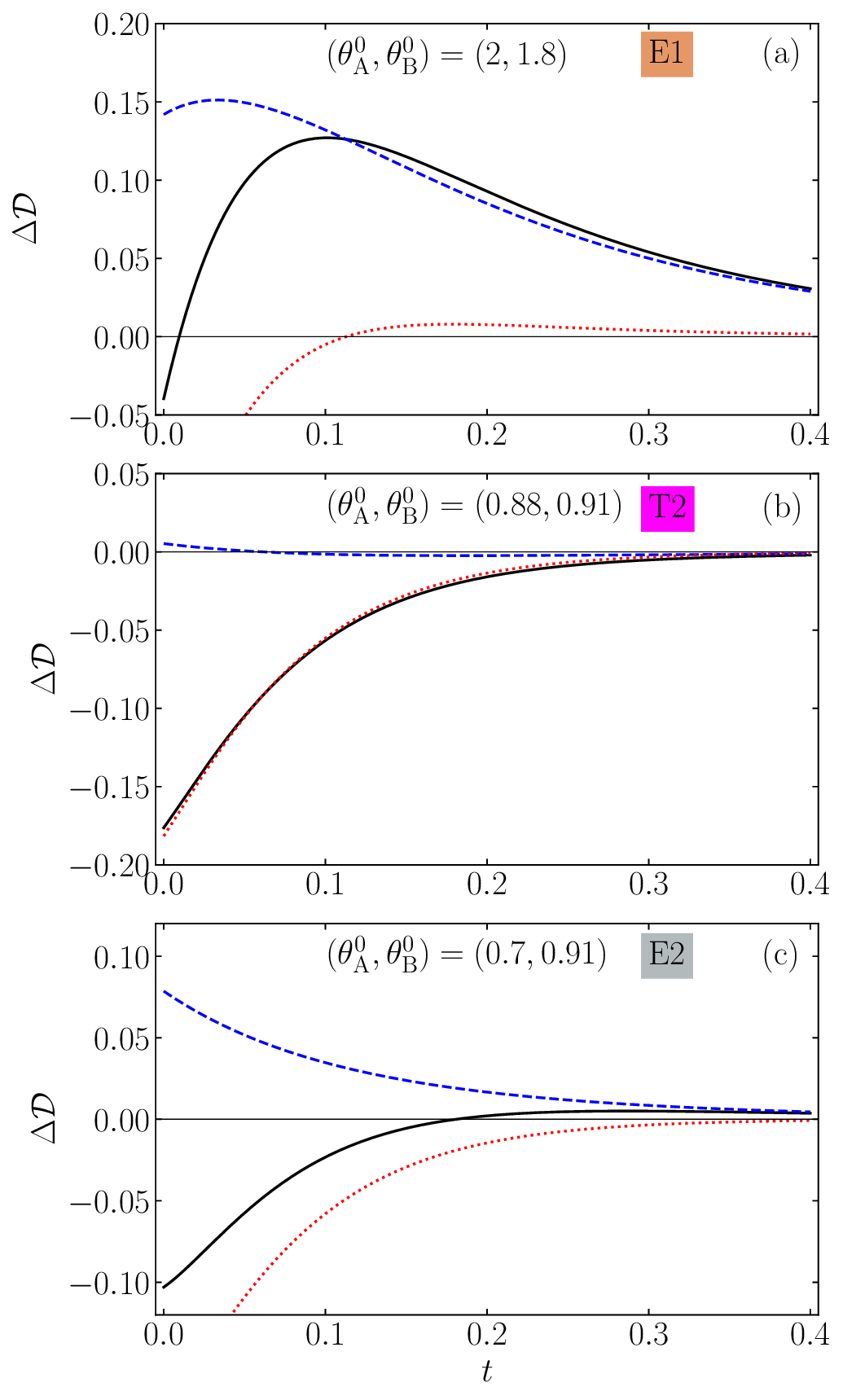}
    \caption{Same as in Fig.~\ref{fig:I}, but now for the representative initial condition IV in Table~\ref{tab:a2Aa2B},  $(a_{2A}^0,a_{2B}^0)=(-0.20,0.50)$. Here, initial temperatures  are: (a) $(\theta_B^0-1,\theta_A^0-\theta_B^0)=(0.8,0.2)$, (b) $(\theta_B^0-1,\theta_A^0-\theta_B^0)=(-0.09,-0.03)$, and (c) $(\theta_B^0-1,\theta_A^0-\theta_B^0)=(-0.09,-0.21)$. Panels (a), (b), and (c) represent examples of cases E1, T2, and E2, respectively [see Fig.~\ref{fig:phase_diag}(d)].
    \label{fig:IV}}
\end{figure}

\subsection{Illustrative examples}\label{sec:illustr-examp}

Let us choose the four representative pairs $(a_{2A}^0,a_{2B}^0)$
presented in Table~\ref{tab:a2Aa2B}. Since the scenarios ET1 and TE1
described in Table~\ref{tab:cases} require $a_{2A}^0>a_{2B}^0$, they
are \emph{in principle} feasible for the pairs I and II. Analogously,
the scenarios ET2 and TE2 might be  possible for the pairs III and
IV. Next, by assuming the initial VDF has the gamma form, Eq.~\eqref{eq:Gam-dist}, we have $\KLD^{\kin,0}_B>\KLD^{\kin,0}_A$ for the
pairs I and IV, but not for the pairs II and III; in view of
Eq.~\eqref{eq:locus-KLD}, we conclude that, \emph{in principle}, cases
T1, E1, and E2 are possible for pair I and cases T2, E1, and E2 for pair IV.

The phase diagrams predicted by the LBSA are shown in Fig.~\ref{fig:phase_diag} for $\zs=1$, $\gamma=0.1$, and  $d=3$.  We observe that, at least for that choice of the parameters, case TE1 is absent for the class of initial conditions I, while cases ET2 and TE2 are absent for the class of initial conditions IV.
This confirms that the initial conditions shown in the third column of Table \ref{tab:cases} for the cases ET1, TE1, ET2, and TE2 represent necessary---but not sufficient---conditions for their occurrence, the actual realization of those scenarios depending on the evolution of $a_{2A}(t)$ and $a_{2B}(t)$.

The time evolution of the differences
\begin{equation}
    \label{eq:KLD-differences}
    \Delta\KLD\equiv \KLD_A-\KLD_B, \quad \Delta\KLD^\LE\equiv \KLD_A^\LE-\KLD_B^\LE
\end{equation}
for the representative points indicated in Fig.~\ref{fig:phase_diag}
are displayed in Figs.~\ref{fig:I}--\ref{fig:IV}, where the difference $\Delta\mathcal{D}^{\text{kin}}=\mathcal{D}^{\text{kin}}_A-\mathcal{D}^{\text{kin}}_B=\Delta\mathcal{D}-\Delta\mathcal{D}^{\text{LE}}$ is also included. The change of sign
of $\KLD_A^\LE-\KLD_B^\LE$  and $\KLD_A-\KLD_B$ during their evolution
signals the presence of the TME and EME, respectively. Here, we made
an extra ansatz to evaluate $\KLD^{\kin}(\ts)$. As we are working with
an initial gamma distribution and the final equilibrium state is a
particular case of such a distribution---with $a_2=0$, we have assumed
that the VDF during its time evolution is sufficiently close to a
gamma distribution so as to estimate $\KLD^{\kin}(\ts)$  by Eq.~\eqref{eq:deltaKLDgamma} with an excess kurtosis $a_2(\ts)$ given by Eq.~\eqref{eq:a2-relax-explicit}.

Figures~\ref{fig:I}(b) and \ref{fig:I}(c) are both examples of the scenario T1 for the class of initial conditions I. In Fig.~\ref{fig:I}(b), where $(\theta_B^0-1,\theta_A^0-\theta_B^0)=(0.8,0.2)$ [see Fig.~\ref{fig:phase_diag}(a)], the difference $\KLD_A^0-\KLD_B^0$ presents a negative local maximum. When moving horizontally in Fig.~\ref{fig:phase_diag}(a) to the point $(\theta_B^0-1,\theta_A^0-\theta_B^0)=(0.3,0.2)$, however, Fig.~\ref{fig:I}(c) shows that the local maximum of $\KLD_A^0-\KLD_B^0$ becomes positive and $\KLD_A^0-\KLD_B^0$ vanishes twice during the time evolution. While interesting, this does not qualify as an EME because, as already said above Eq.~\eqref{eq:locus-KLD}, $\KLD_B>\KLD_A$ both initially and for asymptotically long times. Next, moving vertically in Fig.~\ref{fig:phase_diag}(a) to the point $(\theta_B^0-1,\theta_A^0-\theta_B^0)=(0.3,0.5)$, the local maximum observed in Fig.~\ref{fig:I}(d) is again positive but there is a single crossing $\KLD_A^0-\KLD_B^0=0$, which results in the E1 scenario.

\section{Overshoot Mpemba effect}
\label{sec:OME}

In Sec.~\ref{sec:TME_EME}, we have assumed that, even though the evolution of $\theta(\ts)$ may not be monotonic, $\theta(\ts)-1$ does not change sign, i.e., the temperature does not overshoot the equilibrium value. However, such an overshoot $\theta(\ts_O)=1$ at a finite time $\ts_O$ is possible. In general, $a_2(\ts_O)\neq 0$, and Eq.~\eqref{eq:thetadot} shows that $\dot{\theta}(\ts_O)/\zs\gamma=-2(d+2) a_2(\ts_O)\neq 0$. As a consequence,  starting from $\theta^0>1$, $\theta(\ts)-1$ develops a hump with a negative minimum if $a_2(\ts_O)>0$; analogously, starting from $\theta^0<1$ and if $a_2(\ts_O)<0$, $\theta(\ts)-1$ develops a hump with a positive maximum, reminiscent of the Kovacs effect~\cite{K63,KAHR79,PB10,BL10,PT14,TP14,LVPS19,PSP21}. We will refer to this crossover $\theta(\ts_O)=1$ and subsequent hump, either positive or negative, as an overshoot phenomenon.

Given the fact that the relaxation of $a_2(\ts)$ is generally much faster than that of $\theta(\ts)$, at least if $\theta^0=O(1)$ \cite{PSP21}, it is reasonable to expect that the overshoot effect requires initial values $|\theta^0-1|\ll 1$, unless  $|a_2^0|$ is unphysically large. This suggests a theoretical treatment based on the LBSA \eqref{eq:relax-explicit} with $\theta_r\to 1$, i.e.,
\begin{subequations}
\label{eq:relax-equilibrium}
\begin{align}
\label{eq:relax-equilibrium1}
    \theta(\ts) =& 1 +\left[\bar{A}_{11}\left(\theta^0-1 \right)-\bar{A}_{12}a_2^0 \right]e^{-\bar{\lambda}_{-}\ts}\nonumber \\
    &- \left[\left(\bar{A}_{11}-1\right)\left(\theta^0-1 \right)-\bar{A}_{12}a_2^0 \right]e^{-\bar{\lambda}_{+}\ts},
\end{align}
\begin{align}
    a_2(\ts) =& \left[\bar{A}_{22}a_2^0-\bar{A}_{21}\left(\theta^0-1 \right) \right]e^{-\bar{\lambda}_{-}\ts}\nonumber \\
    &- \left[\left(\bar{A}_{22}-1\right)a_2^0-\bar{A}_{21}\left(\theta^0-1 \right)\right]e^{-\bar{\lambda}_{+}\ts},
\end{align}
\end{subequations}
where overlined quantities refer to their values at $\theta_r=1$. Following the same methodology as in Eqs.~\eqref{t_theta} and \eqref{eq:t_theta2}, we find
\begin{equation}
\label{t_O}
\ts_O=\frac{1}{\bar{\lambda}_+-\bar{\lambda}_-}\ln\left[1+\frac{\bar{A}_{11}^{-1}}{\bar{R}^0_{\max}{a_{2}^0}/(\theta^0-1)-1}\right],
\end{equation}
where
\begin{equation}
\label{eq:t_O2}
\bar{R}^0_{\max}\equiv\frac{\bar{A}_{12}}{\bar{A}_{11}}.
\end{equation}
Therefore, according to the LBSA, the overshoot effect appears if
\begin{equation}
\label{eq:condition-Kovacs}
    0<\frac{\theta^0-1}{a_{2}^0}<\bar{R}^0_{\max}.
\end{equation}

It is interesting to look into the possible change of the ME phenomenology brought about by the overshoot-induced humps. As we show below, the existence of humps may make it necessary to change the preconception of considering the TME present only when the evolution curves of the temperature of the two samples intersect.  To be more specific, we consider, as before, samples A and B with A being the initially hotter, i.e., $\theta^0_A>\theta^0_B>1$.   Let us assume that the colder sample B fulfills condition~\eqref{eq:condition-Kovacs} but the hotter sample does not. In that case, $\theta_B(\ts)$  might not be crossed by the  curve $\theta_A(\ts)$, which remains always above the equilibrium temperature, and yet relax more slowly to equilibrium than A but from below. We could then say that  a (direct) TME is present without the existence of a standard crossover time $\ts_\theta$, provided that  a crossover between the LE KLD curves $\KLD^{\mathrm{LE}}_A(\ts)$ and $\KLD^{\mathrm{LE}}_B(\ts)$ occurs at a certain time $\ts_{\KLD^{\mathrm{LE}}}$. A completely analogous situation is possible for the inverse ME, i.e., when $\theta^0_A<\theta^0_B<1$.  We will refer to this phenomenon, where $\KLD^{\mathrm{LE}}_A$ and $\KLD^{\mathrm{LE}}_B$ intersect but $\theta_A(\ts)$ and $\theta_B(\ts)$ do not, as the overshoot ME (OME).
This phenomenon is reminiscent of the ME observed in Ref.~\cite{B11} in supercooled water.
\begin{figure}
    \includegraphics[width=0.47\textwidth]{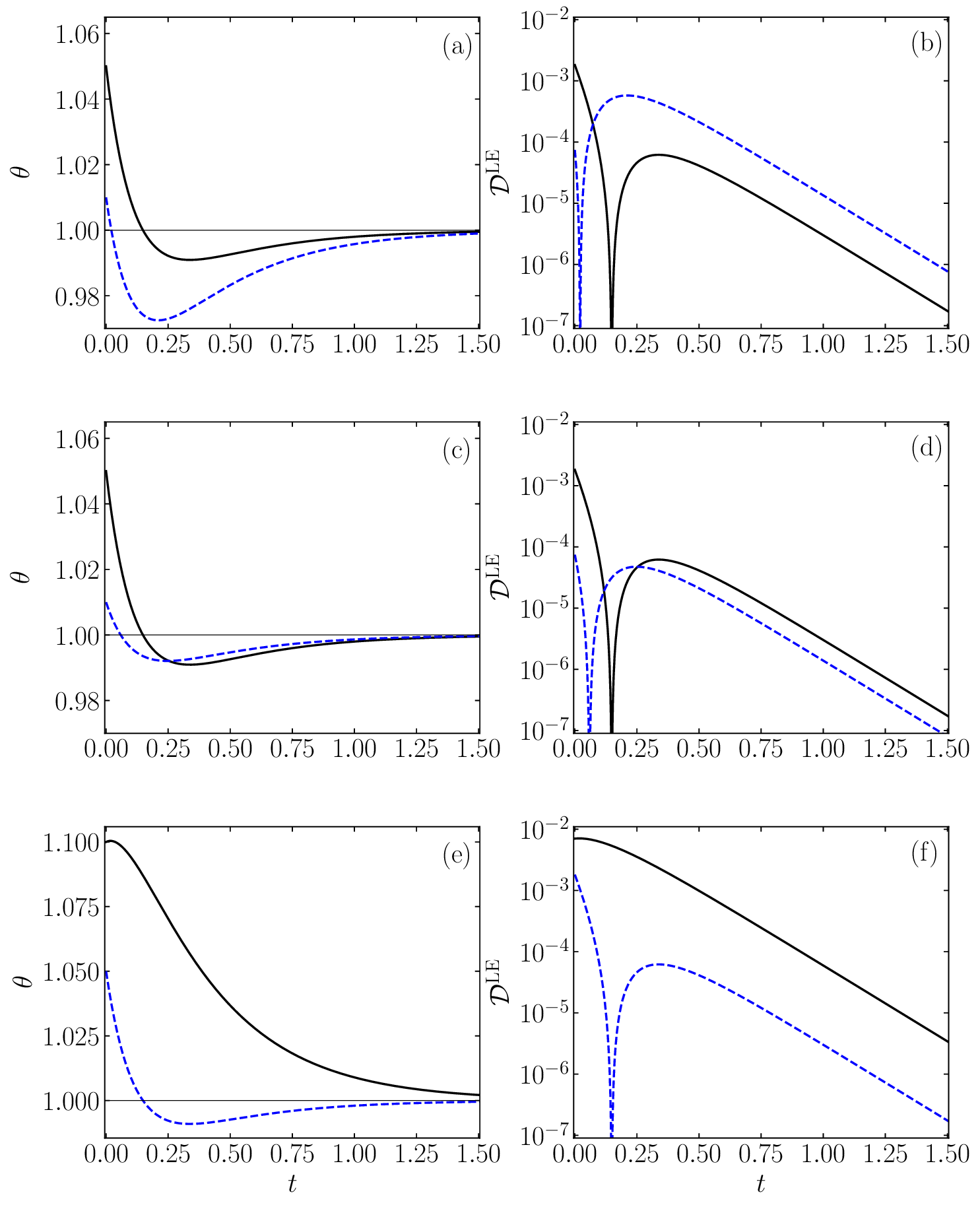}
    \caption{Time evolution of the temperature and LE relative entropy of samples A and B for the direct case, $\theta_A^0>\theta_B^0>1$. Specifically, we show  $\{\theta_A,\KLD^\LE_A\}$ (solid lines) and $\{\theta_B,\KLD^\LE_B\}$ (dashed lines), as obtained from Eq.~\eqref{eq:relax-equilibrium1}. Initial conditions are [(a) and (b)] $(\theta_A^0,\theta_B^0)=(1.05,1.01)$ and $(a_{2A}^0,a_{2B}^0)=(0.5,0.5)$, [(c) and (d)] $(\theta_A^0,\theta_B^0)=(1.05,1.01)$ and $(a_{2A}^0,a_{2B}^0)=(0.5,0.2)$, and [(e) and (f)] $(\theta_A^0,\theta_B^0)=(1.1,1.05)$ and $(a_{2A}^0,a_{2B}^0)=(-0.35,0.5)$. In all cases, $\zs=1$, $\gamma=0.1$, and $d=3$.}
    \label{fig:OME_cases}
\end{figure}
\begin{figure}
    \includegraphics[width=0.47\textwidth]{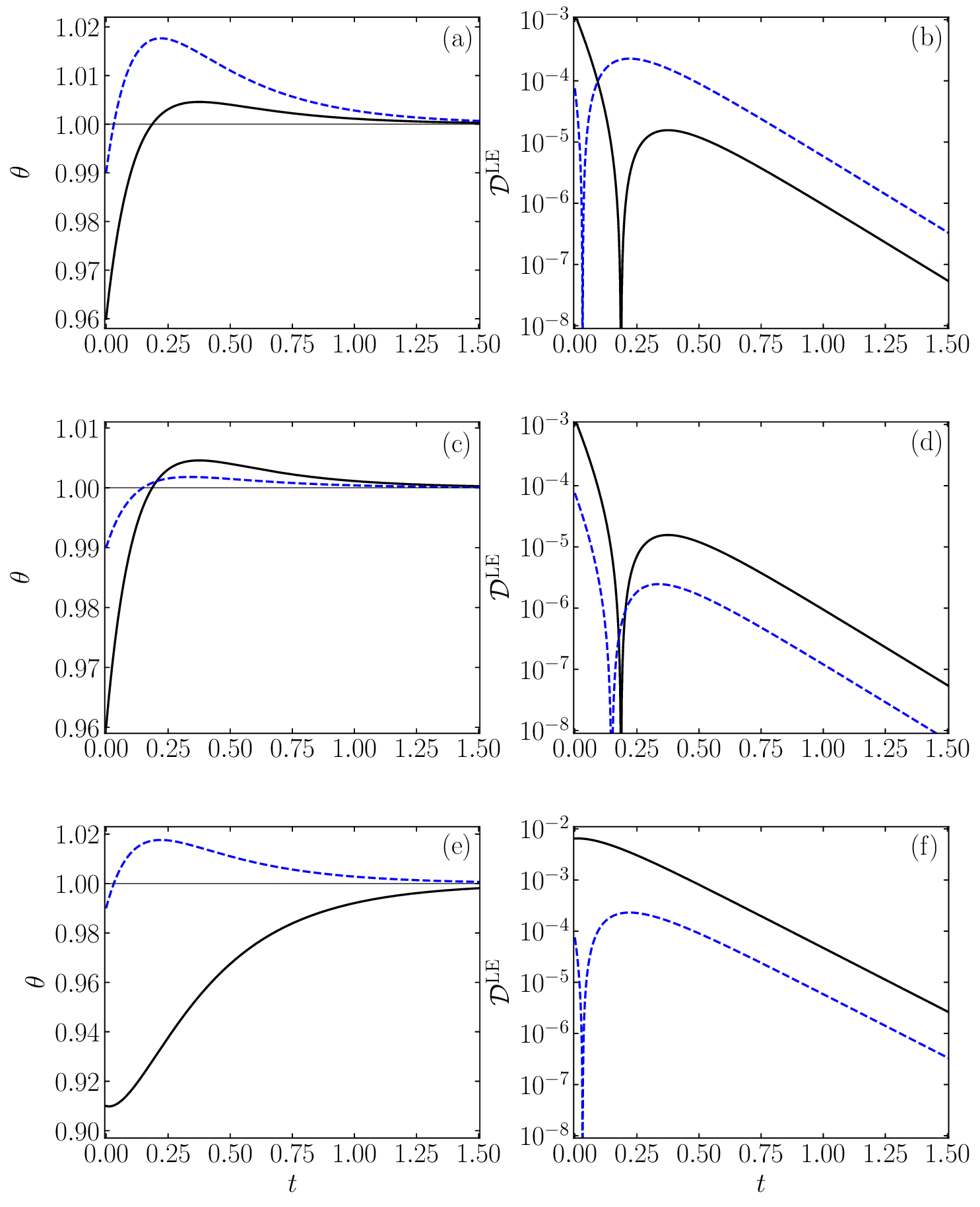}
    \caption{Same as in Fig.~\ref{fig:OME_cases}, but now for the inverse case $\theta_A^0<\theta_B^0<1$.  Initial conditions are  [(a) and (b)] $(\theta_A^0,\theta_B^0)=(0.96,0.99)$ and $(a_{2A}^0,a_{2B}^0)=(-0.35,-0.35)$, [(c) and (d)] $(\theta_A^0,\theta_B^0)=(0.96,0.99)$ and $(a_{2A}^0,a_{2B}^0)=(-0.35,-0.1)$, and [(e) and (f)] $(\theta_A^0,\theta_B^0)=(0.91,0.99)$ and $(a_{2A}^0,a_{2B}^0)=(0.3,-0.35)$.}
    \label{fig:OME_cases_b}
\end{figure}

The different scenarios where overshoot-induced humps appear are illustrated in Fig.~\ref{fig:OME_cases}  for  direct  preparations,  i.e., $\theta_A^0>\theta_B^0>1$. In Fig.~\ref{fig:OME_cases}(a), $\theta_A(\ts)$ and $\theta_B(\ts)$ do not cross each other, but they both exhibit  humps, the one in system B being stronger than in system A. This makes the latter system relax to equilibrium earlier than the former, which physically qualifies as a direct TME. While in the thermal scheme there is no crossing, the positiveness of $\KLD^{\mathrm{LE}}$ forces an intersection between A and B curves, as observed in Fig.~\ref{fig:OME_cases}(b). This is the essence of the OME.

On the other hand, the existence of humps or of a finite crossover
time $\ts_\theta$  does not ensure the existence of TME. In fact, in
Fig.~\ref{fig:OME_cases}(c) there is a crossing between the thermal
curves, but the overshoot-induced humps make  the initially hotter system
relax later to the equilibrium state, thus frustrating the TME. This
is signaled by a pair of intersections in the $\KLD^{\mathrm{LE}}$
curves of Fig.~\ref{fig:OME_cases}(d), so the OME is absent. The
third different scenario is reflected in Figs.~\ref{fig:OME_cases}(e)
and \ref{fig:OME_cases}(f), where there is no crossover either in
the thermal evolution or in $\KLD^{\mathrm{LE}}$, even though sample B exhibits a thermal hump.

The analogous cases for inverse preparations $\theta_A^0<\theta_B^0<1$ are illustrated in Fig.~\ref{fig:OME_cases_b}.

To summarize, the OME is characterized by a single crossing $\KLD^\LE_A=\KLD^\LE_B$ at a certain time $\ts_{\KLD^\LE}$,  without any crossing between $\theta_A$ and $\theta_B$. In order to establish the conditions under which this may happen, let us assume again $|\theta-1|\ll 1$ and approximate $\ln \theta \approx \theta-1 -\frac{1}{2}(\theta-1)^2$ in Eq.~\eqref{eq:KLD^LE}. Therefore, the condition $\KLD^\LE_A(\ts_{\KLD^\LE})=\KLD^\LE_B(\ts_{\KLD^\LE})$ with $\theta_A(\ts_{\KLD^\LE})\neq\theta_B(\ts_{\KLD^\LE})$ translates into
\begin{equation}
\label{eq:KLD_LE_COND}
    \theta_A(\ts_{\KLD^\LE})-1= 1-\theta_B(\ts_{\KLD^\LE}).
\end{equation}
Making use of  Eq.~\eqref{eq:relax-equilibrium1} entails
\begin{equation}
\label{t_KLD_LE}
\ts_{\KLD^\LE}=\frac{1}{\bar{\lambda}_+-\bar{\lambda}_-}\ln\left(1+\frac{\bar{A}_{11}^{-1}}{\bar{R}^0_{\max}/R_+^0-1}\right),
\end{equation}
where $\bar{R}^0_{\max}$ is defined in Eq.~\eqref{eq:t_O2} and
\begin{equation}
\label{R0+}
R_+^0\equiv\frac{\theta^0_A+\theta_B^0-2}{a_{2A}^0+a_{2B}^0}.
\end{equation}
Note the difference between this parameter $R_+^0$ and the parameter $R^0$ defined before in Eq.~\eqref{eq:t_theta2}. Since $\ts_{\KLD^\LE}$ must be finite in the OME, the corresponding condition on the initial preparation is
\begin{subequations}
\label{eq:OME-condition}
\begin{equation}
\label{eq:OME-condition1}
    0<R^0_{+}<\bar{R}^0_{\max},
\end{equation}
\begin{equation}
\label{eq:OME-condition2}
R^0<0 \text{ or } R^0>\bar{R}^0_{\max}.
\end{equation}
\end{subequations}
The supplementary condition \eqref{eq:OME-condition2} represents the violation of Eq.~\eqref{condition-TME} (with $\theta_r\to 1$) and is needed to exclude any thermal crossing.

According to Eq.~\eqref{eq:condition-Kovacs}, if  both systems A and B exhibit overshoot-induced humps, the condition given by Eq.~\eqref{eq:OME-condition1} is ensured. As a test, note that $R_{\max}^0=0.172$ for all the cases considered in Figs.~\ref{fig:OME_cases} and \ref{fig:OME_cases_b}. The values of $(R_+^0,R^0)$ are $(0.060,\infty)$, $(0.086,0.133)$, and $(1, -0.059)$ in the cases represented in Figs.~\ref{fig:OME_cases}(a) and \ref{fig:OME_cases}(b), Figs.~ \ref{fig:OME_cases}(c) and \ref{fig:OME_cases}(d), and Figs.~\ref{fig:OME_cases}(e) and \ref{fig:OME_cases}(f), respectively. Analogously, $(R_+^0,R^0)$ are $(0.071,\infty)$, $(0.111,0.120)$, and $(2, -0.123)$ in the cases represented in Figs.~\ref{fig:OME_cases_b}(a) and \ref{fig:OME_cases_b}(b), Figs.~\ref{fig:OME_cases_b}(c) and \ref{fig:OME_cases_b}(d), and Figs.~\ref{fig:OME_cases_b}(e) and \ref{fig:OME_cases_b}(f), respectively. Thus, the OME double condition \eqref{eq:OME-condition} is fulfilled only in the cases (a) and (b) of Figs.~\ref{fig:OME_cases} and \ref{fig:OME_cases_b}.

\begin{figure}[htb]
    \includegraphics[width=\columnwidth]{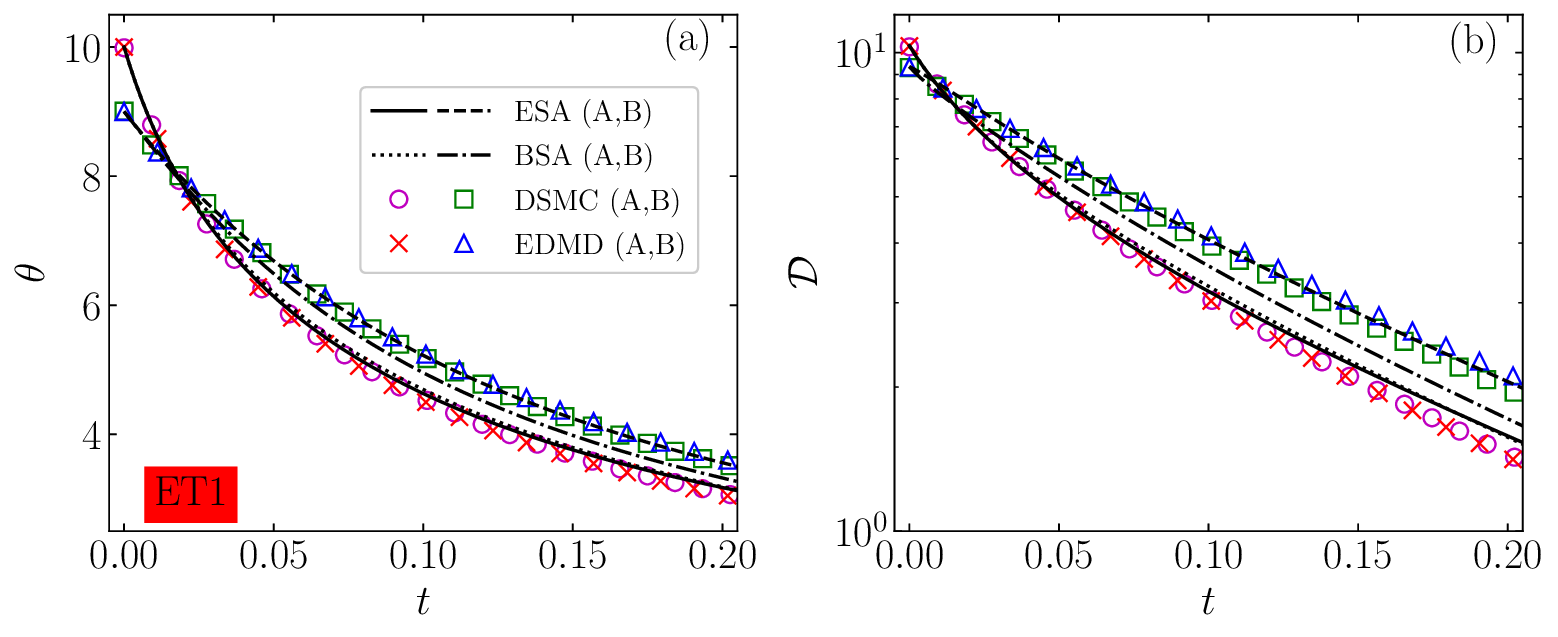}\\
    \includegraphics[width=\columnwidth]{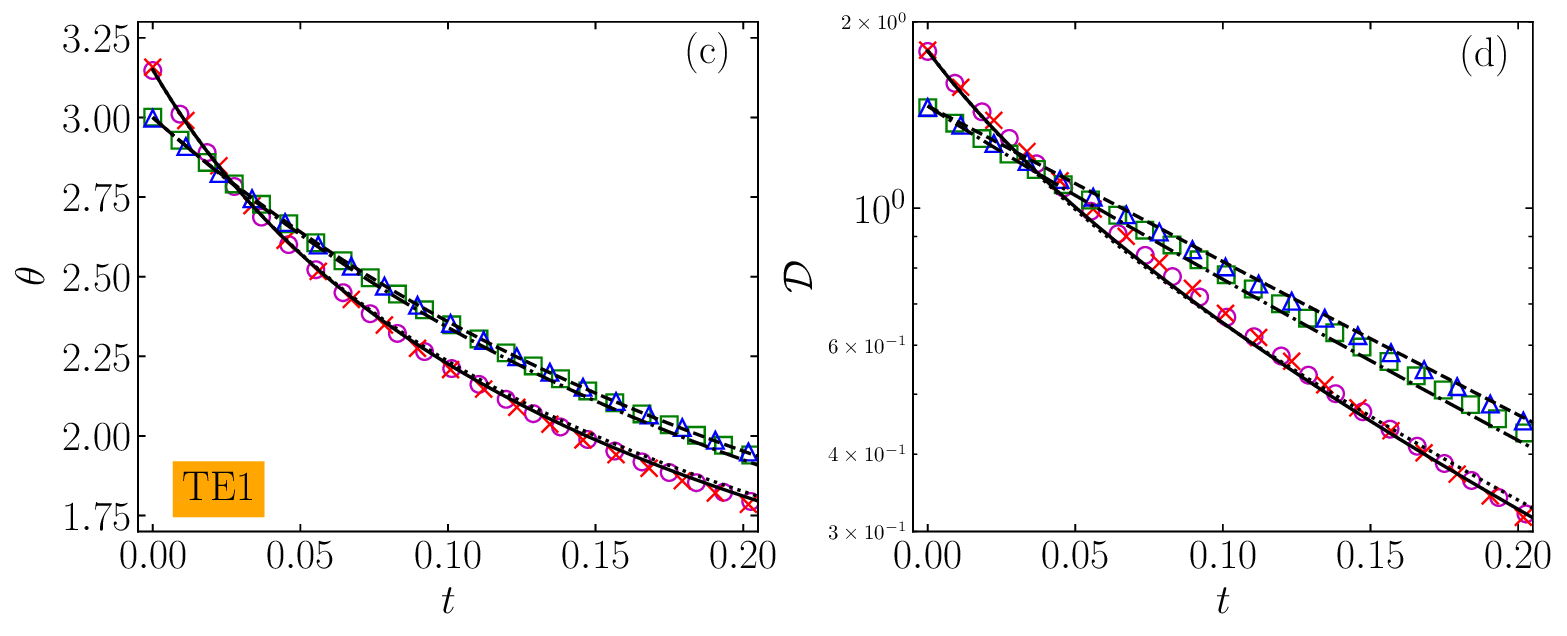}\\
    \includegraphics[width=\columnwidth]{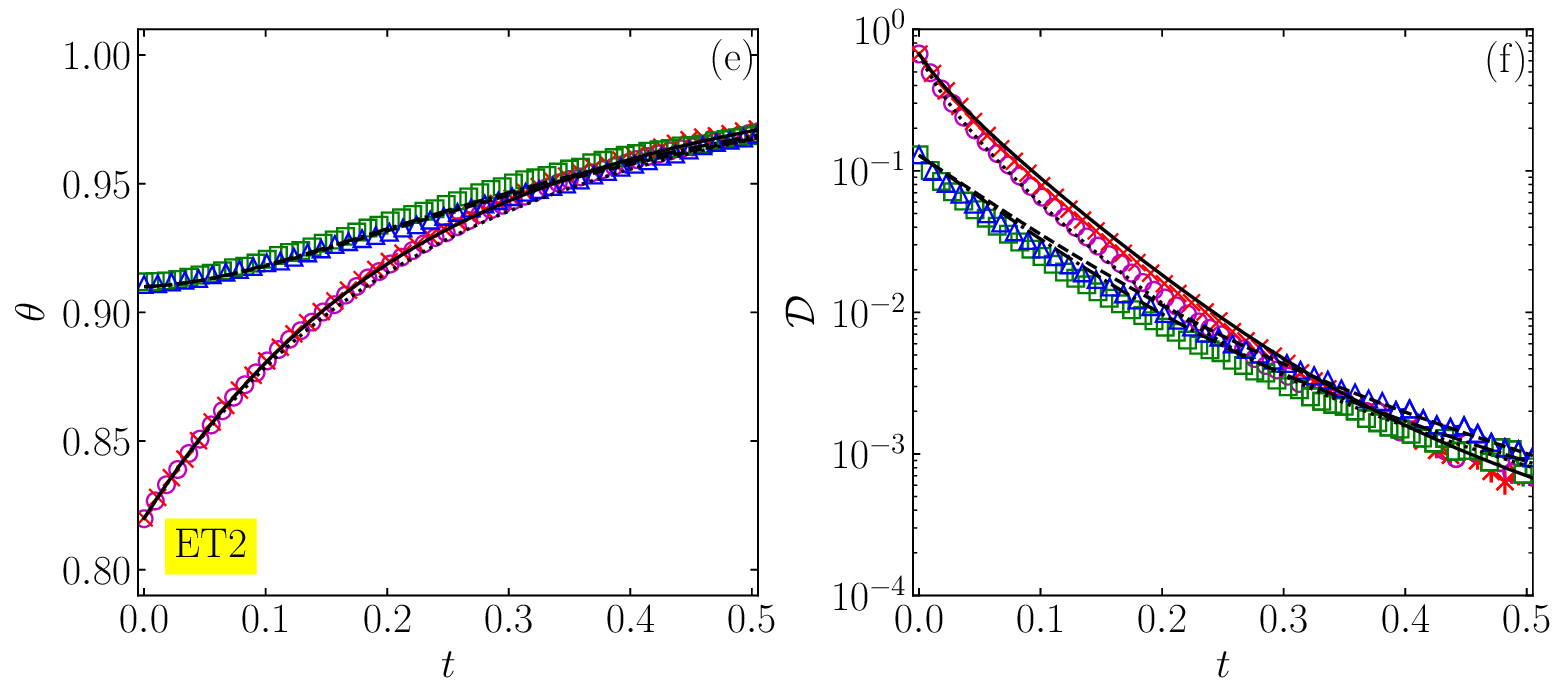}\\
    \includegraphics[width=\columnwidth]{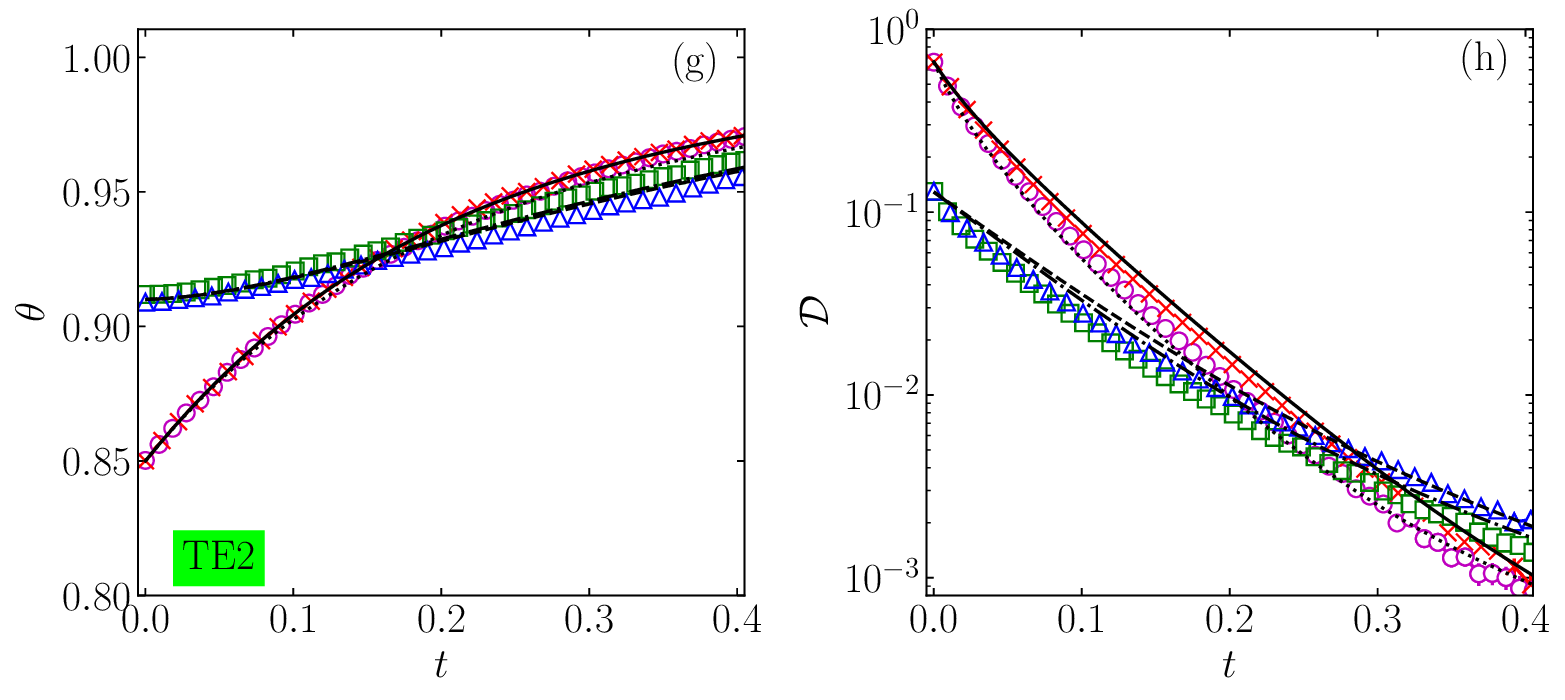}
    \caption{ME for cases when $\theta-1$ does not change its sign. Specifically, we plot the time evolution of the temperature and relative entropy for samples A and B: $\{\theta_A,\KLD_A\}$ (solid and dotted lines, circles, and crosses) and $\{\theta_B,\KLD_B\}$ (dashed and dash-dotted lines, squares, and triangles). Initial conditions are [(a) and (b)] $(\theta_A^0,\theta_B^0)=(10,9)$ and $(a_{2A}^0,a_{2B}^0)=(0.5,-0.35)$, [(c) and (d)] $(\theta_A^0,\theta_B^0)=(3.15,3)$ and $(a_{2A}^0,a_{2B}^0)=(0.5,-0.2)$, [(e) and (f)] $(\theta_A^0,\theta_B^0)=(0.82,0.91)$ and $(a_{2A}^0,a_{2B}^0)=(-0.35,0.3)$, and [(g) and (h)] $(\theta_A^0,\theta_B^0)=(0.85,0.91)$ and $(a_{2A}^0,a_{2B}^0)=(-0.35,0.3)$. Other parameter values are $\zs=1$, $\gamma=0.1$, and $d=3$.}
    \label{fig:ET_TE_12}
\end{figure}

\begin{figure}[htb!]
    \includegraphics[width=\columnwidth]{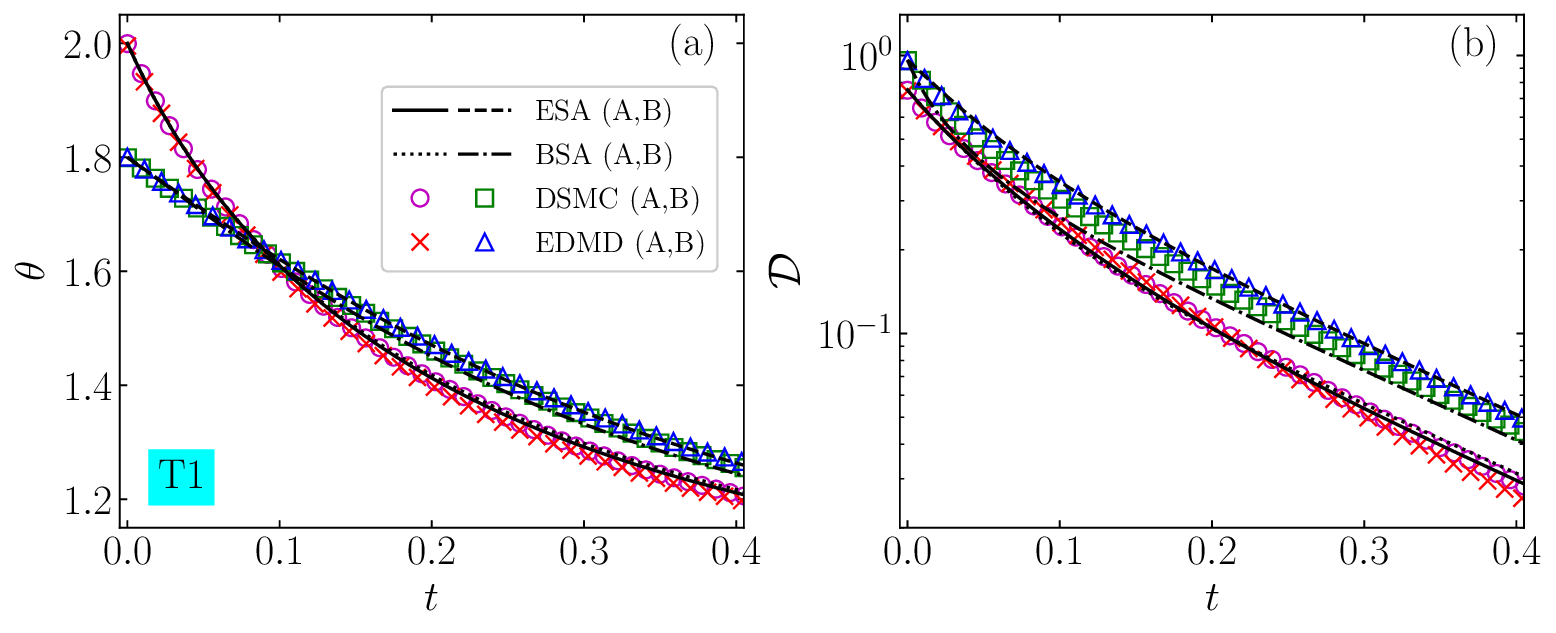}\\
    \includegraphics[width=\columnwidth]{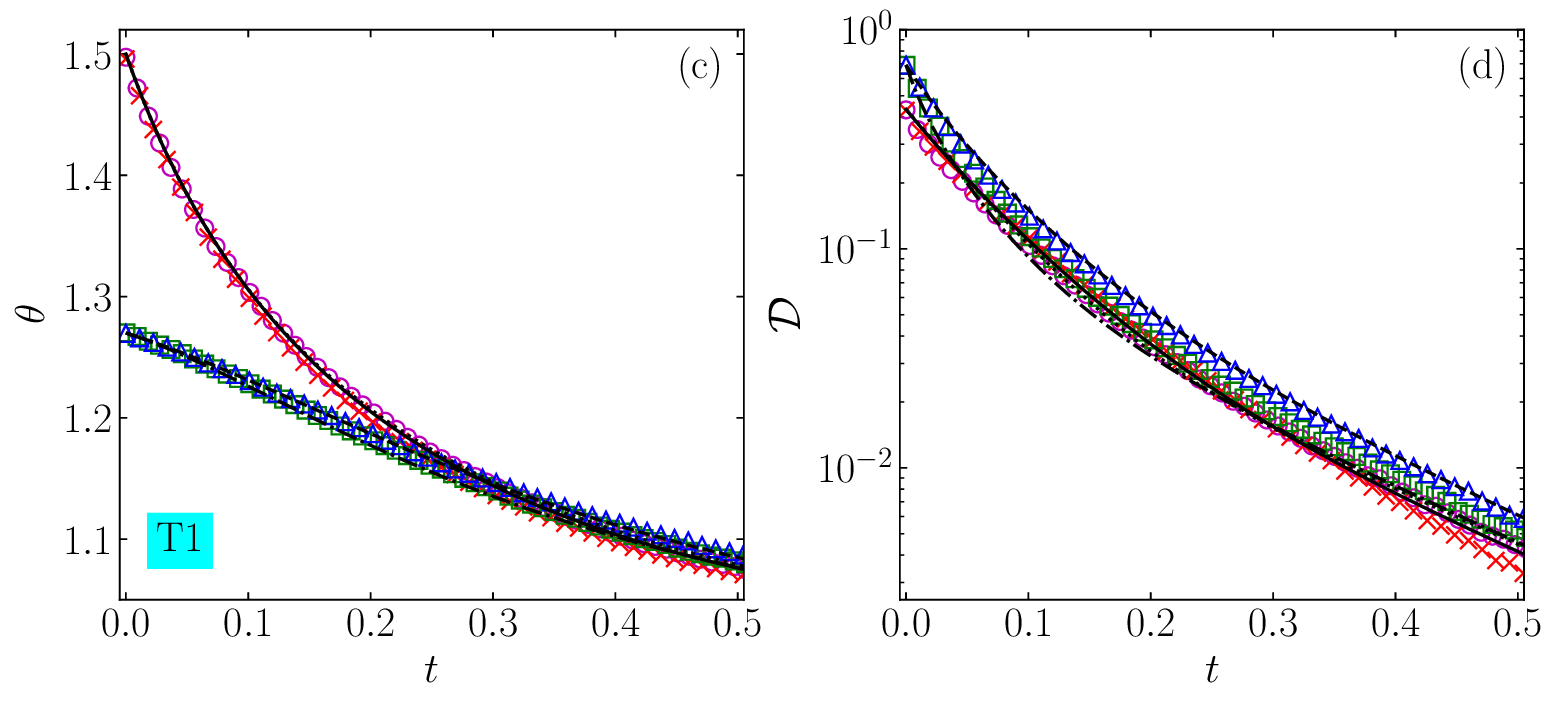}\\
    \includegraphics[width=\columnwidth]{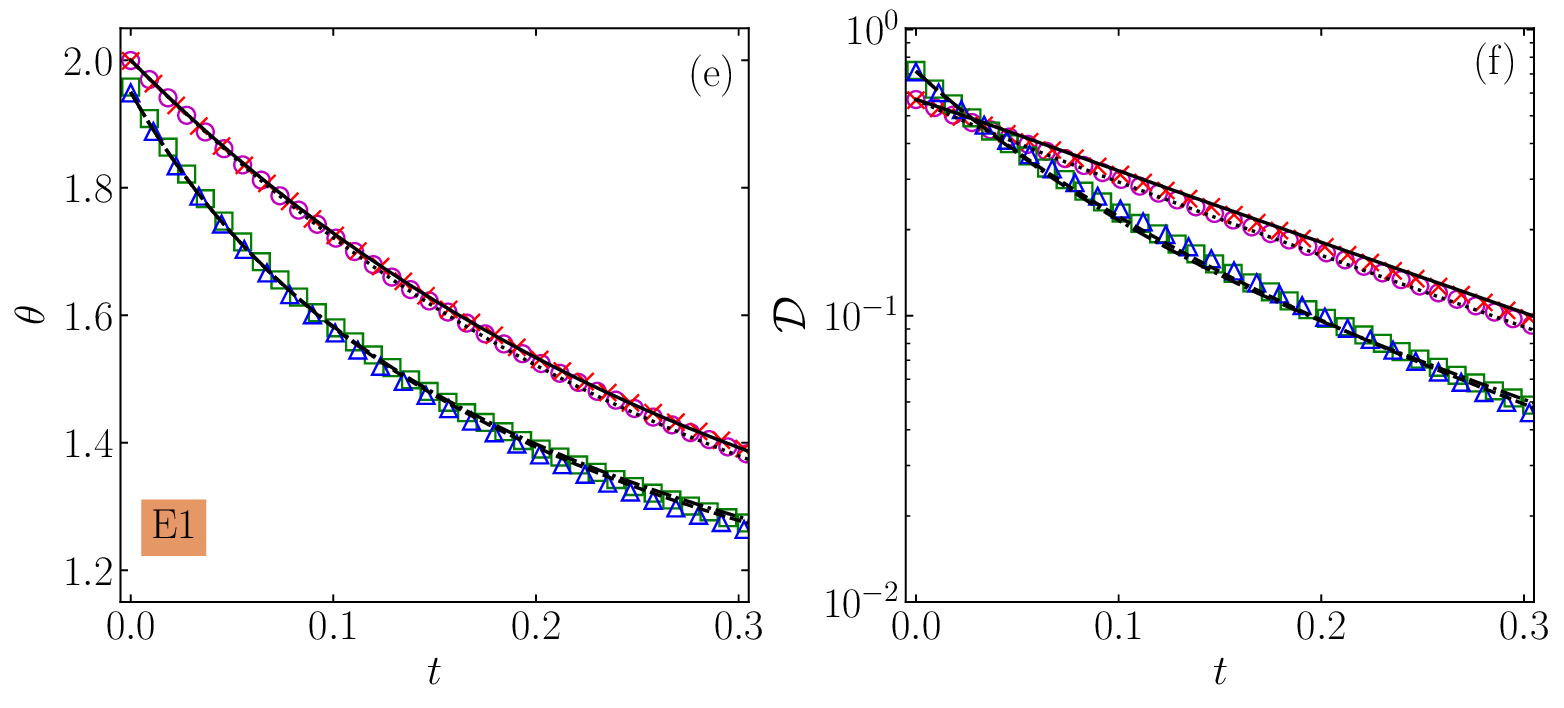}\\
    \includegraphics[width=\columnwidth]{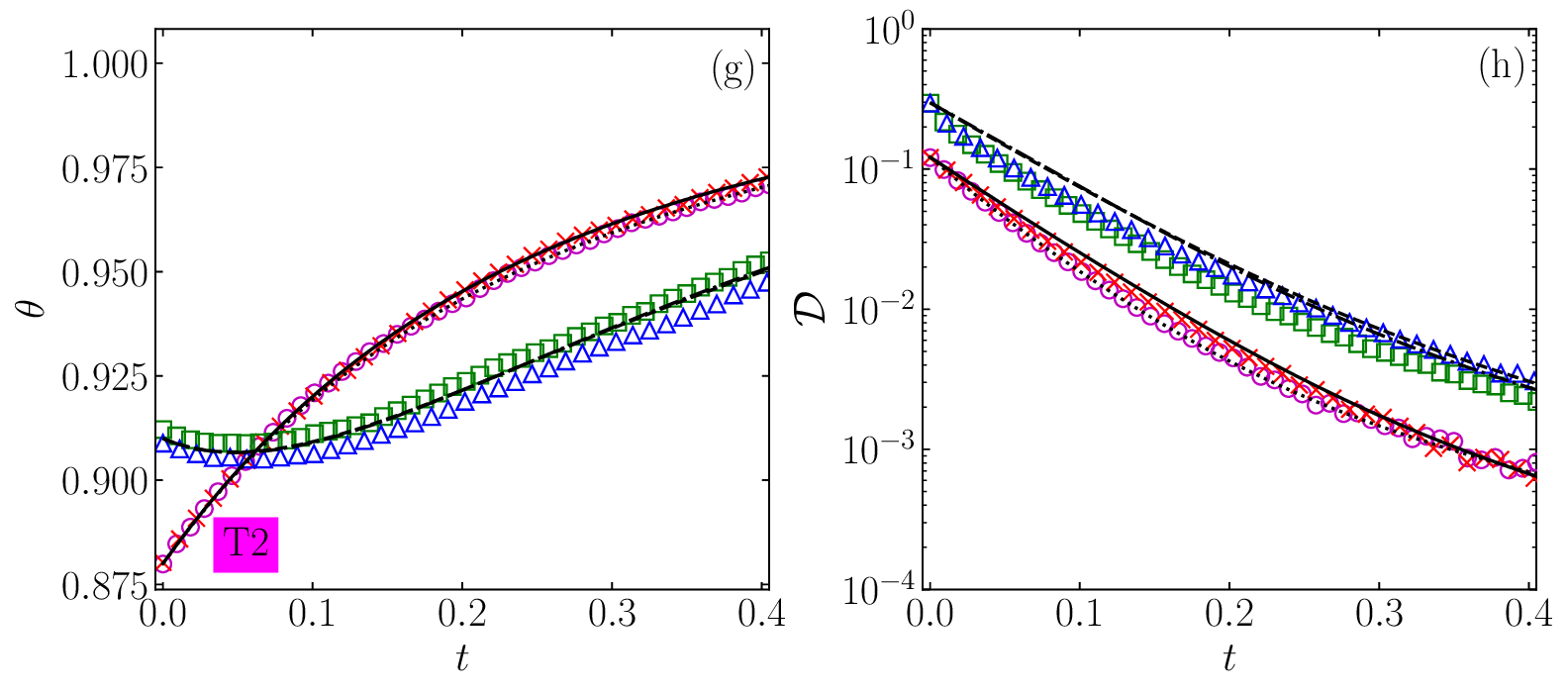}\\
    \includegraphics[width=\columnwidth]{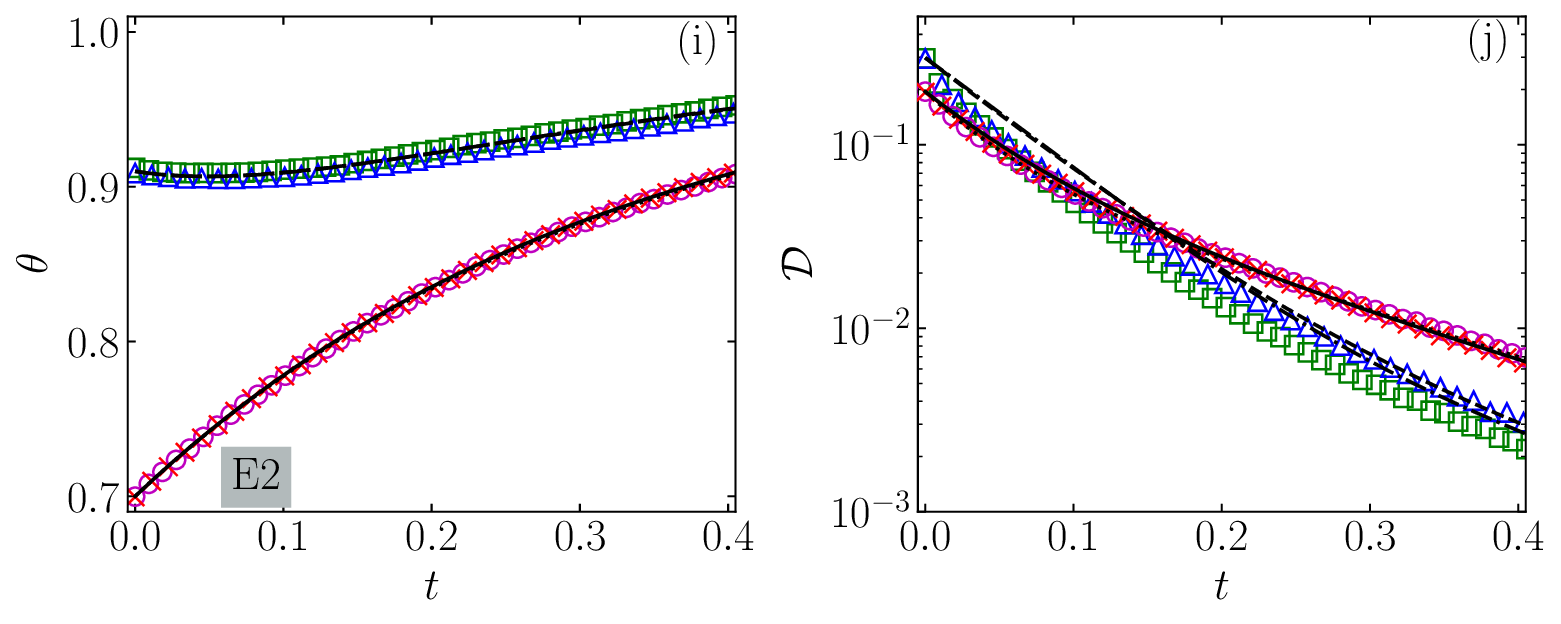}
    \caption{Same as in Fig.~\ref{fig:ET_TE_12}, except that the initial conditions are [(a) and (b)] $(\theta_A^0,\theta_B^0)=(2,1.8)$ and $(a_{2A}^0,a_{2B}^0)=(0.5,-0.35)$, [(c) and (d)] $(\theta_A^0,\theta_B^0)=(1.5,1.27)$ and $(a_{2A}^0,a_{2B}^0)=(0.5,-0.35)$, [(e) and (f)] $(\theta_A^0,\theta_B^0)=(2,1.95)$ and $(a_{2A}^0,a_{2B}^0)=(-0.2,0.5)$, [(g) and (h)] $(\theta_A^0,\theta_B^0)=(0.88,0.91)$ and $(a_{2A}^0,a_{2B}^0)=(-0.2,0.5)$, and [(i) and (j)] $(\theta_A^0,\theta_B^0)=(0.7,0.91)$ and $(a_{2A}^0,a_{2B}^0)=(-0.2,0.5)$.}
    \label{fig:T_E_12}
\end{figure}

\begin{figure*}[htb]
    \includegraphics[width=0.95\textwidth]{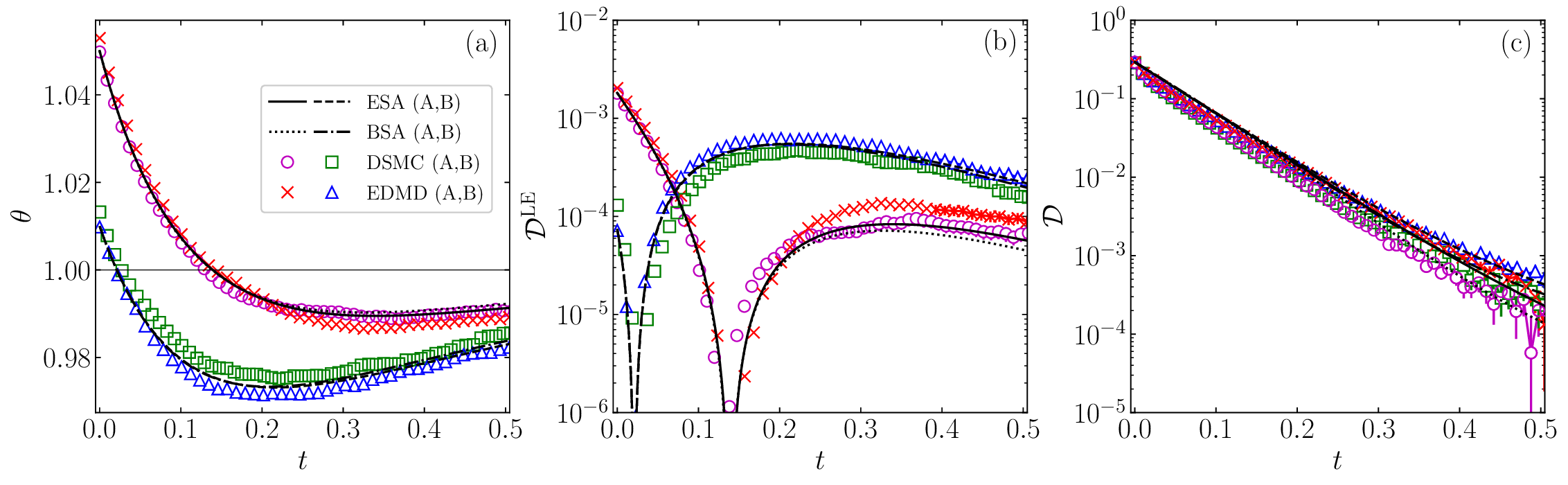}\\
    \includegraphics[width=0.95\textwidth]{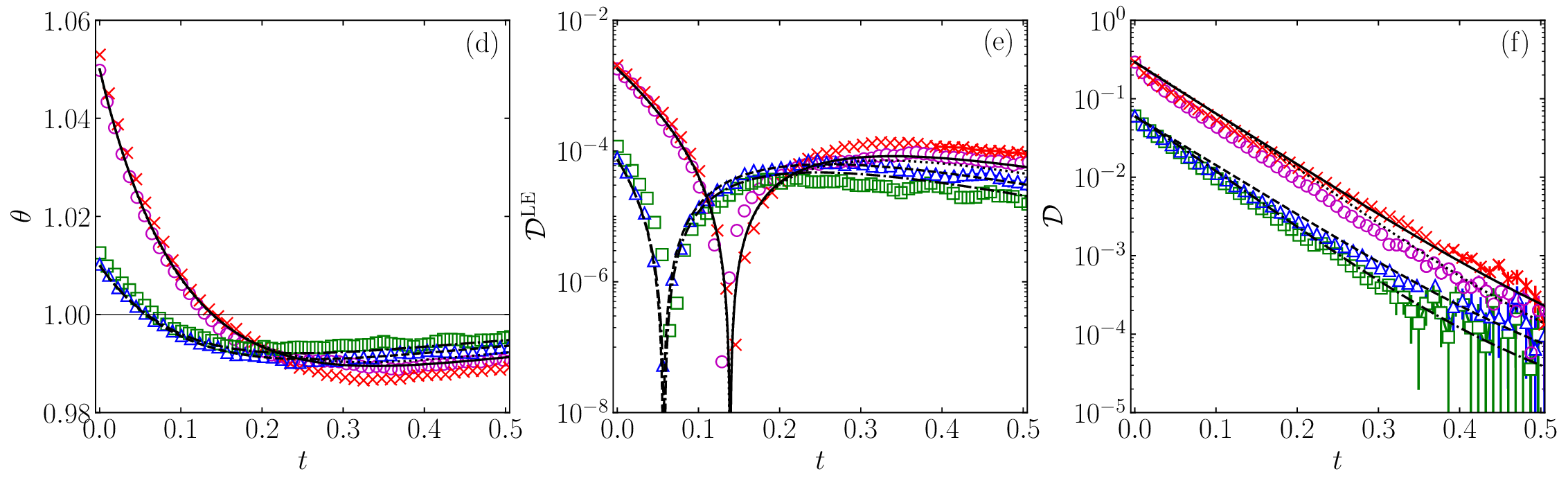}\\
    \includegraphics[width=0.95\textwidth]{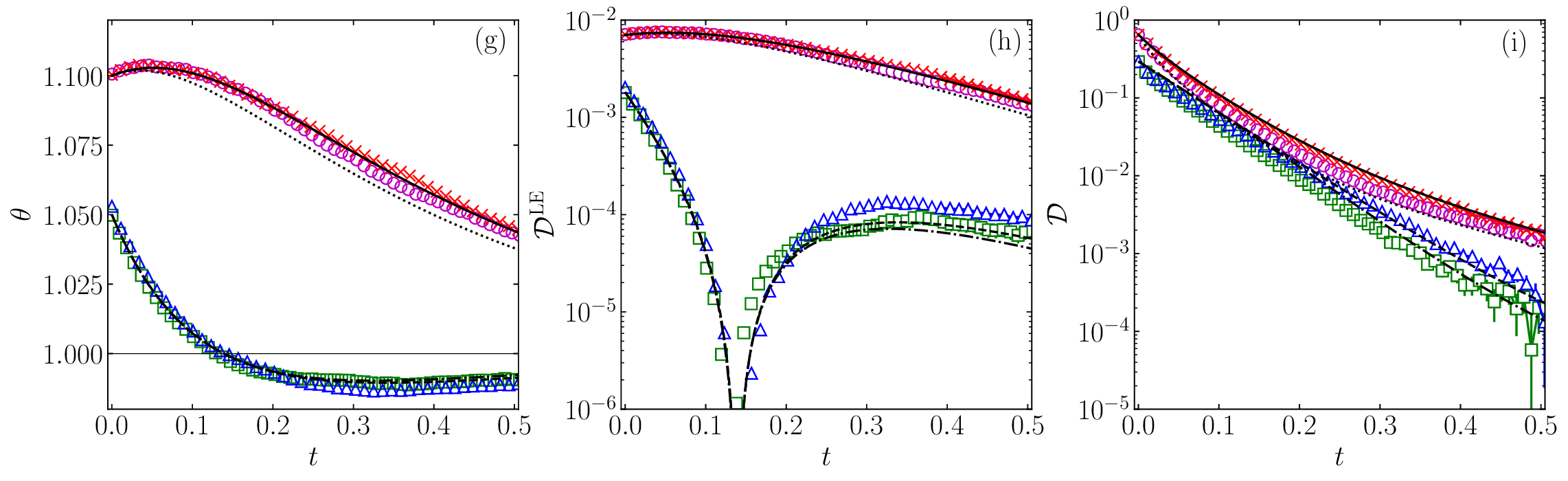}
    \caption{
    ME for cases case in which $\theta-1$ changes sign, i.e., the OME. In this case, we show the time evolution of:  $\{\theta_A,\KLD^\LE_A,\KLD_A\}$ (solid and dotted lines, circles, and crosses) and $\{\theta_B,\KLD^\LE_B,\KLD_B\}$ (dashed and dash-dotted lines, squares, and triangles). Parameter values and initial conditions correspond to those in Fig.~\ref{fig:OME_cases}. }
    \label{fig:OME_I}
\end{figure*}

\begin{figure*}
    \includegraphics[width=0.95\textwidth]{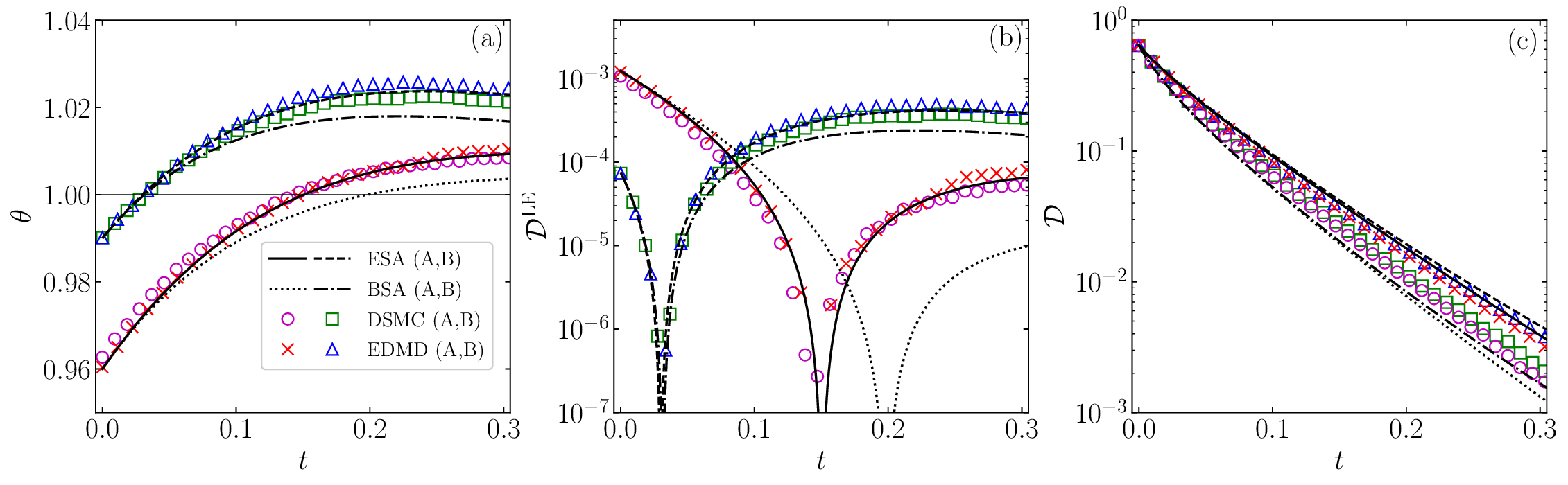}\\
    \includegraphics[width=0.95\textwidth]{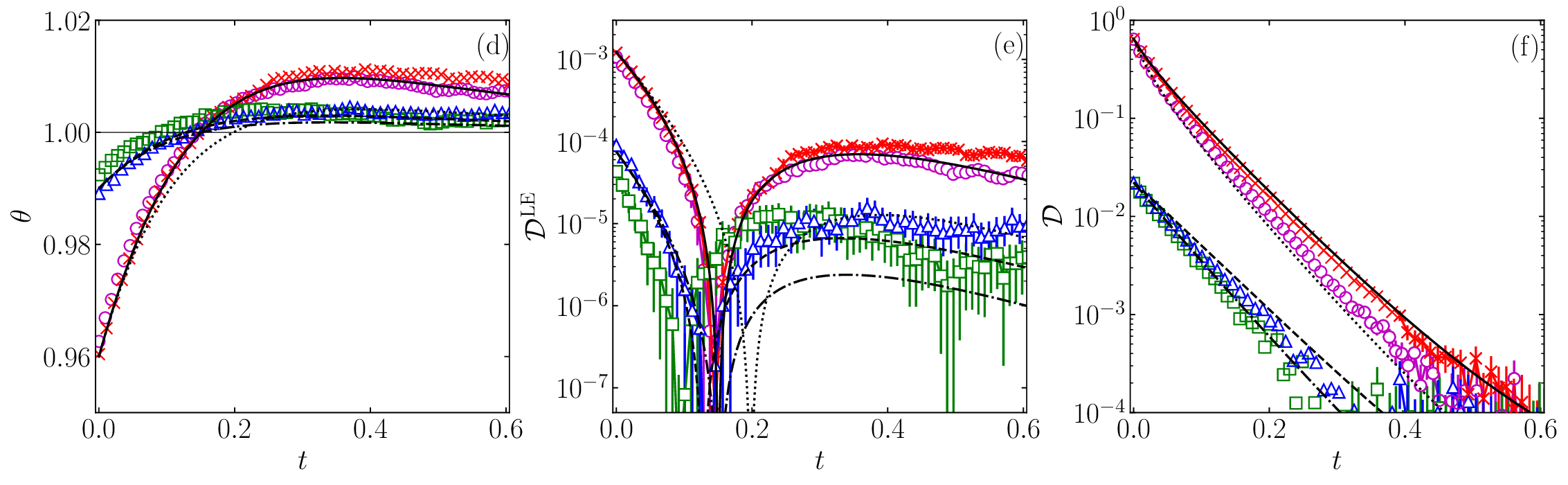}\\
    \includegraphics[width=0.95\textwidth]{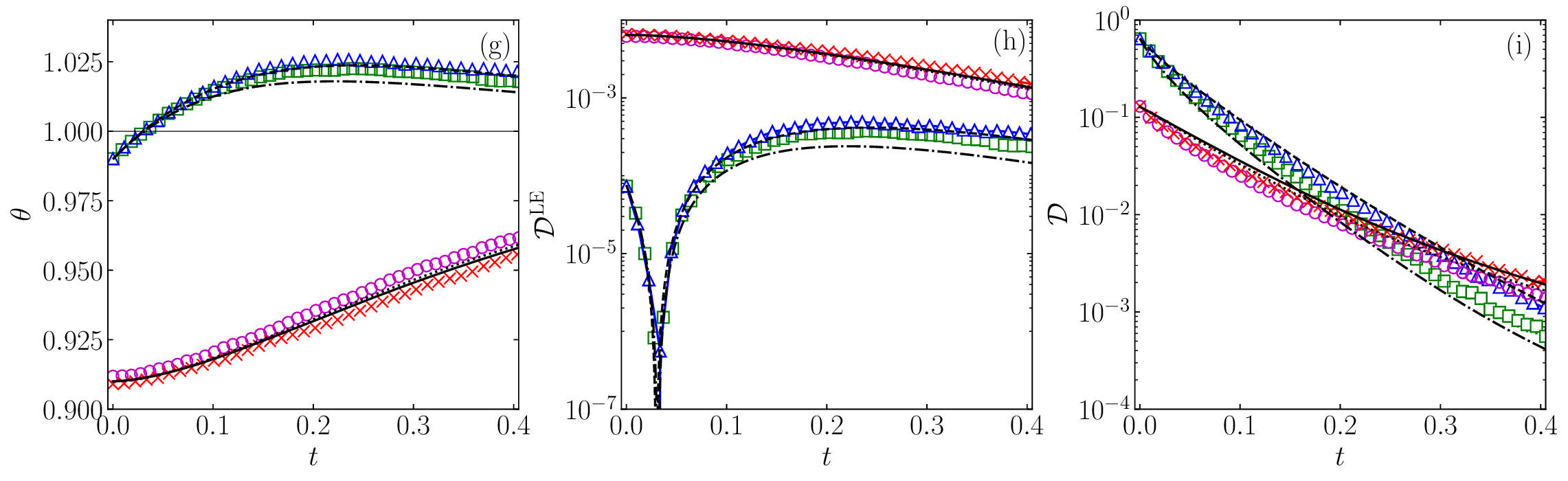}
    \caption{
    Same as in Fig.~\ref{fig:OME_I}, but now for the initial conditions in Fig.~\ref{fig:OME_cases_b}.}
    \label{fig:OME_II}
\end{figure*}

\section{Simulation results}
\label{sec:RESULTS}

In this section, our simulation results are used to test the  theoretical predictions stemming from the numerical solutions of: (i) the (nonlinear) BSA, Eqs.~\eqref{eq:thetadot} and \eqref{eq:a2-evol-Sonine} with $a_3\to 0$, and ESA, Eqs.~\eqref{eq:thetadot} and \eqref{eq:cumul-evol-Sonine}. The theoretical results for the KLD $\KLD(\ts)$ and $\KLD^\LE(\ts)$ are constructed by introducing the theoretical $\theta(\ts)$ and $a_2(\ts)$ in Eqs.~\eqref{eq:KLD^LE} and \eqref{eq:deltaKLDgamma}, respectively.

The employed computer simulation schemes, DSMC and EDMD, to build the simulation curves are presented in Appendix \ref{app:simul}. In all cases, the (reduced) initial VDF has been taken as the gamma distribution given by Eq.~\eqref{eq:Gam-dist} with the chosen value of the initial excess kurtosis $a_2^0$, as  previously  done in Refs.~\cite{LVPS17,MS20,MS21}. The KLD from simulations is computed as described in Refs.~\cite{MS20,MS21}.

Figures~\ref{fig:ET_TE_12} and \ref{fig:T_E_12} contain the
theoretical and simulation results of the time evolution of the
temperature ratio $\theta$ and the KLD $\KLD$. The graphs for the
cumulants $a_2$ and $a_3$ are presented in
Appendix~\ref{app:cum}. Samples A and B are prepared with the
representative values of the cumulants in Table~\ref{tab:a2Aa2B},
namely pairs I and II (III and IV) for the direct (inverse) ME, and
different values of the initial temperatures $\theta_A^0$ and
$\theta_B^0$. Pairs $(\theta_A^0,\theta_B^0)$ are chosen to illustrate
the different cases summarized in Table~\ref{tab:cases}. Specifically,
we present the cases ET1 in Figs.~\ref{fig:ET_TE_12}(a) and \ref{fig:ET_TE_12}(b), TE1 in
Figs.~\ref{fig:ET_TE_12}(c) and \ref{fig:ET_TE_12}(d), ET2 in Figs.~\ref{fig:ET_TE_12}(e) and \ref{fig:ET_TE_12}(f),
TE2 in Figs.~\ref{fig:ET_TE_12}(g) and \ref{fig:ET_TE_12} (h), T1 in Figs.~\ref{fig:T_E_12}(a) and
\ref{fig:T_E_12}(b), T1 with a double crossing in $\KLD$ in Figs.~\ref{fig:T_E_12}(c)
\ref{fig:T_E_12}(d), E1 in Figs.~\ref{fig:T_E_12}(e) and \ref{fig:T_E_12}(f), T2 in
Figs.~\ref{fig:T_E_12}(g) and \ref{fig:T_E_12}(h), and E2 in Figs.~\ref{fig:T_E_12}(i)
and \ref{fig:T_E_12}(j). Note that the cases in Figs.~\ref{fig:ET_TE_12}(a) and \ref{fig:ET_TE_12}(b),
\ref{fig:ET_TE_12}(c) and \ref{fig:ET_TE_12}(d), \ref{fig:ET_TE_12}(e) and (f),
\ref{fig:T_E_12}(a) and \ref{fig:T_E_12}(b), \ref{fig:T_E_12}(g) and \ref{fig:T_E_12}(h), and
\ref{fig:T_E_12}(i) \ref{fig:T_E_12}(j) are the same as in Figs.~\ref{fig:I}(a),
\ref{fig:II}(b), \ref{fig:III}(b), \ref{fig:I}(b), \ref{fig:IV}(b),
and \ref{fig:IV}(c), respectively. Moreover, the case in Figs.~\ref{fig:T_E_12}(c) and \ref{fig:T_E_12}(d) is close to the case in Fig.~\ref{fig:I}(c).

It must be remarked that the classification of the case in Figs.~\ref{fig:ET_TE_12}(e) and \ref{fig:ET_TE_12}(f) as  ET2 is less clear than expected. The LBSA theory predicts the ET2 behavior with a wide difference between $\ts_\theta$ and $\ts_\KLD$, as observed in the inset of Fig.~\ref{fig:III}(b). Still, nonlinearities reduce the time difference $\ts_\theta-\ts_\KLD$. Moreover, the double crossing in $\KLD$  predicted by the LBSA for the case of Figs.~\ref{fig:T_E_12}(c) and \ref{fig:T_E_12}(c)(d) is not actually observed in the simulations. The shallow positive maximum of $\KLD_A-\KLD_B$ predicted by the LBSA, as seen in Fig.~\ref{fig:I}(c), is washed out by nonlinear contributions---at least in the case represented in Figs.~\ref{fig:T_E_12}(c) and \ref{fig:T_E_12}(d).

Let us now turn to the OME predicted by the LBSA, which we have
discussed in Sec.~\ref{sec:OME}. Figures~\ref{fig:OME_I} and
\ref{fig:OME_II} show the time evolution of $\theta$, $\KLD^\LE$, and
$\KLD$ for the same cases as considered in Figs.~\ref{fig:OME_cases}
and \ref{fig:OME_cases_b}, respectively. Again, the graphs for $a_2$
and $a_3$ can be found in Appendix \ref{app:cum}. We see that the
overshoot behavior and the OME phenomenology are indeed
present. The crossover characterizing the TME is accounted for by the
intersection of the $\KLD^\LE$ curves in Figs.~\ref{fig:OME_I}(a)--\ref{fig:OME_I}(c)
and \ref{fig:OME_II}(a)--\ref{fig:OME_II}(c), though the corresponding temperature
curves never cross.

The figures in this section show that our theoretical predictions for
both $\theta(\ts)$ and $\KLD(\ts)$ are generally in very good
agreement with DSMC and EDMD simulation results. This is especially
true for the ESA, which still gives a good account of the behavior of
the fourth cumulant $a_2(\ts)$ and a fair account of the behavior of
the sixth cumulant $a_3(\ts)$ (see
Figs.~\ref{fig:ET_TE_12_cum}--\ref{fig:OME_II_cum}), consistently
with the results reported in Ref.~\cite{PSP21}. The improvement of the
ESA over the BSA can be understood by noticing that the values of the
cumulant $a_3$ are typically of the same magnitude as those of $a_2$.
It is also worth highlighting the generally good agreement between the
simulation results for the relative entropy $\KLD$ and those obtained
from Eqs.~\eqref{eq:KLD^LE} and \eqref{eq:deltaKLDgamma} when $\theta$
and $a_2$ are given by either the BSA or the ESA. This means that the
gamma distribution in Eq.~\eqref{eq:Gam-dist} represents a convenient
proxy of the unknown time-dependent VDF. This ansatz is further
confirmed by the generally fair agreement (not shown) between the
simulation data for $a_3$ and the right-hand side of
Eq.~\eqref{a3_Gamma} when plugging the simulation data of $a_2$,
especially in the cases with $a_2<0$.

Finally, we note that small deviations between EDMD and DSMC
simulation are observed (especially for the subtler quantities $\KLD$
and $a_3$), despite the low density of the systems. This might be a
consequence of the approximations carried out in the numerical
implementation of the Langevin dynamics in the \emph{approximate Green function} (AGF) algorithm explained
in Appendix \ref{subsec:EDMD}. Nevertheless, there is a good agreement
for the collisional scheme, as tested in Appendix
\ref{subsec:moments_test}.

\section{Conclusions}
\label{sec:Conclusions}

In this paper, we have analyzed in depth the relaxation to equilibrium of a dilute gas of elastic hard spheres subjected to a nonlinear drag and the associated stochastic force. We have particularly focused on two versions of the ME, namely, the TME and the EME. Our analysis combines theory and simulation. The theoretical approach is based on a Sonine expansion of the solution of the EFPE, Eq.~\eqref{eq:EFP}. The simulation approach comprises both DSMC results, which integrate numerically Eq.~\eqref{eq:EFP}, and EDMD results.

We have employed the Kullback--Leibler divergence (or relative entropy) $\KLD$, defined in Eq.~\eqref{eq:KLD-kinetic}, to measure the distance to equilibrium and monitor the possible emergence of the EME. It must be remarked that other distances to equilibrium have been employed in the literature, as long as they share some common properties of monotonicity, convexity, etc. However, the choice of the distance function does not impinge on the existence of the EME---for a thorough discussion of this issue, see Ref.~\cite{LR17}.
The KLD choice for the distance function is quite natural due, to its connection to the nonequilibrium entropy, and especially convenient for comparing the TME and EME, since $\KLD$ can be  decomposed into two summands, see Eqs.~\eqref{eq:KLD_decomposition} and \eqref{eq:KLD_decomposition_2}. First, the hydrodynamic LE contribution $\KLD^\LE$, which only depends on the temperature and, second,  the kinetic-stage correction $\KLD^{\kin}$, which depends on the whole VDF. To obtain an approximate expression for the latter within the Sonine approximation, we have employed the gamma distribution function, Eq.~\eqref{eq:deltaKLDgamma}.

For given values of the drag force, i.e., given values of $(\zs,\gamma)$, the emergence of ME---either direct or inverse---depends on the initial preparations of the two samples (A, whose initial temperature is farther from  equilibrium,  and B, whose initial temperature is closer to equilibrium). The simplest approach, based on heuristic arguments, is the LBSA given by Eqs.~\eqref{eq:relax-explicit}.  Therein, both the temperature ratio $\theta(\ts)$ and the excess kurtosis $a_2(\ts)$ are a linear superposition of two exponentials.

When the difference $\theta-1$ keeps its initial sign during the relaxation process---i.e., when the temperature does not cross its equilibrium value at a finite time, we have the most usual, standard situation. Bringing to bear that the kinetic-stage contribution $\KLD^{\kin}$ is expected to decay to zero over a shorter timescale than that of the local equilibrium contribution $\KLD^\LE$, we have argued that the existence of TME implies that of EME (and vice versa) if $\KLD_A^0>\KLD_B^0$. There are two possibilities: either the thermal crossover occurs earlier than the entropic one (scenarios TE1 and TE2 for direct and inverse effects, respectively) or it occurs later than the entropic crossover (scenarios ET1 and ET2 for direct and inverse effects, respectively).

Interestingly, even though $\theta_A^0$ departs from the equilibrium value $1$ more than $\theta_B^0$, one may have $\KLD_A^0<\KLD_B^0$ due to the kinetic contribution $\KLD^{\kin}$ to the entropic distance. This gives rise to the existence of TME without entropy crossover (scenarios T1 and T2 for direct and inverse effects, respectively) or, reciprocally, the existence of EME without thermal crossover (scenarios E1 and E2 for direct and inverse effects, respectively).

A summary of all the possible scenarios above (assuming a constant sign of $\theta-1$) is provided by Table~\ref{tab:cases}. The corresponding phase diagrams in the plane $\theta_A^0-\theta_B^0$ vs $\theta_B^0-1$ are depicted in Fig.~\ref{fig:phase_diag} for a few representative choices of the initial excess kurtoses $a_{2A}^0$ and $a_{2B}^0$, given in Table~\ref{tab:a2Aa2B}. Those scenarios are modified when the condition of constant sign of $\theta(\ts)-1$ is violated, as explained below.

Nonmonotonic evolutions of $\theta$ with a crossing of the equilibrium line $\theta=1$ induce the appearance of overshoot-induced humps. Sample B may relax to equilibrium later than sample A when the temperature of the former overshoots the equilibrium value, a fact that sample A can take advantage of. Even though $\theta_A(\ts)$ and $\theta_B(\ts)$ do not intersect, the corresponding curves of $\KLD^\LE$ do intersect. We have termed this class of ME as OME. Simple conditions for its existence, Eqs.~\eqref{eq:OME-condition}, have been derived by adapting the LBSA to this situation.

The different scenarios for the ME outlined above for the nonlinear fluid, emerging in the extremely simplistic LBSA, have been tested and confirmed by computer simulations (both DSMC and EDMD).  These numerical results have also been compared with the more complex nonlinear BSA and ESA. The inclusion of the additional cumulant $a_3$ in the set of coupled evolution equations allows the ESA to improve over the BSA.  Moreover, DSMC and EDMD results are practically indistinguishable, with small discrepancies that can be traced back to the approximations in the EDMD scheme during the free streaming stage, see Appendix \ref{subsec:EDMD}. On the other hand, the collisional schemes are tested in Appendix \ref{subsec:moments_test}, with good results.

The ME effect is brought about by the nonlinearity in the drag force, which makes the time-evolution of the kinetic temperature be coupled to that of higher cumulants---specifically, to that of the excess kurtosis $a_2$ for the quadratic dependence of the drag coefficient in Eq.~\eqref{eq:4}. The nonlinear drag force is also responsible for the algebraic nonexponential relaxation after a temperature quench and for the emergence of Kovacs-like response~\cite{PSP21}. It is important to remark that these behaviors, and also the ME, survive in the limit $\zeta_0^*\to\infty$, in which the EFPE reduces to the Fokker--Planck equation---which, interestingly, successfully models mixtures of ultracold atoms~\cite{HKLMSLW17}.

The nonmonotonic relaxation of the kinetic temperature observed in the OME entails the necessity of revising the conventional definition of TME. Provided that both initial temperatures are either above (direct case) or below (inverse case) equilibrium, the TME is not necessarily characterized by the crossover of the nonequilibrium temperature but by the crossover of the associated positive-definite quantity $\KLD^\LE$.  Within this generalized scheme and for a general complex system, we propose the following, more dependable, definition of the TME based on the idea of local equilibrium: The TME exists in a pair of different initially prepared setups if there are an odd number of crossings between their LE relative entropy $\KLD^\LE$ curves. On another note, the definition of the EME is not affected, due to the monotonic decay of the whole relative entropy $\KLD$ to equilibrium. The EME exists in a pair of samples if their relaxation curves for $\KLD$ present an odd number of crossings.

It is relevant to stress that the splitting of $\KLD$ into  ``kinetic" and  ``local-equilibrium" contributions can be done on quite a general basis, not only for the molecular fluid we are analyzing in this paper. Moreover, this allows for defining a nonequilibrium temperature $T(t)$, even in systems for which the kinetic temperature makes no sense. Let us consider a general system with Hamiltonian $H(\bm{x})$, in which $\KLD(t)$ is given by Eq.~\eqref{eq:KLD-def-intro}. The system is initially prepared in a certain state with average energy $\mean{H}^0$ and is put in contact with a thermal bath at temperature $T_\bb$. Thus the probability distribution function $P(\bm{x},t)$ relaxes toward the equilibrium distribution $P^\eq(\bm{x})=\exp\left[-H(\bm{x})/\kb T_\bb\right]/Z(T_\bb)$, where $Z(T_\bb)$ is the partition function. One can always introduce a LE distribution with the canonical form $P^{\LE}(\bm{x},T(t))\equiv \exp\left[-H(\bm{x})/\kb T(t)\right]/Z(T(t))$, with $T(t)$ being determined self-consistently by the condition
\begin{equation}
    \langle H\rangle(t)=\int d\bm{x}\, H(\bm{x}) P(\bm{x},t)=\int d\bm{x}\, H(\bm{x}) P^{\LE}(\bm{x},T(t)).
\end{equation}
In this way, $T(t)$ corresponds to the temperature that a system would  have at equilibrium if it had an average energy equal to the instantaneous value $\langle H\rangle(t)$. With such a definition of the nonequilibrium temperature $T(t)$, is it easily shown that
\begin{equation}
    \KLD(t)=\KLD^{\kin}(t)+\KLD^{\LE}(T(t)),
\end{equation}
where $\KLD^{\kin}$ and $\KLD^{\LE}$ are given by
\begin{subequations}
\label{eq:KLD_decomposition_general}
\begin{align}
\KLD^{\kin}(t)=&\int d\bm{x}\, P(\bm{x},t)\ln\frac{P(\bm{x},t)}{P^\LE(\bm{x},T(t))},\\
\KLD^\LE(T(t))=&\int d\bm{x}\, P^{\LE}(\bm{x},T(t))\ln\frac{P^{\LE}(\bm{x},T(t))}{P^\eq(\bm{x})}.
\end{align}
\end{subequations}
Note that $T(t)$ may in general overshoot its equilibrium value $T_\bb$, leading to an OME, but $\KLD^\LE$ is positive definite and makes it possible to introduce the more reliable definition of the TME explained in the previous paragraph~\cite{note_22_05_1}.

We expect this work can motivate the experimental investigation, making use of a suitable aging protocol to prepare the initial samples, of the whole variety of ME phenomenology described in this work. Specifically, our predicting and observing the OME---as a novel unexpected behavior---in this molecular gas driven by a nonlinear drag opens the door to its finding in other complex systems. Also, we  plan to employ the theoretical and computational framework developed here to study the relaxation times of pairs of temperature quenches thermodynamically equidistant  from equilibrium~\cite{LG20,VH21}, one above and the other one below.

\acknowledgments A.M.\ and A.S.\ acknowledge financial support from
Grant No.~PID2020-112936GB-I00 funded by MCIN/AEI/10.13039/501100011033,
and from Grants No.~IB20079 and No.~GR21014 funded by Junta de Extremadura
(Spain) and by ERDF ``A way of making Europe.'' A.P.\ acknowledges
financial support from Grant No.~PGC2018-093998-B-I00 funded by
MCIN/AEI/10.13039/501100011033/ and by ERDF ``A way of making
Europe.'' A.M.\ is grateful to the Spanish Ministerio de Ciencia,
Innovaci\'on y Universidades for a predoctoral fellowship
FPU2018-3503. We are grateful to the computing facilities of the Instituto de Computaci\'on Cient\'ifica Avanzada of the University of Extremadura (ICCAEx), where our simulations were run.

\appendix
\section{Langevin Equation under Nonlinear Drag}
\label{sec:Langevin}
In this Appendix, we discuss the Langevin equation associated with the left-hand side of Eq.~\eqref{eq:EFP}, since the effect of interparticle collisions represented by the right-hand side is well known.

Let us start writing the Langevin equation as
\begin{equation}\label{eq:LE}
    \dot{\vv}(t) = -\zeta(v(t))\vv_i(t)+\xi(v(t))\boldsymbol{\eta}(t),
\end{equation}
where $\boldsymbol{\eta}(t)$ is a Gaussian white-noise stochastic term with the statistical properties
\begin{equation}
  \langle \bm{\eta}(t)\rangle_{\text{noise}}=0,\quad \langle \bm{\eta}(t)\bm{\eta}(t')\rangle_{\text{noise}}=\mathsf{I}\delta(t-t'),
\end{equation}
where  $\mathsf{I}$ is the $d\times d$ unit tensor and $\langle  \cdot\rangle_{\text{noise}}$ reads for an average over different realizations. Let us define a Wiener process $W(t)$ with elemental increment $\dif W(t)=\xi(v(t))\boldsymbol{\eta}(t)\dif t$. This is the case of a multiplicative noise and, therefore, there is no a unique way of interpreting the proper time within a given interval $[t,t+h]$ at which the process $W(t)$ must be evaluated \cite{MM12}. In general, one can choose a time $t+\epsilon h$ parameterized by $0\leq\epsilon\leq 1$. Hence, the associated Fokker--Planck equation is~\cite{MM12}
\begin{equation}\label{eq:FP1}
    \partial_t f(\vv) - \frac{\partial}{\partial \vv}\cdot \left[\zeta(v)\vv+\frac{\xi^{2\epsilon}(v)}{2}\frac{\partial}{\partial \vv}\xi^{2(1-\epsilon)}(v) \right]f(\vv)=0.
\end{equation}
The specific choices $\epsilon=0$, $\frac{1}{2}$, and $1$ correspond to the It\^o \cite{vK07}, Stratonovich \cite{vK07}, and Klimontovich \cite{K94} interpretations, respectively.

The (differential) fluctuation-dissipation relation stemming from Eq. \eqref{eq:FP1} turns out to be
\begin{equation}
    \zeta(v) = \frac{m\xi^2(v)}{2\kb T_\bb}-\frac{1-\epsilon}{2v}\frac{\partial\xi^2(v)}{\partial v}.
\end{equation}
Only in the Klimontovich interpretation ($\epsilon=1$) does one
recover the conventional fluctuation-dissipation relation,
Eq.~\eqref{eq:FDR}, holding for constant drag coefficient and additive
noise. In that case, the left-hand sides of Eqs.~\eqref{eq:EFP} and
\eqref{eq:FP1} coincide.

On the other hand, from a simulation point of view, the It\^o interpretation ($\epsilon=0$)  is the simplest one to implement. Fortunately, even if $\epsilon\neq 0$ (as happens in the Stratonovich and Klimontovich interpretations), one can always apply the It\^o interpretation to the Langevin equation, provided that the original drag coefficient $\zeta(v)$ is replaced by an effective one $\zeta_{\mathrm{eff}}$ (``spurious drift'').
Note first the mathematical identity
\begin{equation}
\xi^{2\epsilon}(v)\frac{\partial}{\partial \vv} \xi^{2(1-\epsilon)}(v)f(\vv)=\frac{\partial}{\partial \vv} \xi^{2}(v)f(\vv)-\epsilon \frac{\partial \xi^2(v)}{\partial \vv}f(\vv).
\end{equation}
Inserting this into Eq.~\eqref{eq:FP1}, one gets
\begin{equation}
    \partial_t f(\vv) - \frac{\partial}{\partial \vv}\cdot \left[\zeta_{\mathrm{eff}}(v)\vv+\frac{\partial}{\partial \vv}\frac{\xi^{2}(v)}{2} \right]f(\vv)=0,
\end{equation}
where
\begin{align}\label{eq:zeta_eff}
    \zeta_{\mathrm{eff}}(v) \equiv& \zeta(v)-\frac{\epsilon}{2v}\frac{\partial \xi^2(v)}{\partial v}\nonumber \\
    = & \frac{m\xi^2(v)}{2\kb T_\bb}-\frac{1}{2v}\frac{\partial \xi^2(v)}{\partial v}.
\end{align}

In the particular case of Eq.~\eqref{eq:4} and $\epsilon=1$, the effective drag coefficient becomes
\begin{equation}
    \zeta_{\mathrm{eff}}(v) = \zeta(v)-2\zeta_0\gamma.
\end{equation}
Thus, the original Langevin equation, Eq.~\eqref{eq:LE}, in the Klimontovich interpretation is equivalent to the Langevin equation
\begin{equation}\label{eq:LE_eff}
    \dot{\vv}(t) = -\zeta_{\mathrm{eff}}(v(t))\vv(t)+\xi(v(t))\boldsymbol{\eta}(t)
\end{equation}
in the It\^o interpretation.

\section{Derivation of the evolution equations}\label{sec:evol-eqs-derivation}

To write the hierarchy of moment equations, it is convenient to introduce dimensionless quantities~\cite{SP20}. First, we define a rescaled velocity $\cc$ as
\begin{equation}
\label{eq:8}
    \cc\equiv \frac{\vv}{v_{\thm}(t)},\quad v_{\thm}(t)\equiv \sqrt{\frac{2 \kb T(t)}{m}},
\end{equation}
in which $v_{\thm}(t)$ is the thermal velocity at time $t$. Analogously, the dimensionless VDF is introduced as
\begin{equation}
\label{eq:9}
    \phi(\cc,t) \equiv \frac{v_{\thm}^d(t)}{n}f(\vv,t).
\end{equation}

In terms of these reduced quantities, the EFPE, Eq.~\eqref{eq:EFP}, can be rewritten as
\begin{align}
\label{eq:phi}
    \partial_{\ts} \phi(\cc,\ts)=&\frac{1}{2\theta}\partial_\cc\cdot\left[\dot{\theta}\cc+\zs\left(1+\gamma\theta c^2 \right)\left(2\theta\cc+\partial_\cc \right)\right]\phi(\cc,\ts) \nonumber\\
    & + K_d\sqrt{\theta} I[\cc|\phi,\phi],
\end{align}
where $\theta$ is the temperature ratio---as defined in Eq.~\eqref{eq:KLD^LE}--- and
\begin{align}
    I[\cca|\phi,\phi] = &\int\dif  \ccb\, \int_{+}\dif\ssig\, \ccab\cdot\ssig \nonumber\\
    & \times \left[\phi(\cca^\prime,\ts)\phi(\ccb^\prime,\ts)-\phi(\cca,\ts)\phi(\ccb,\ts)\right]
    \label{eq:I-reduced}
\end{align}
is the reduced collision operator with $\ccab\equiv\cca-\ccb$. In Eq.~\eqref{eq:phi}, and consistently with the main text, dimensionless variables are used---recall that the stars on the dimensionless time $t^*$ and the zero-velocity drag coefficient $\zeta_0^*$ are dropped.

Multiplying both sides of Eq.~\eqref{eq:phi} by $c^\ell$ and defining the reduced moments
\begin{equation}\label{eq:Ml-def}
    M_\ell(t) \equiv \langle c^\ell\rangle=\int d\cc \,c^\ell \phi(\cc,t),
\end{equation}
one obtains the hierarchy of equations~\cite{SP20}
\begin{align}\label{eq:Ml-evol}
    \frac{\dot{M}_\ell}{\zs} = &\ell\left\{ \left[(\ell-2)\gamma+(d+2)\gamma\theta(1+a_2)-\frac{1}{\theta}\right]M_\ell \right.\nonumber\\
    &\left.-2\gamma \theta M_{\ell+2}+\frac{d+\ell-2}{2}\frac{M_{\ell-2}}{\theta}\right\}-\frac{K_d}{\zs}\sqrt{\theta}\mu_\ell,
\end{align}
where we have introduced the collisional moments $\mu_\ell$ as
\begin{equation}\label{eq:13}
    \mu_\ell\equiv -\int \dif\cc \,c^\ell I[\cc|\phi,\phi].
\end{equation}
Note that $M_0=1$, $M_2=\frac{d}{2}$, and $M_4=\frac{d(d+2)}{4}(1+a_2)$ [see Eq.~\eqref{a2}]. Conservation of mass and energy imply that $\mu_0=\mu_2=0$, so that Eq.~\eqref{eq:Ml-evol} is obviously consistent with $\dot{M}_0=\dot{M}_2=0$.

Making use of the explicit form of the collision operator, it is possible to express the collisional moments as two-particle averages of the form
\begin{equation}
    \mu_\ell = \int\dif\cca\int\dif\ccb\, \phi(\cca)\phi(\ccb)\Phi_\ell(\cca,\ccb).
\end{equation}
In particular, $\Phi_2 = 0$ and, after some algebra, one gets
\begin{subequations}
\label{Phi4Phi6}
\begin{equation}
\label{Phi4}
    \Phi_4(\cca,\ccb) = \frac{2\pi^{\frac{d-1}{2}}}{\Gamma\left(\frac{d+5}{2}\right)}c_{12}\left[d (\COMc\cdot \ccab)^2-c_{12}^2C^2 \right],
\end{equation}
\begin{equation}
\label{Phi6}
    \Phi_6 (\cca,\ccb) = 3\Phi_4(\cca,\ccb)\left(C^2+\frac{c_{12}^2}{4}\right),
\end{equation}
\end{subequations}
where $\COMc\equiv \frac{1}{2}(\cca+\ccb)$ is the center-of-mass  reduced velocity.

\subsection{Sonine approximation}
\label{sec:Sonine}

Let us first consider the case of linear drag force, i.e., $\gamma=0$. In that case, the LE state defined in Eq.~\eqref{eq:f-LE} is an exact solution of the EFPE, Eq.~\eqref{eq:EFP}. Equivalently, in reduced variables,
\begin{equation}
\label{eq:VDF-LE-reduced}
  \phi^\LE(\cc)=\pi^{-d/2}e^{-c^2} \implies  M_{2k}^\LE=\frac{[d+2(k-1)]!!}{2^k},
  \end{equation}
becomes an exact stationary solution to Eqs.~\eqref{eq:phi} and \eqref{eq:Ml-evol}, because of the properties $I[\cc|\phi^\LE,\phi^\LE]=0$, $\mu_\ell^\LE=0$. Moreover, the solution to Eq.~\eqref{eq:thetadot} is simply $\theta(\ts)=1+\left[\theta(0)-1\right]e^{-2\zs\ts}$, as stated in the main text. Thus, if $\gamma=0$ and the system is initially prepared in an equilibrium state with a temperature $T(0)$, its coupling to a bath at temperature $T_\bb$ makes the temperature evolve toward  $T_\bb$ but otherwise the system remains always in local equilibrium, i.e., the VDF is Maxwellian with the time-dependent temperature.

Going back to the nonlinear case $\gamma\neq 0$,  the VDF can be represented by the Sonine expansion
\begin{equation}
\label{eq:Sonine}
\phi(\cc{;t})=\phi^\LE(\cc)\left[1+\sum_{j=2}^\infty a_j{(t)} L_j^{(\frac{d-2}{2})}(c^2)\right],
\end{equation}
where $L_j^{(\frac{d-2}{2})}(c^2)$ are generalized Laguerre (or Sonine) polynomials and the coefficients $a_j=\left[j!\Gamma(\frac{d}{2})/\Gamma\left(j+\frac{d}{2}\right)\right]\mean{L_j^{(\frac{d-2}{2})}(c^2)}$ are the cumulants of the nonequilibrium VDF. The associated velocity moments are
\begin{equation}\label{M2k}
    M_{2k}= M_{2k}^\LE\left[1+\sum_{j=2}^k(-1)^j \binom{k}{j}a_j\right], \quad k\geq 2.
\end{equation}
In particular,
\begin{subequations}
   \begin{equation}
   M_6=\frac{d(d+2)(d+4)}{8}(1+3a_2-a_3),
  \end{equation}
  \begin{equation}
   M_8=\frac{d(d+2)(d+4)(d+6)}{16}(1+6a_2-4a_3+a_4).
  \end{equation}
\end{subequations}
The infinite moment hierarchy, Eq.~\eqref{eq:Ml-evol}, cannot be solved in an exact way  in general. This is even the case for linear drag, $\gamma=0$, when the initial state is not a Maxwellian. As a consequence, the exact evolution of the temperature ratio $\theta(\ts)$ cannot be obtained from Eq.~\eqref{eq:thetadot} if $\gamma\neq 0$.

Let us suppose, however, that the initial condition and subsequent evolution are sufficiently close to the LE state as to assume both the cumulants in Eq.~\eqref{eq:Sonine}  beyond $a_3$ and the quadratic terms in $a_2$, $a_3$ (i.e., those proportional to $a_2^2$, $a_3^2$, and $a_2 a_3$) being negligible. In that case, the two first nontrivial collisional moments become~\cite{BP06}
\begin{subequations}
\label{eq:mu4-mu6}
\begin{align}
   \mu_4\approx& \frac{2(d-1)}{K_d}\left(a_2-\frac{a_3}{4} \right), \label{eq:17a}\\
    \mu_6\approx& \frac{3(d-1)(2d+9)}{2K_d}\left(a_2-\frac{3a_3}{4} \right).\label{eq:17b}
\end{align}
\end{subequations}
Within this scheme, Eq.~\eqref{eq:Ml-evol} with $\ell=4$ and $\ell=6$ yields Eq.~\eqref{eq:cumul-evol-Sonine} in the main text.

\section{Parameters in Eqs.~\eqref{eq:relax-explicit}}
\label{appB}
The parameters $\lambda_\pm$, $B_i$, and $A_{ij}$ are given by
\begin{subequations}
\label{4.3}
\begin{equation}
\label{4.5}
\lambda_\pm=\frac{\Lambda_{11}+\Lambda_{22}\pm\sqrt{(\Lambda_{11}-\Lambda_{22})^2+4\Lambda_{12}\Lambda_{21}}}{2},
\end{equation}
\begin{equation}
\label{Bs}
B_1=\theta_r+\frac{\Lambda_{22}C_1-\Lambda_{12}C_2}{\Lambda_{11}\Lambda_{22}-\Lambda_{12}\Lambda_{21}},\quad
B_2=\frac{\Lambda_{11}C_2-\Lambda_{21}C_1}{\Lambda_{11}\Lambda_{22}-\Lambda_{12}\Lambda_{21}},
\end{equation}
\begin{equation}
A_{11}=\frac{\lambda_+-\Lambda_{11}}{\lambda_+-\lambda_-},\quad A_{22}=\frac{\lambda_+-\Lambda_{22}}{\lambda_+-\lambda_-},
\end{equation}
\begin{equation}
A_{12}=\frac{\Lambda_{12}}{\lambda_+-\lambda_-},\quad A_{21}=\frac{\Lambda_{21}}{\lambda_+-\lambda_-},
\end{equation}
\end{subequations}
where
\begin{subequations}
\begin{equation}
\label{4.3a}
{\Lambda}_{11}=2\zs\left[1+(d+2)\gamma\left(2\theta_r-1\right)\right],
\end{equation}
\begin{equation}
\label{4.3d}
{\Lambda}_{22}=\zs\left[\frac{4}{\theta_r}-8\gamma+4(d+8)\gamma\theta_r\right]
+\frac{8(d-1)}{d(d+2)}{\sqrt{\theta_r}},
\end{equation}
\begin{equation}
\label{4.3b}
\Lambda_{12}= 2\zs(d+2)\gamma\theta_r^2,\quad
{\Lambda}_{21}=8\zs\gamma,
\end{equation}
\begin{equation}
C_1=2\zs\left(1-\theta_r\right)\left[1+(d+2)\gamma\theta_r\right],\quad C_2=8\zs\gamma \left(1-\theta_r\right).
\end{equation}
\end{subequations}

\begin{figure*}[ht]
    \includegraphics[width=0.47\textwidth]{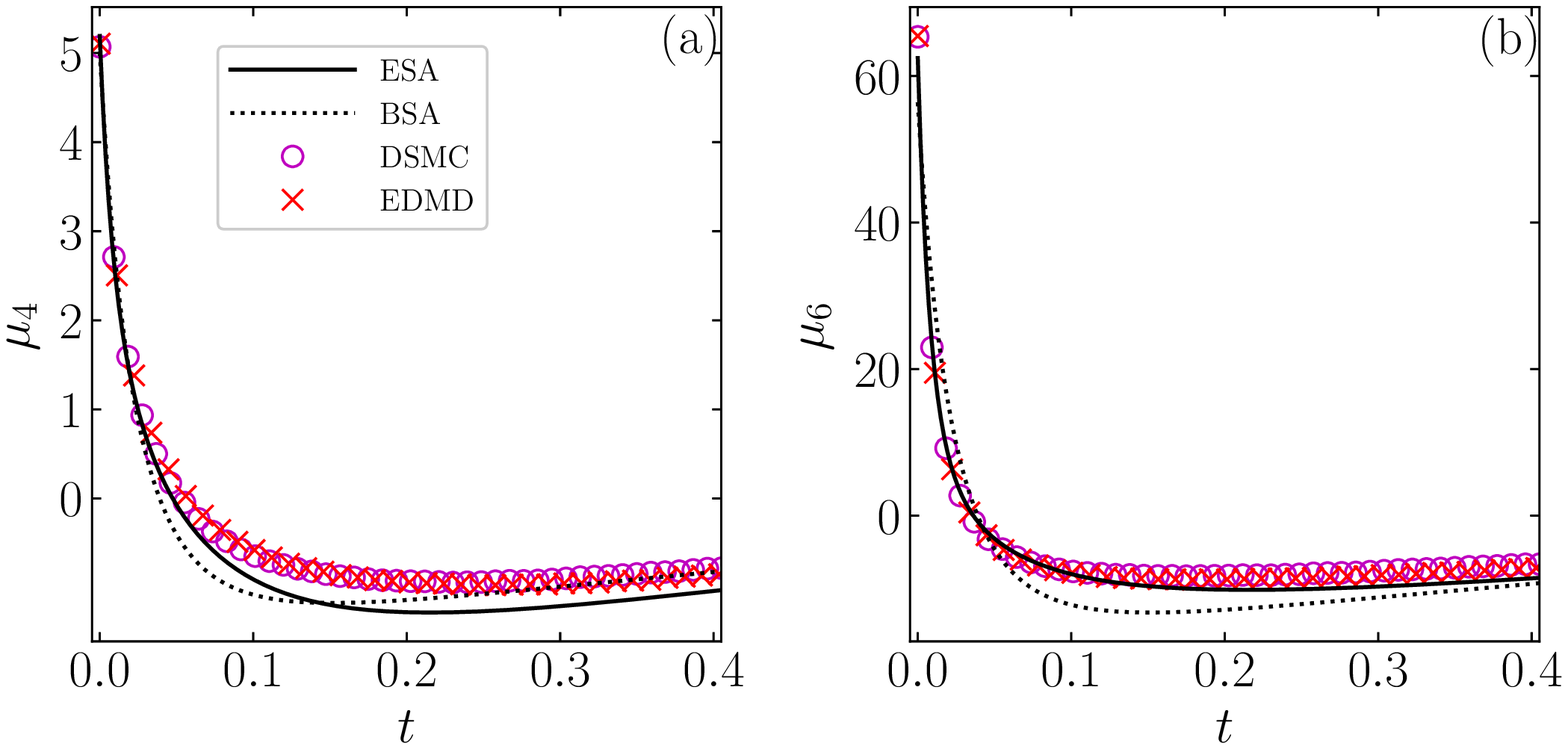}
    \includegraphics[width=0.47\textwidth]{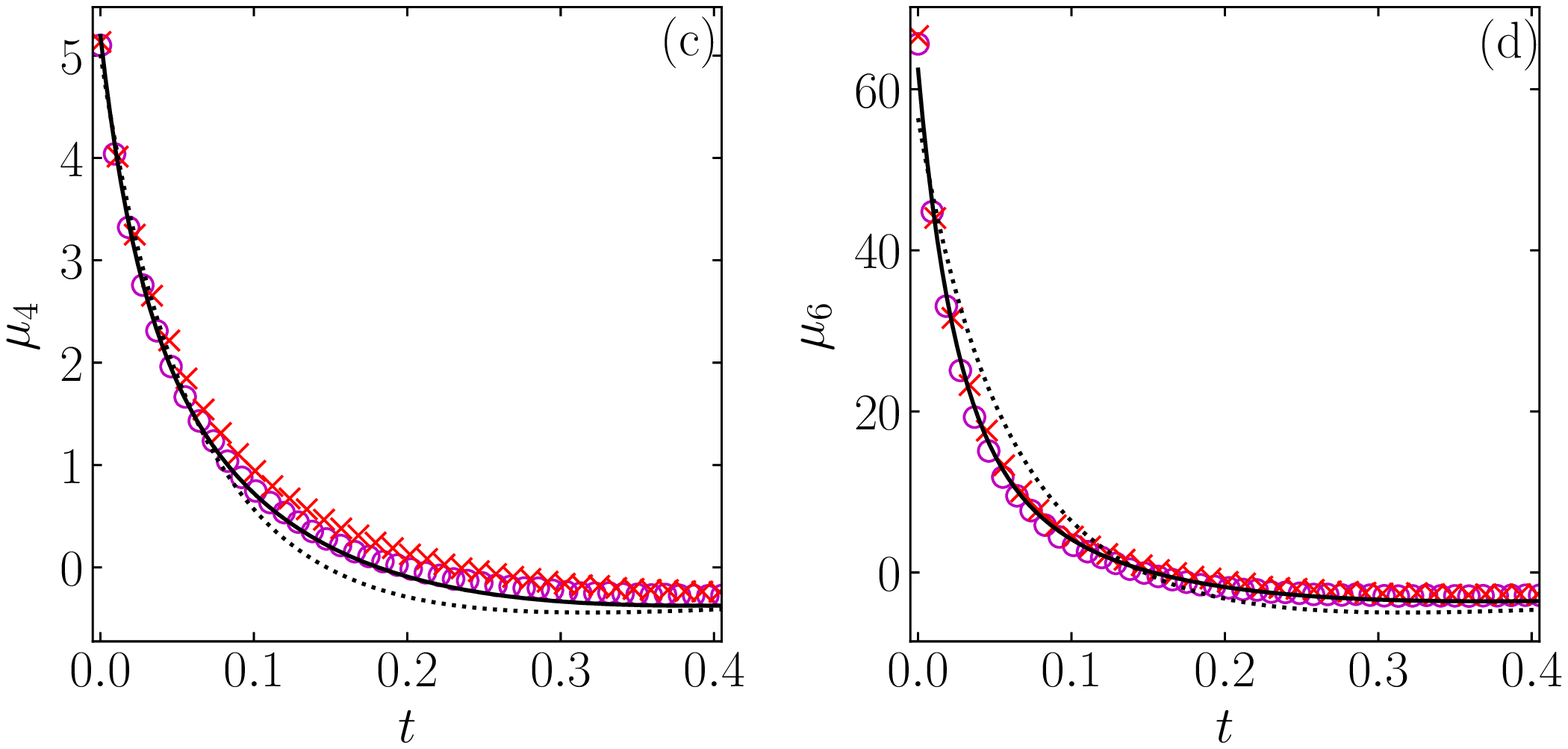}\\
    \includegraphics[width=0.47\textwidth]{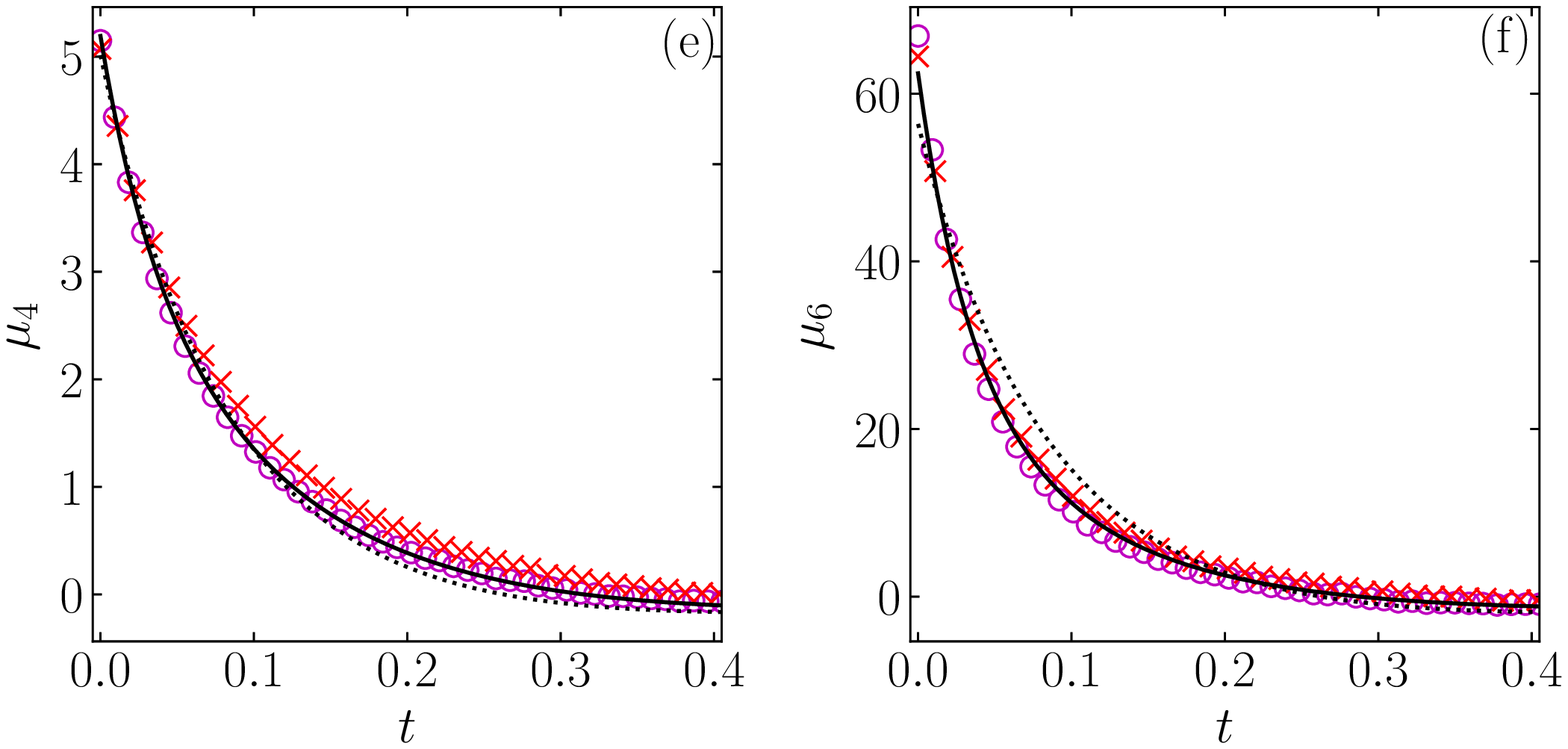}
    \includegraphics[width=0.47\textwidth]{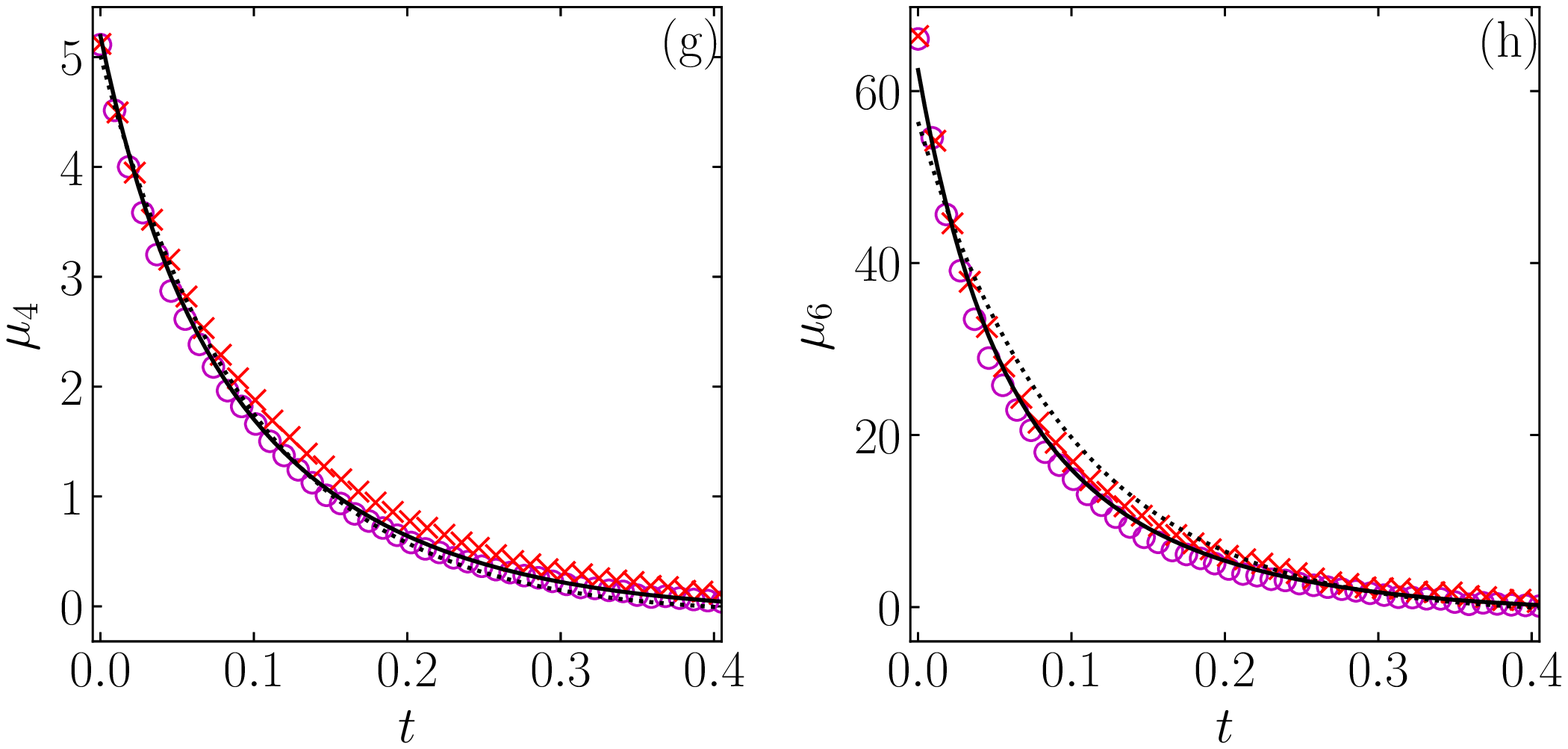}\\
    \includegraphics[width=0.47\textwidth]{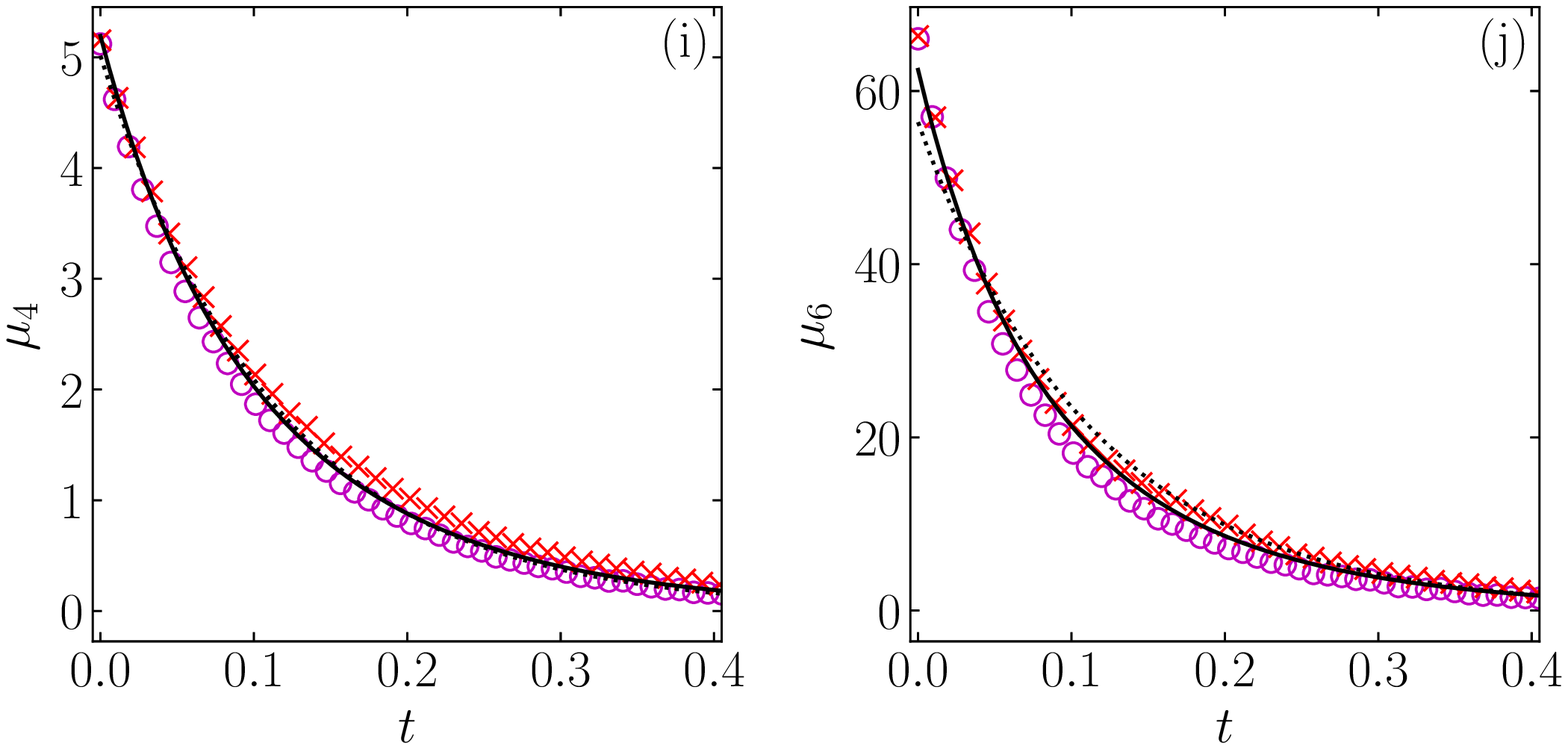}
    \includegraphics[width=0.47\textwidth]{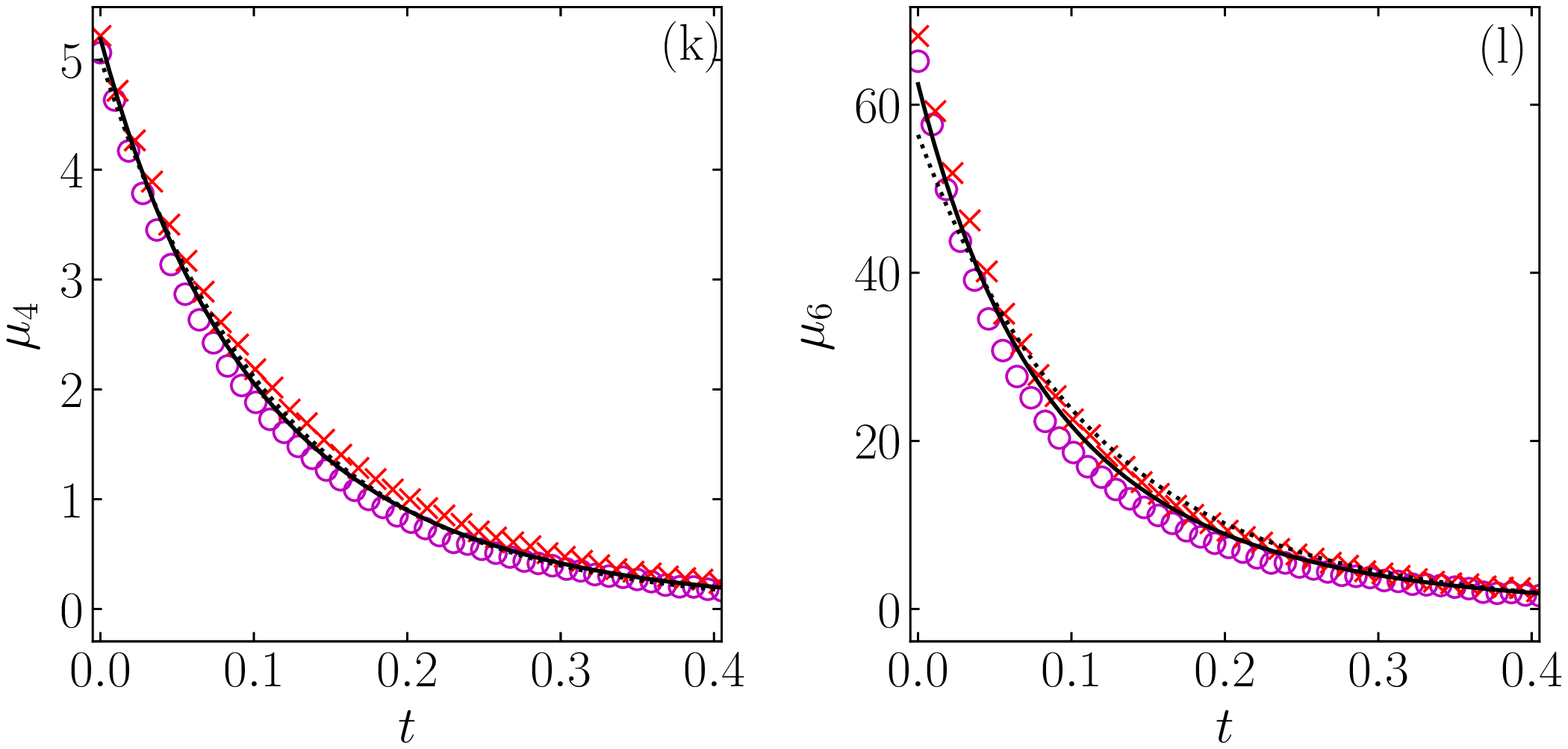}\\
    \includegraphics[width=0.47\textwidth]{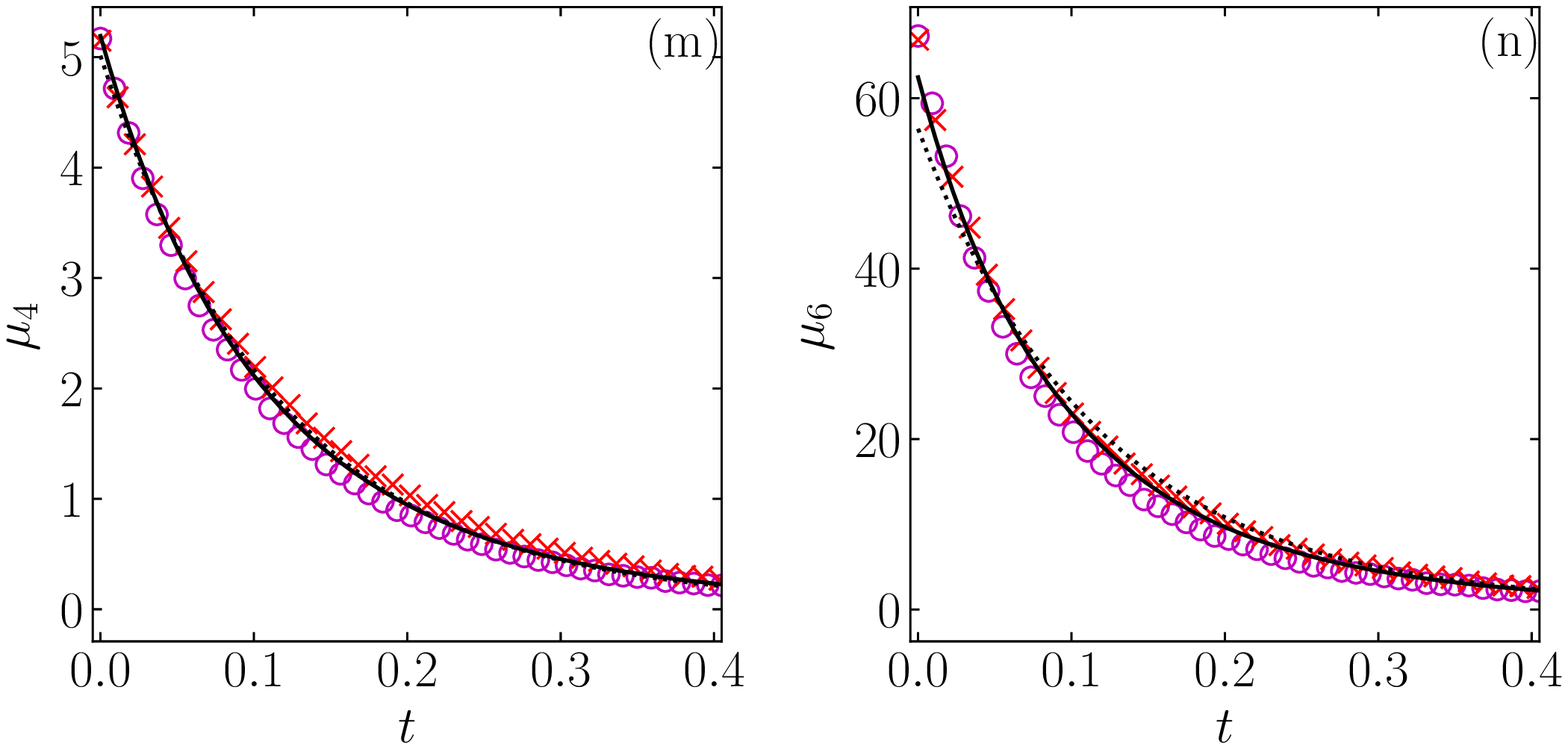}
    \caption{Evolution of the collisional moments $\mu_4$ and $\mu_6$ for $\zs=1$, $\gamma=0.1$, $d=3$,  and initial conditions $a_2^0=0.5$ and [(a) and (b)] $\theta^0=10$, [(c) and (d)] $\theta^0=3.15$,  [(e) and (f)]  $\theta^0=2$, [(g) and (h)]
     $\theta^0=1.5$, [(i) and (j)] $\theta^0=1.05$,  [(k) and (l)]  $\theta^0=1.01$, and [(m) and (n)] $\theta^0=0.91$. Symbols correspond to simulation data for DSMC ($\circ$) and EDMD ($\times$) schemes, while lines represent the theoretical predictions for ESA (---) and BSA ($\cdots$).}
    \label{fig:a2Plus_I}
\end{figure*}

\begin{figure*}[ht]
    \includegraphics[width=0.47\textwidth]{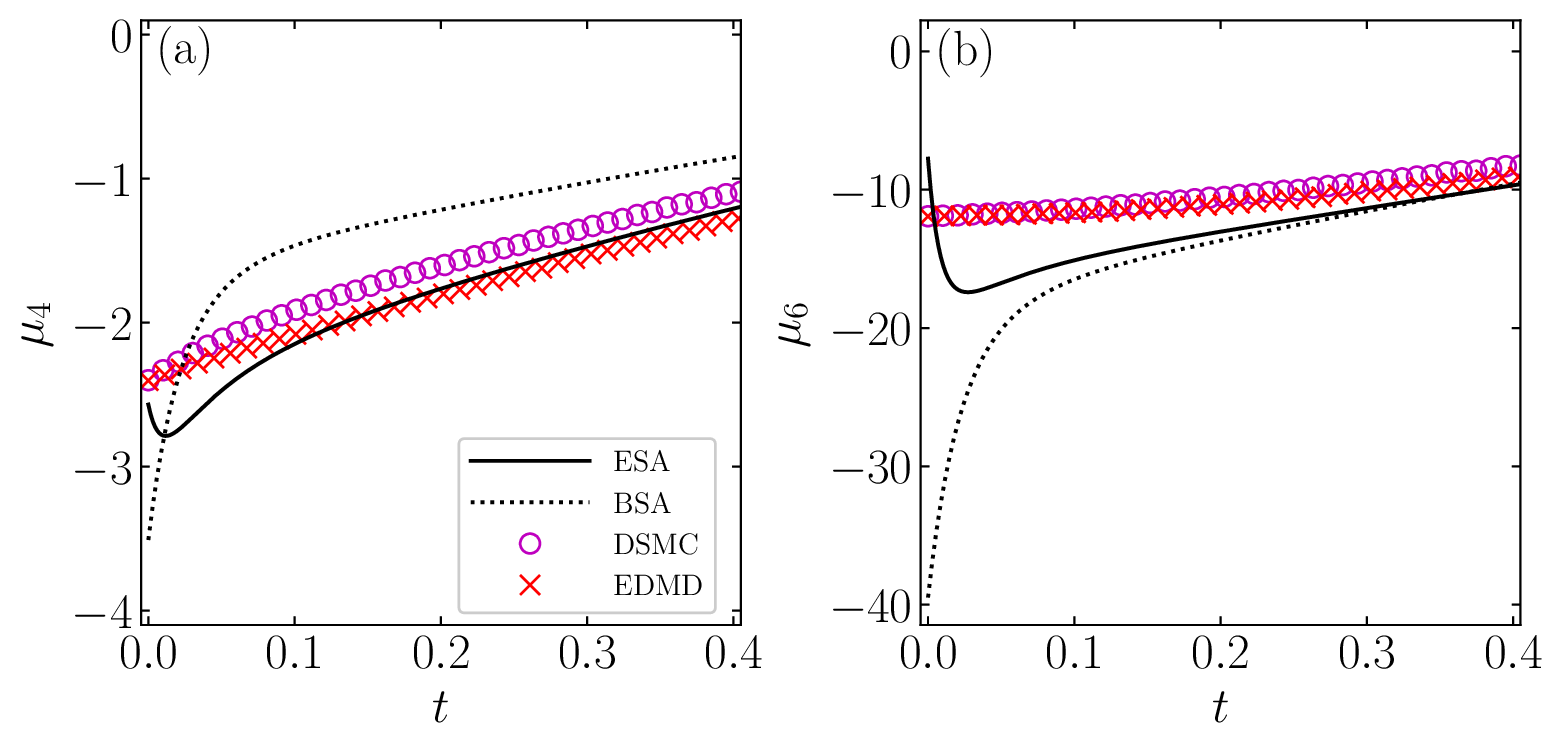}
    \includegraphics[width=0.47\textwidth]{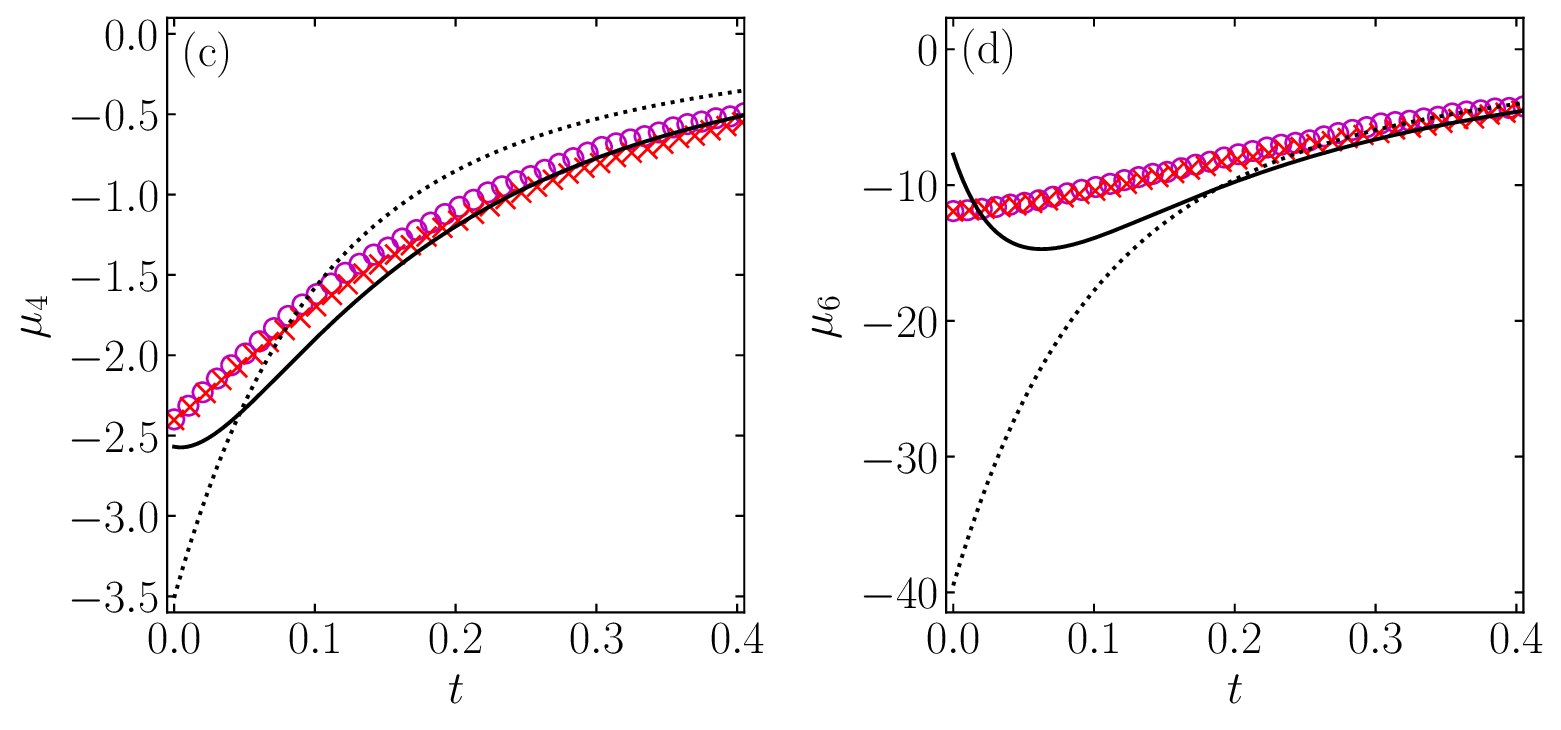}\\
    \includegraphics[width=0.47\textwidth]{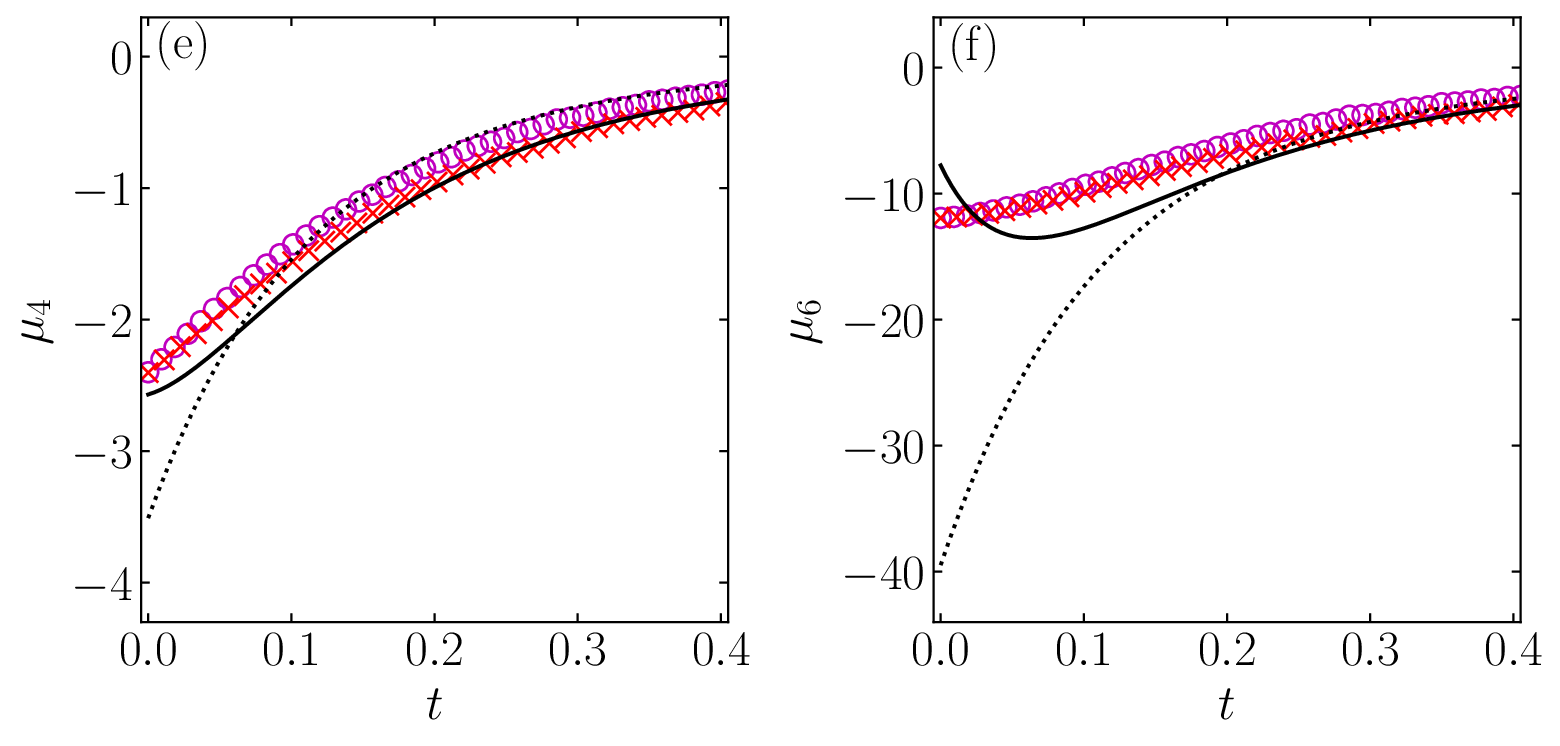}
    \includegraphics[width=0.47\textwidth]{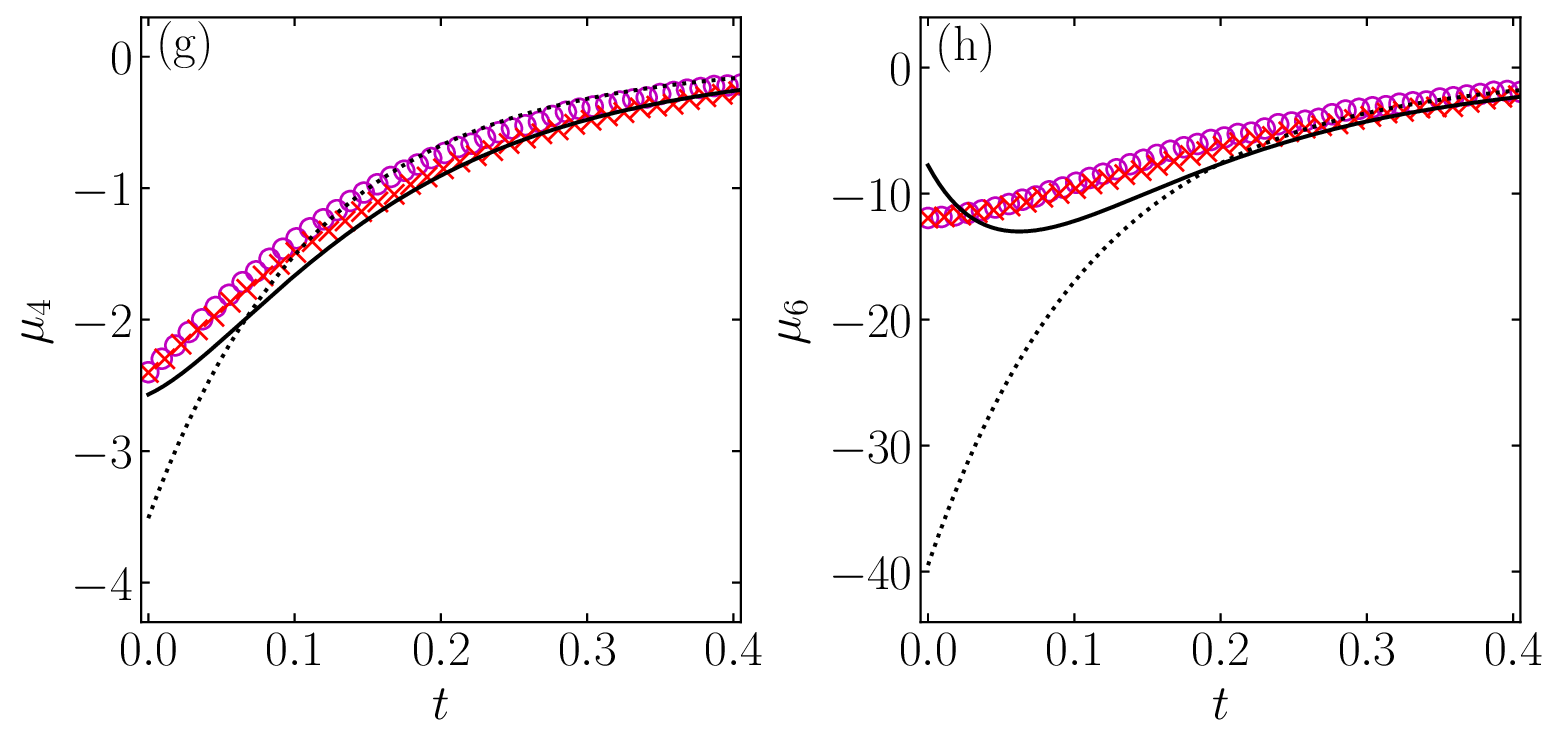}\\
    \includegraphics[width=0.47\textwidth]{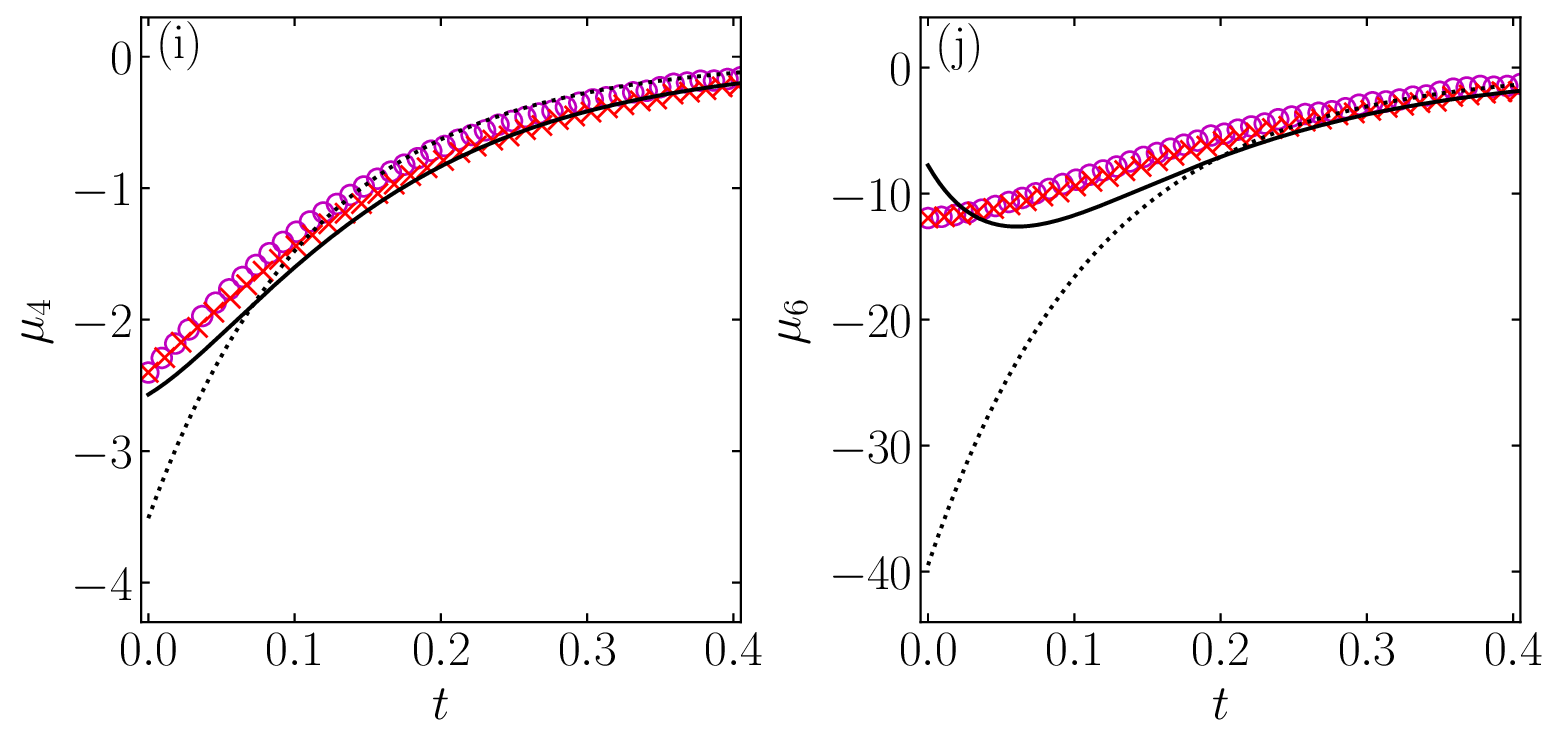}
    \includegraphics[width=0.47\textwidth]{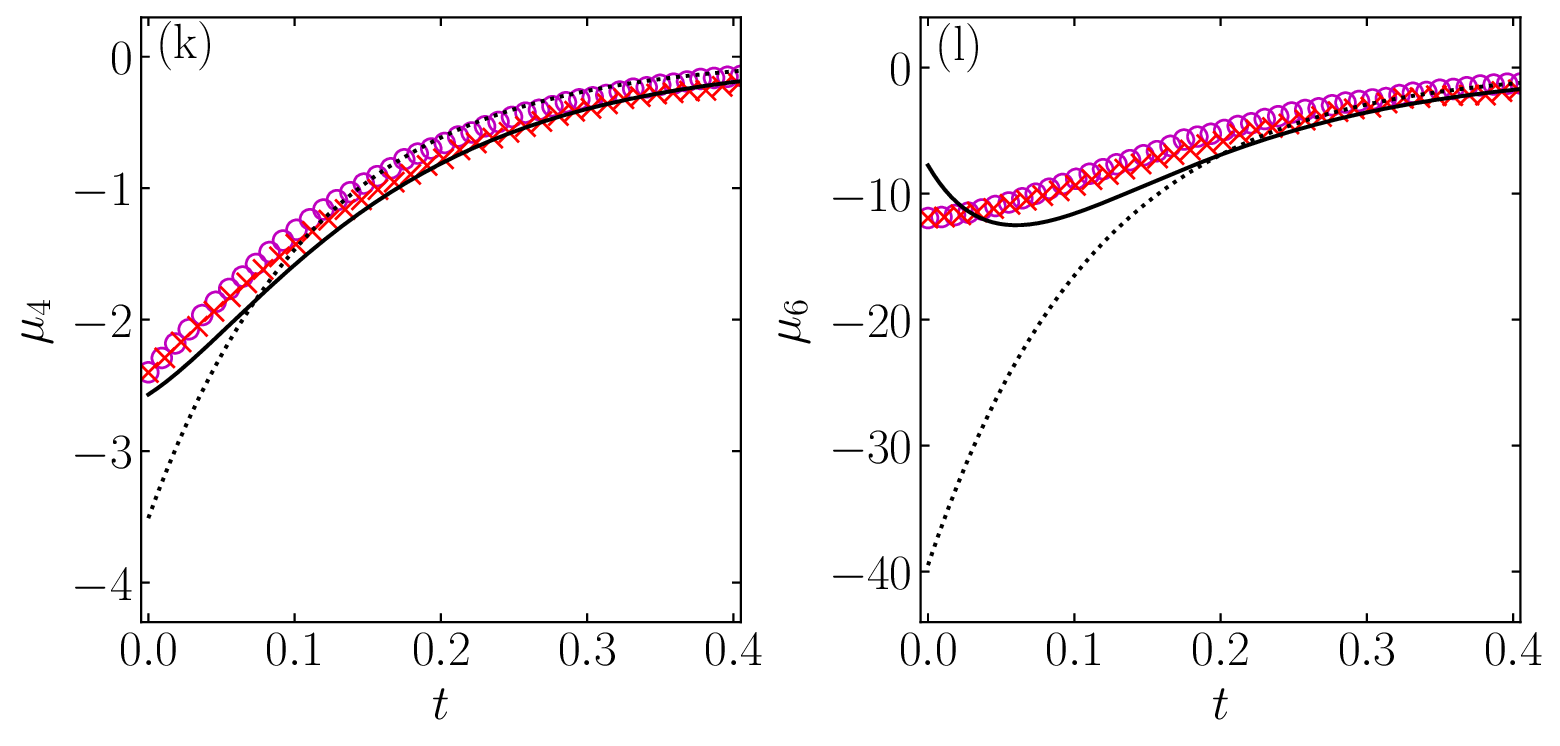}\\
    \includegraphics[width=0.47\textwidth]{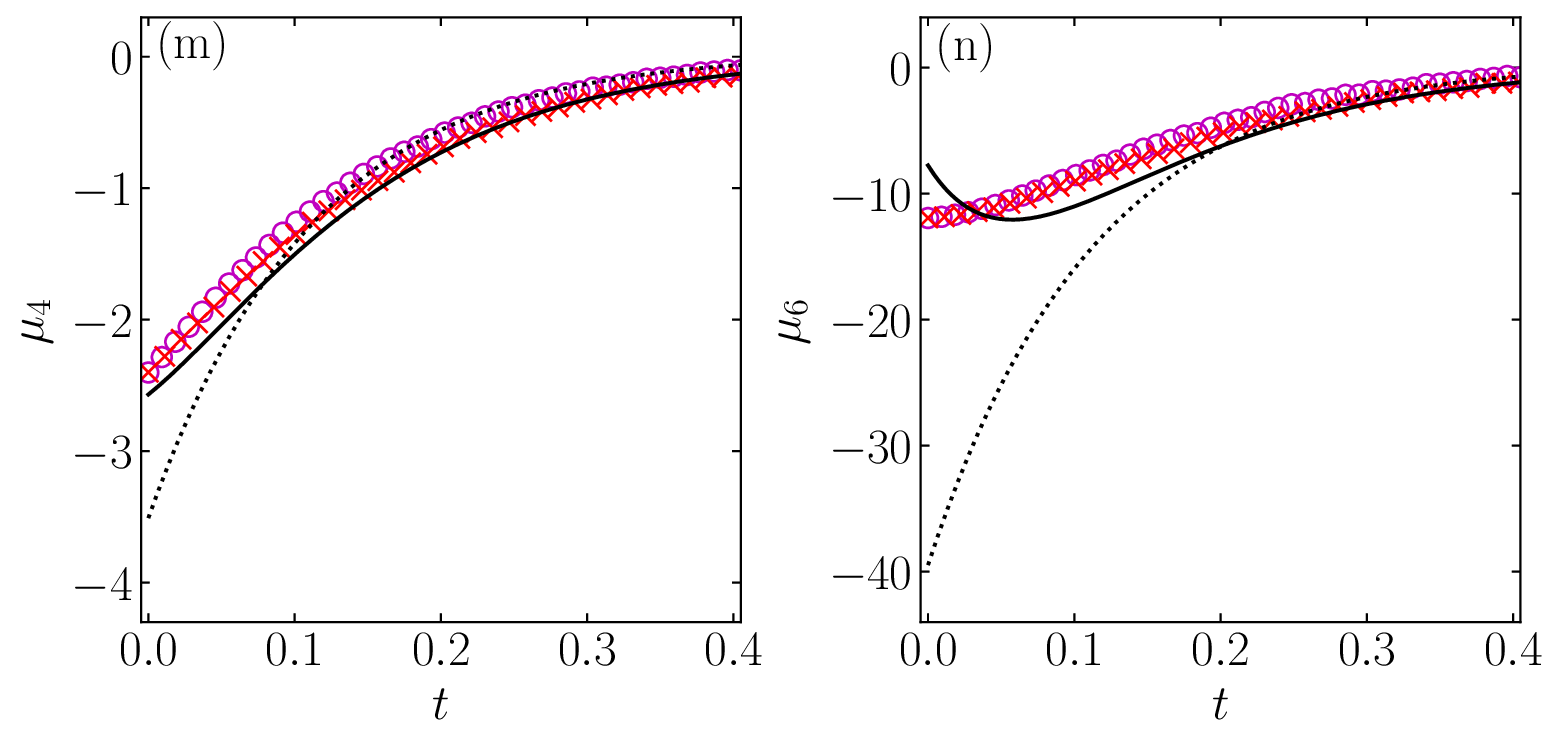}
    \includegraphics[width=0.47\textwidth]{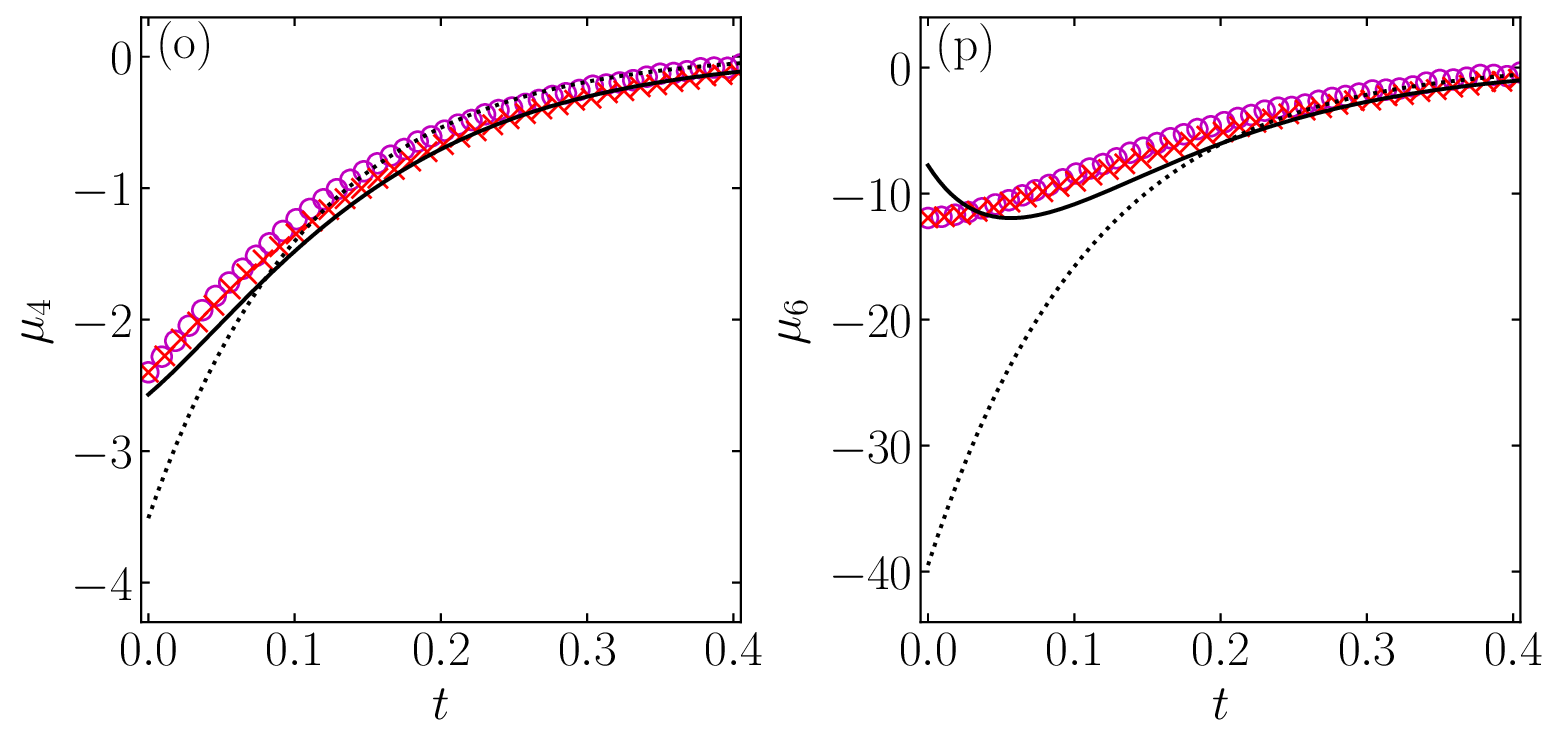}
    \caption{Evolution of the collisional moments $\mu_4$ and $\mu_6$ for $\zs=1$, $\gamma=0.1$, $d=3$, and initial conditions $a_2^0=-0.35$ and [(a) and (b)] $\theta^0=9$, [(c) and (d)] $\theta^0=1.8$,  [(e) and (f)]  $\theta^0=1.27$, [(g) and (h)] $\theta^0=1.1$, and [(i) and (j)] $\theta^0=0.99$, [(k) and (l)] $\theta^0=0.96$, [(m) and (n)] $\theta^0=0.85$, and [(o) and (p)] $\theta^0=0.82$. Symbols correspond to simulation data for DSMC ($\circ$) and EDMD ($\times$) schemes, while lines represent the theoretical predictions for ESA (---) and BSA ($\cdots$).}
    \label{fig:a2Minus_I}
\end{figure*}

\section{Computer Simulation Schemes}
\label{app:simul}

We have performed DSMC and EDMD simulations of the model to test the theoretical predictions. In both schemes, the nonlinear drag is implemented at the level of stochastic equations of motion by using Eq.~\eqref{eq:LE_eff} and applying the associated Wiener process at time $t$ within the interval $[t,t+h]$.

\subsection{Direct simulation Monte Carlo}
The implementation of DSMC for this system is based on the pioneering
work by Bird~\cite{B94,B13}, except for our taking into account of both the nonlinear drag and white-noise forces during the free-streaming stage of the algorithm~\cite{MS00}.

Let us assume a homogeneous system of $N$ particles, where their
dynamics is just controlled by their velocities $\{\vv_i\}$ with
$i=1,\dots,N$, and positions are obviated. The discrete VDF of such a
system of particles is given by
\begin{equation}
    n^{-1}f(\vv,t) = \frac{1}{N}\sum_{i=1}^{N} \delta(\vv_i(t)-\vv).
\end{equation}
At the initialization of the system, the squared moduli of the
particles velocities $\{v_i^2(0)\}$ are drawn from a gamma distribution
parameterized by $\langle v^2(0)\rangle=dk_BT^0/m$ and
$\langle v^4(0)\rangle=d(d+2)(k_BT^0/m)^2(1+a_2^0)$. Next, the
velocity vectors $\{\vv_i(0) \}$ are constructed from these moduli, with
random directions. To enforce a vanishing initial total momentum, the
velocity of every particle is subsequently subtracted by the amount
$N^{-1}\sum_i\vv_i(0)$.

After initialization, velocities are updated from $t$ to $t+h $ (where
the time step $h$ is much smaller than the mean free time) by splitting
the algorithm into two different stages: collisions and free
streaming.

During the collision stage, a number $\frac{1}{2}N\omega_{\max}h$ of
pairs are randomly chosen with uniform probability, where
$\omega_{\mathrm{max}}$ is an upper bound estimate for the collision
rate of one particle.
Then, given a pair $ij$, the collision is accepted
(acceptance-rejection Metropolis criterion) with probability
$\Theta(\vv_{ij}\cdot\widehat{\bm{\sigma}}_{ij})\omega_{ij}/\omega_{\max}$,
where $\widehat{\bm{\sigma}}_{ij}$ a random vector in the unit
$d$-sphere and
$\omega_{ij} = \Omega_d \sigma^{d-1}n
g_c|\vv_{ij}\cdot\widehat{\bm{\sigma}}_{ij}|$ with
$\Omega_d = 2\pi^{d/2}/\Gamma(d/2)$ being the $d$-dimensional solid
angle. If the collision is accepted for the given pair, then
postcollisional velocities are assigned following the collisional
rules for elastic hard spheres, namely
$\vv_{i,j}\to
\vv_{i,j}\mp(\vv_{ij}\cdot\widehat{\bm{\sigma}}_{ij})\widehat{\bm{\sigma}}_{ij}$.
If $\omega_{ij}>\omega_{\max}$ in one of the sampled pairs, then the
collision is accepted and the estimate is updated as
$\omega_{\max}=\omega_{ij}$.

During free streaming, velocities are updated according to the scheme
given by \eqref{eq:LE_eff}, namely
\begin{align}
\label{eq:LE_num}
    \vv_i(t)\to \vv_i(t+h)\approx &\vv_i(t)-\zeta_{\mathrm{eff}}(v_i(t))\vv_i(t) h \nonumber\\
    &+\xi(v_i(t))\sqrt{h}\mathbf{Y}_i+  \mathcal{O}(h^{3/2}),
    \end{align}
where $\mathbf{Y}_i$ is a random vector drawn from the Gaussian probability distribution
\begin{equation}
\label{P(Y)}
    P(\mathbf{Y}) = (2\pi)^{-d/2}e^{-Y^2/2}.
\end{equation}

In our DSMC algorithm we took $N=10^4$ three-dimensional particles ($d=3$) and chose a time step $h=10^{-2}\lambda/\sqrt{2k_BT_\bb/m}$, where $\lambda=(\sqrt{2}\pi n\sigma^2)^{-1}$ is the mean free path.

\subsection{Event-driven molecular dynamics}
\label{subsec:EDMD}

EDMD algorithms are based on the evolution driven by events which can
be particle-particle collisions, boundary effects, or other more
complex interactions. Between two consecutive events, there is a free
streaming of particles. Again, the stochastic and drag forces directly
influence the particle dynamics. Whereas in DSMC positions were not
required, in EDMD they are essential and are affected by the
nonlinear noise, as explained below.

In order to implement the effect of the Langevin dynamics in our EDMD
simulations, we have followed the AGF algorithm proposed in Ref.~\cite{S12}.  Since
$\dot{\rr}_i(t)=\vv(t)$, Eq.~\eqref{eq:LE_num} must be supplemented
by \cite{S12}
\begin{align}
\rr_i(t)\to\rr_i(t+h)\approx & \rr_i(t)+\vv_i(t)h\left[1-\frac{\zeta_{\mathrm{eff}}(v_i(t))}{2}h \right] \nonumber \\
&+\frac{1}{2}\xi(v_i(t))h^{3/2}\mathbf{W}_i+\mathcal{O}(h^{5/2}), \end{align}
with \begin{equation}
\mathbf{W}_i=\mathbf{Y}_i+\sqrt{\frac{5}{3}}\bar{\mathbf{Y}}_i, \end{equation}
where we have particularized the algorithm to the three-dimensional
geometry, $\mathbf{Y}_i$ is the random vector appearing in Eq.~\eqref{eq:LE_num}, and $\bar{\mathbf{Y}}_i$ is an independent random
vector, also drawn from the Gaussian distribution \eqref{P(Y)}.
Notice that, since the equation for $\vv_i(t)$ is expanded up to
$\mathcal{O}(h^{3/2})$ and $\rr_i(t+h)-\rr_i(t) \sim \vv_i(t) h$, then
the equation for $\rr_i(t)$ needs to be expanded up to
$\mathcal{O}(h^{5/2})$.

In our EDMD simulations, we deal with a system of $N=1.065\times 10^4$
spheres and reduced number density $n\sigma^3=10^{-3}$. The time step
is $h=10^{-3}\lambda/\sqrt{2k_BT_\bb/m}$, and periodic
boundary conditions are used.

\subsection{Test of the time evolution of the collisional moments $\mu_4$ and $\mu_6$}
\label{subsec:moments_test}

Let us present now a comparison between the collisional moments
$\mu_4$ and $\mu_6$ measured in simulations and those provided by the
ESA and the BSA. In the ESA, those collisional moments are given by
Eqs.~\eqref{eq:mu4-mu6}, complemented with the numerical solution of
Eqs.~\eqref{eq:thetadot} and
\eqref{eq:cumul-evol-Sonine}. Analogously, in the BSA, the collisional
moments are given by Eqs.~\eqref{eq:mu4-mu6} with $a_3\to 0$,
complemented with the numerical solution of Eqs.~\eqref{eq:thetadot}
and \eqref{eq:a2-evol-Sonine}, again with $a_3\to 0$.

In the simulations (both DSMC and EDMD), the following numerical
scheme has been used to address the computation of the collisional
moments \cite{MS00,SM09}. We randomly choose $N^\prime = 10^6$ pairs
of particles out of the total number $N(N-1)/2=5\times 10^7$ and
approximate the collisional moments as
\begin{equation}
    \mu_\ell = \frac{1}{N^\prime}\sum_{ij}^{N^\prime} \Phi_\ell(\cc_i,\cc_j),
  \end{equation}
where $\Phi_4$ and $\Phi_6$ are given by Eqs.~\eqref{Phi4Phi6}.

Figures~\ref{fig:a2Plus_I} and \ref{fig:a2Minus_I} show the time
evolution of $\mu_4$ and $\mu_6$, as measured in our DSMC and EDMD
simulations and as predicted by the BSA and ESA, for a number of
representative initial conditions. A very good agreement between DSMC
and EDMD is found; the small differences between them could explain the deviations observed in Figs.~\ref{fig:OME_cases_b}--\ref{fig:T_E_12}. There is also a
general good agreement with the ESA results, while the BSA predictions
exhibit important deviations in the initial stage, especially in the
case of $\mu_6$. The deviations of the BSA and ESA are due to the
nonnegligible role played by nonlinear terms of the form $a_2^2$,
$a_3^2$, and $a_2 a_3$, as well as by higher-order
cumulants. Interestingly, those deviations are more relevant for
negative $a_2^0$ than for positive $a_2^0$ and tend to increase as the
initial temperature $\theta^0$ grows---in accordance with the results
in Ref.~\cite{PSP21} for a quench to low temperatures.

\section{Evolution of the cumulants $a_2$ and $a_3$}
\label{app:cum}

\begin{figure}[H]
    \includegraphics[width=\columnwidth]{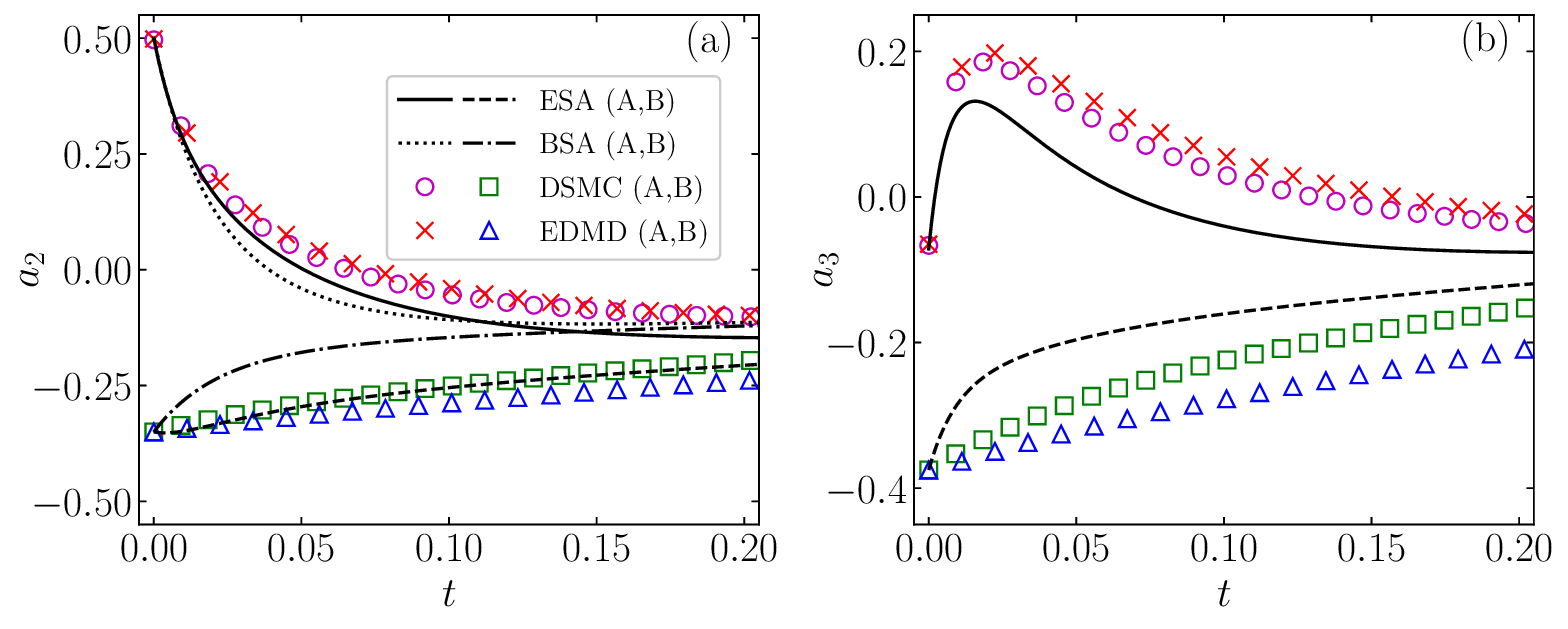}\\
    \includegraphics[width=\columnwidth]{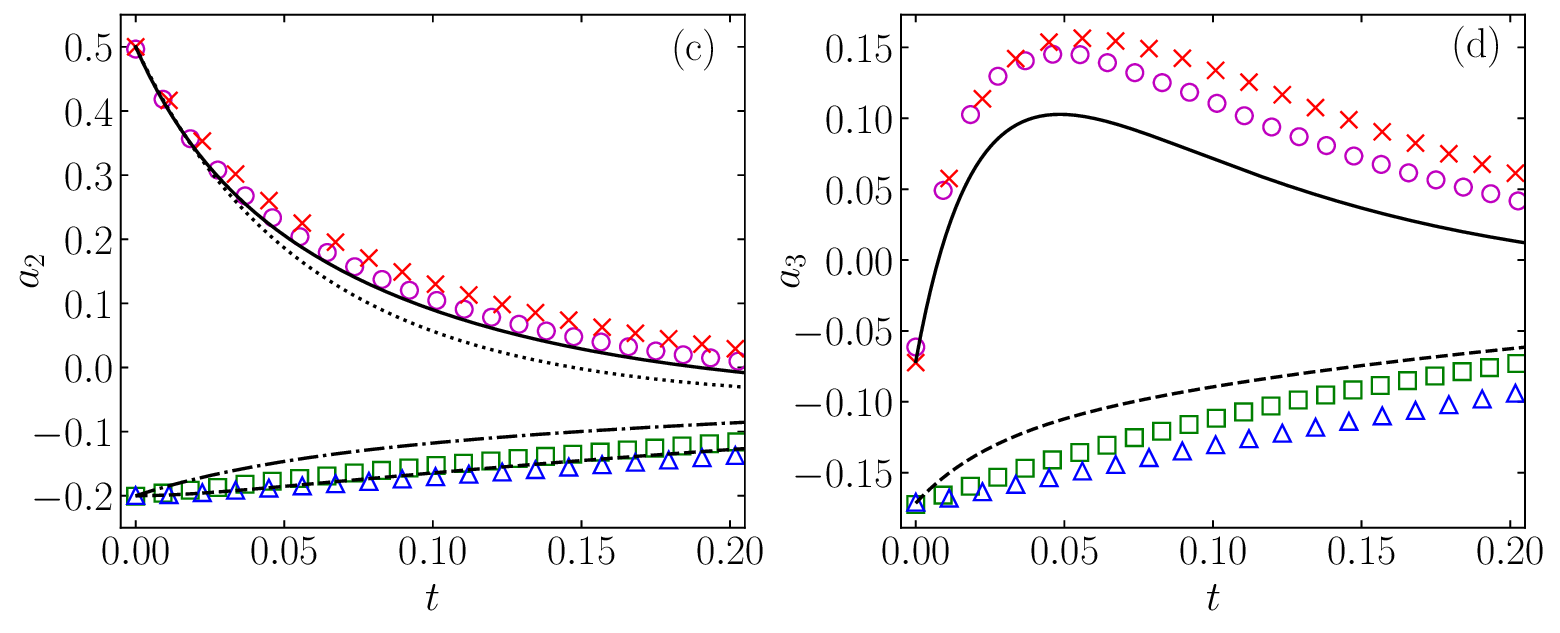}\\
    \includegraphics[width=\columnwidth]{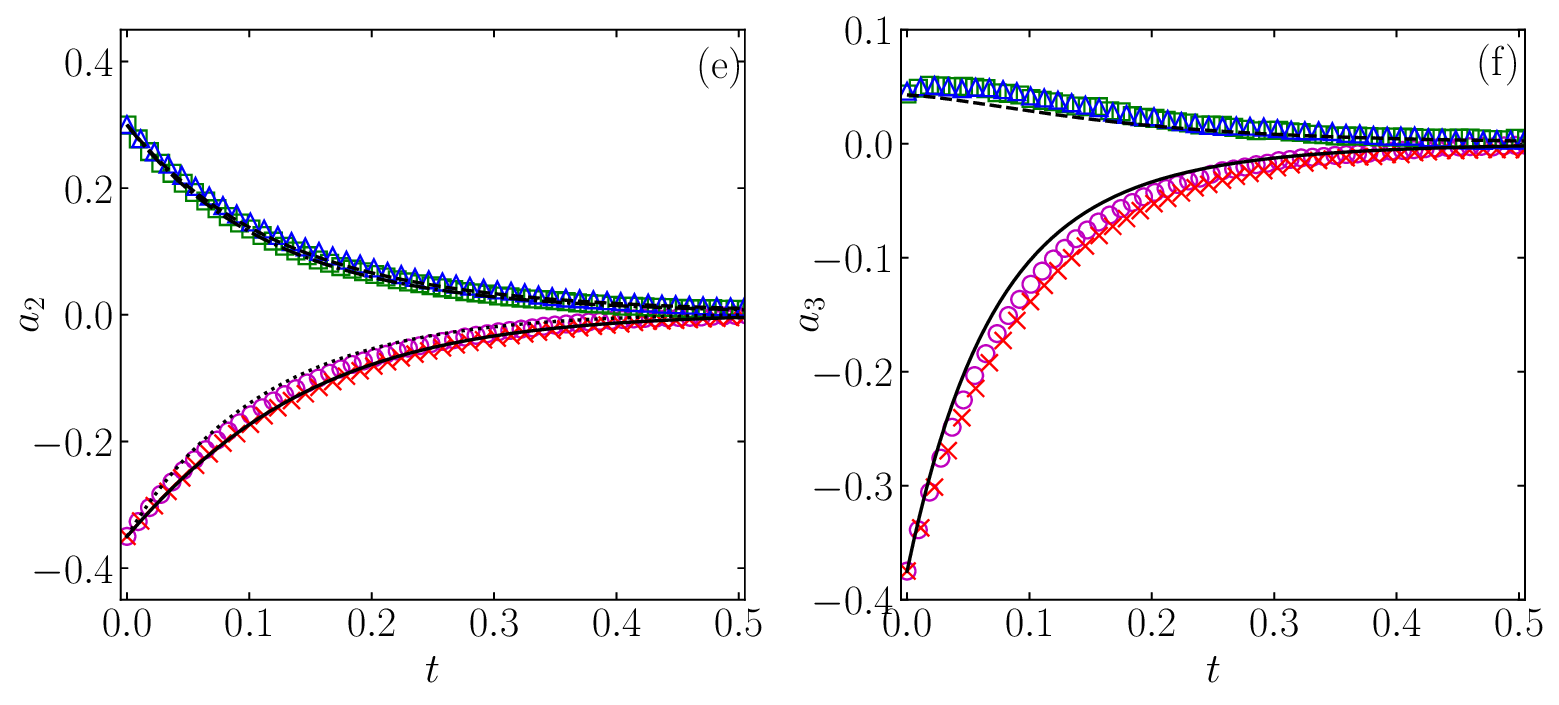}\\
    \includegraphics[width=\columnwidth]{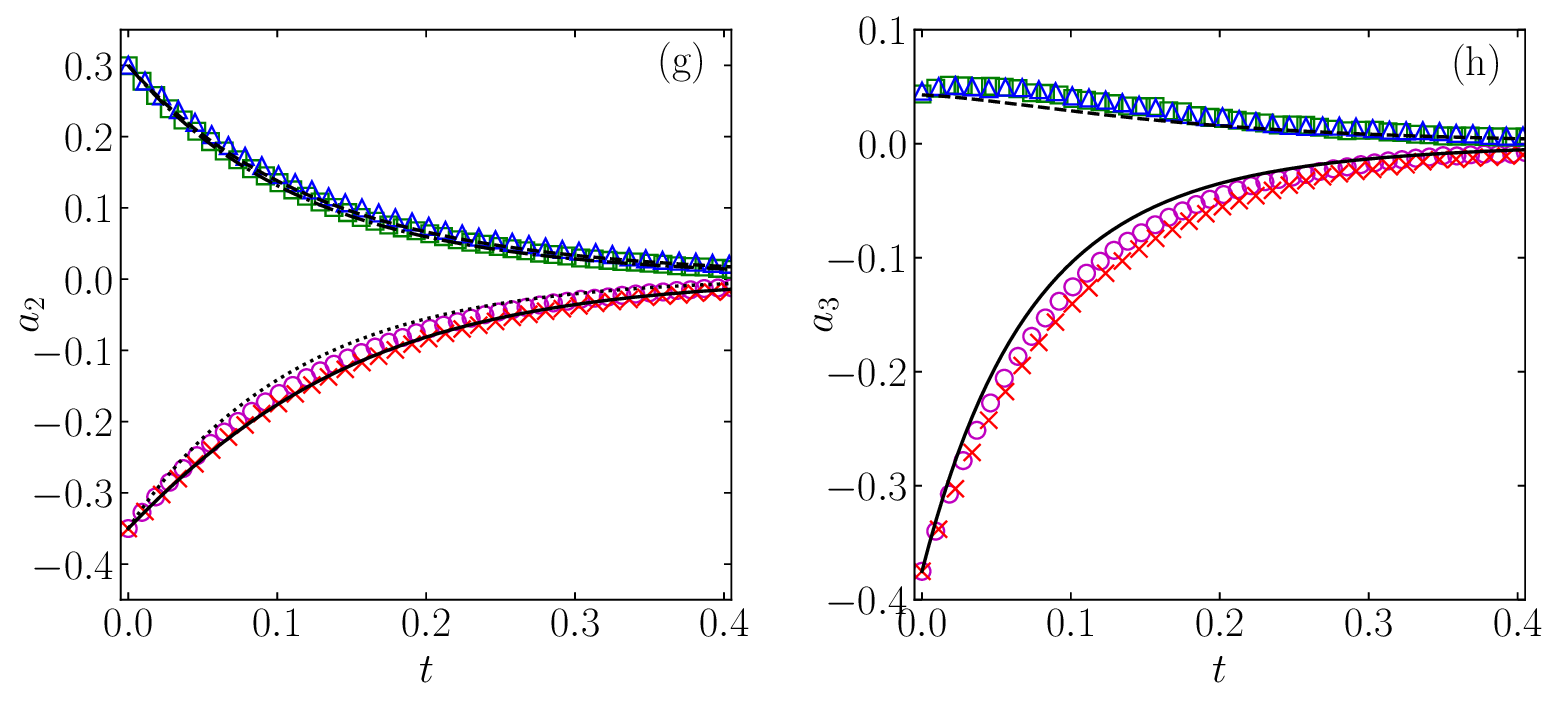}
    \caption{Same as in Fig.~\ref{fig:ET_TE_12}, except that the quantities plotted are $a_2$ and $a_3$.}
    \label{fig:ET_TE_12_cum}
\end{figure}

\begin{figure}[H]
    \includegraphics[width=\columnwidth]{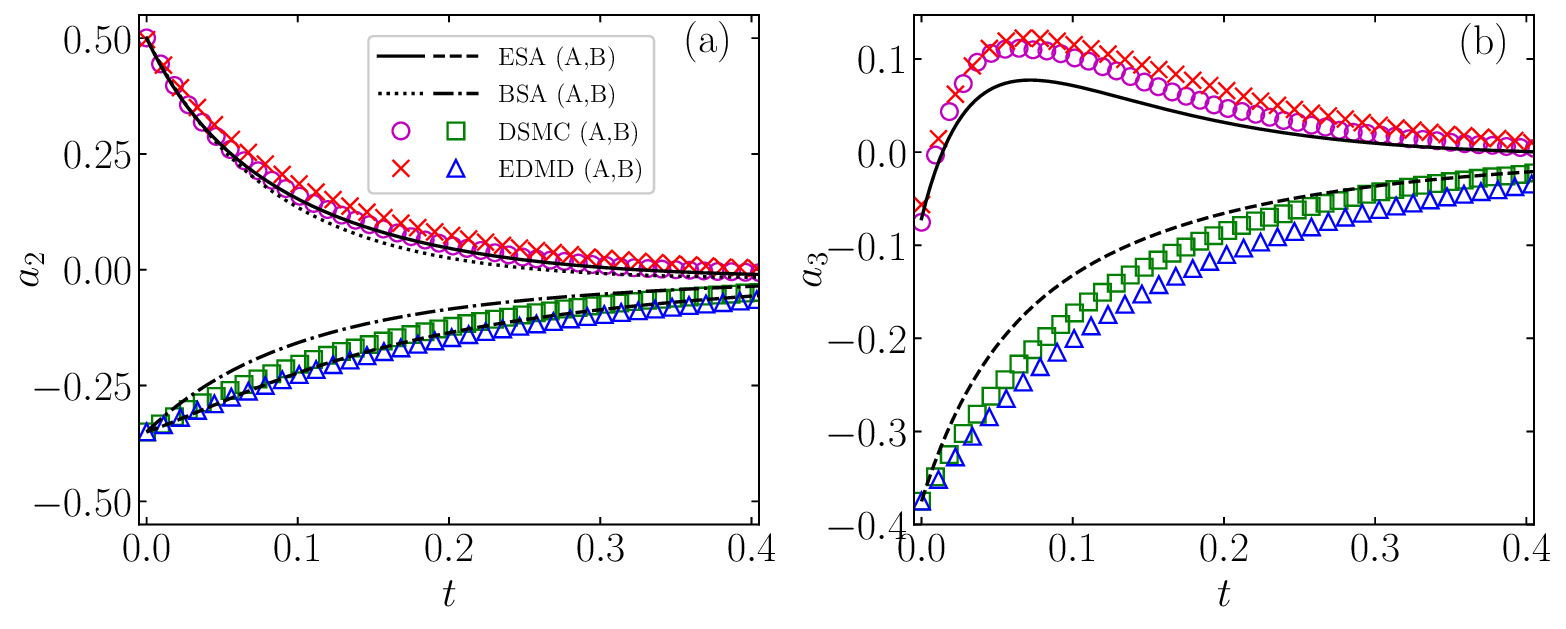}\\
    \includegraphics[width=\columnwidth]{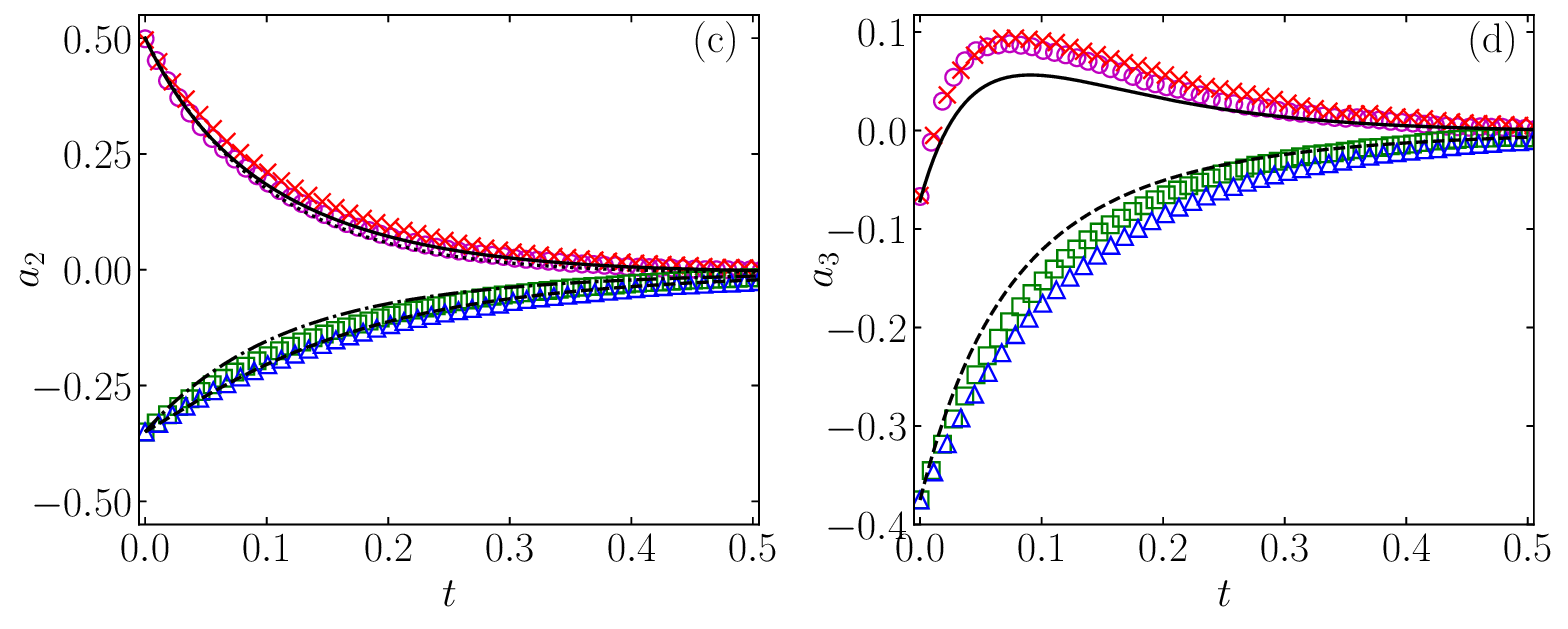}\\
    \includegraphics[width=\columnwidth]{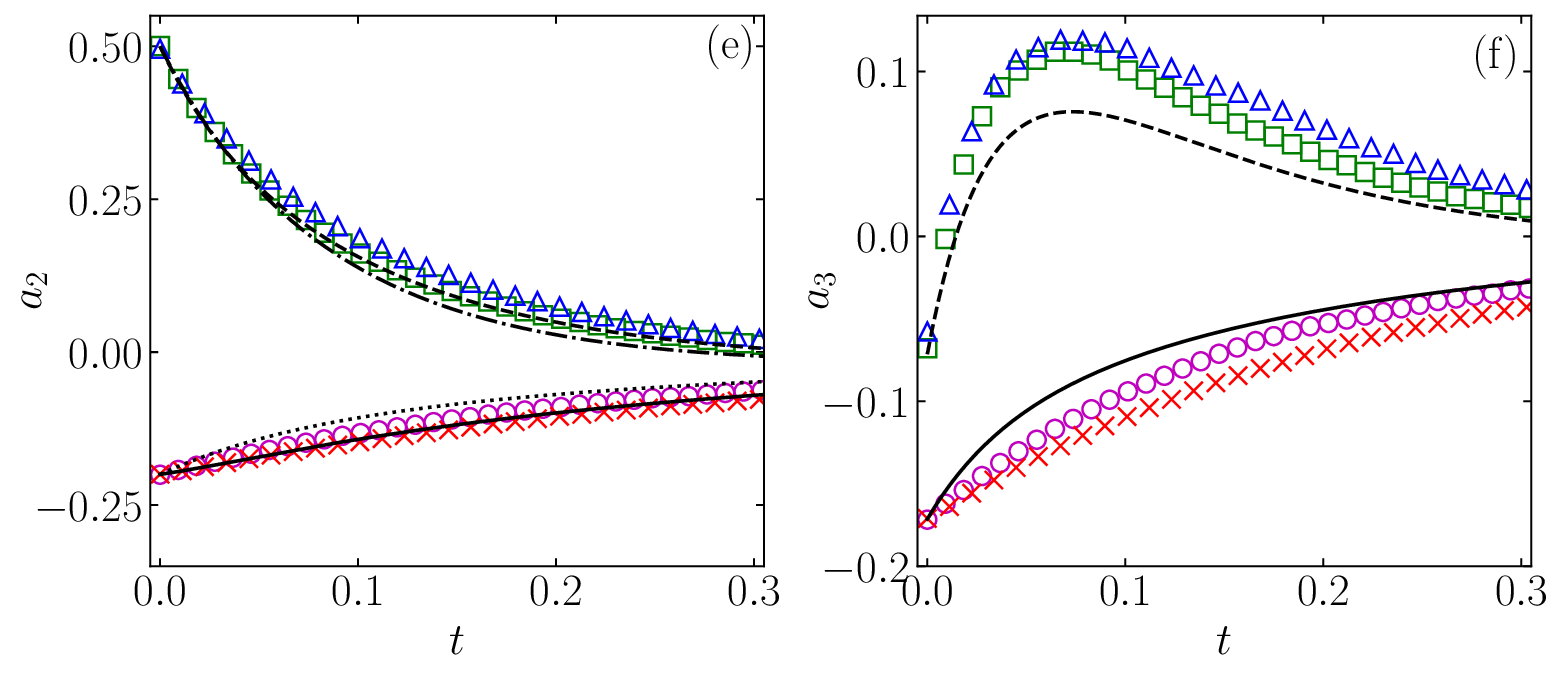}\\
    \includegraphics[width=\columnwidth]{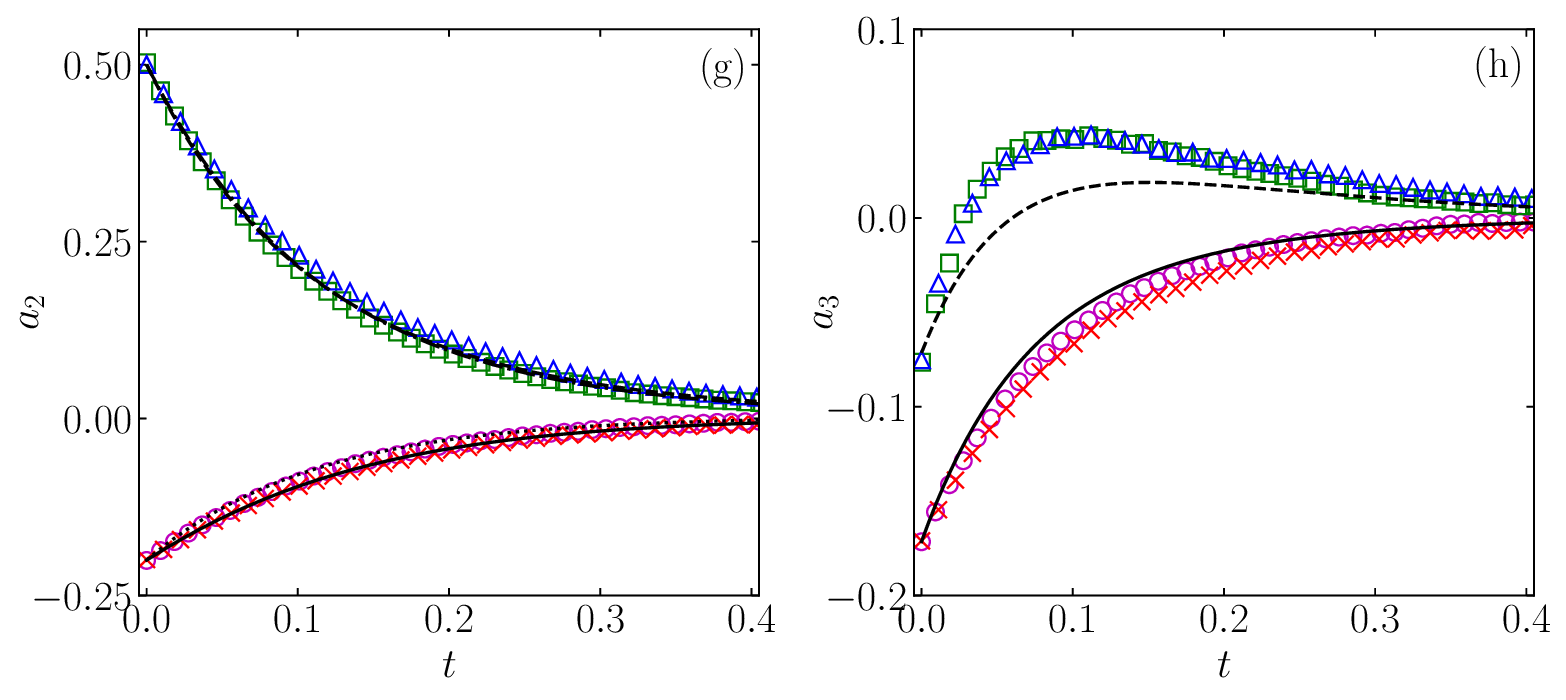}\\
    \includegraphics[width=\columnwidth]{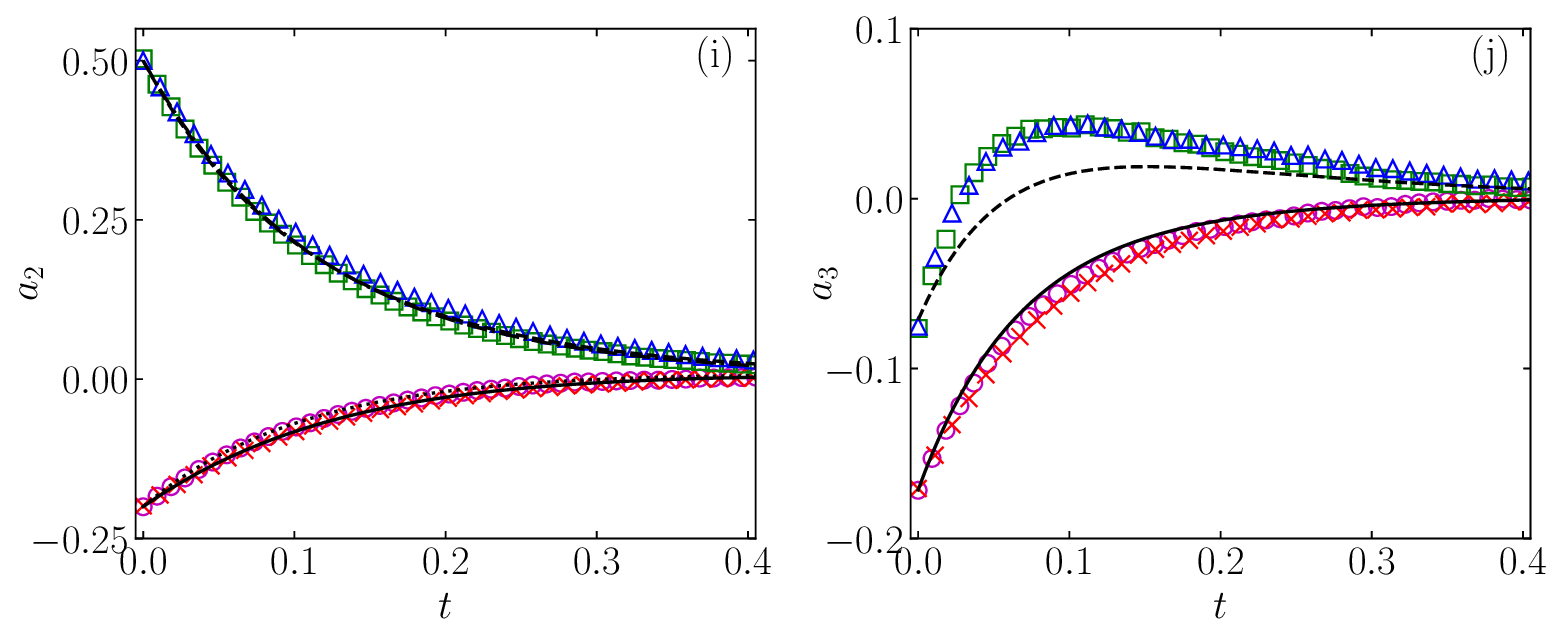}
    \caption{Same as in Fig.~\ref{fig:T_E_12}, except that the quantities plotted are $a_2$ and $a_3$.}
    \label{fig:T_E_12_cum}
\end{figure}

\begin{figure}[ht]
    \includegraphics[width=\columnwidth]{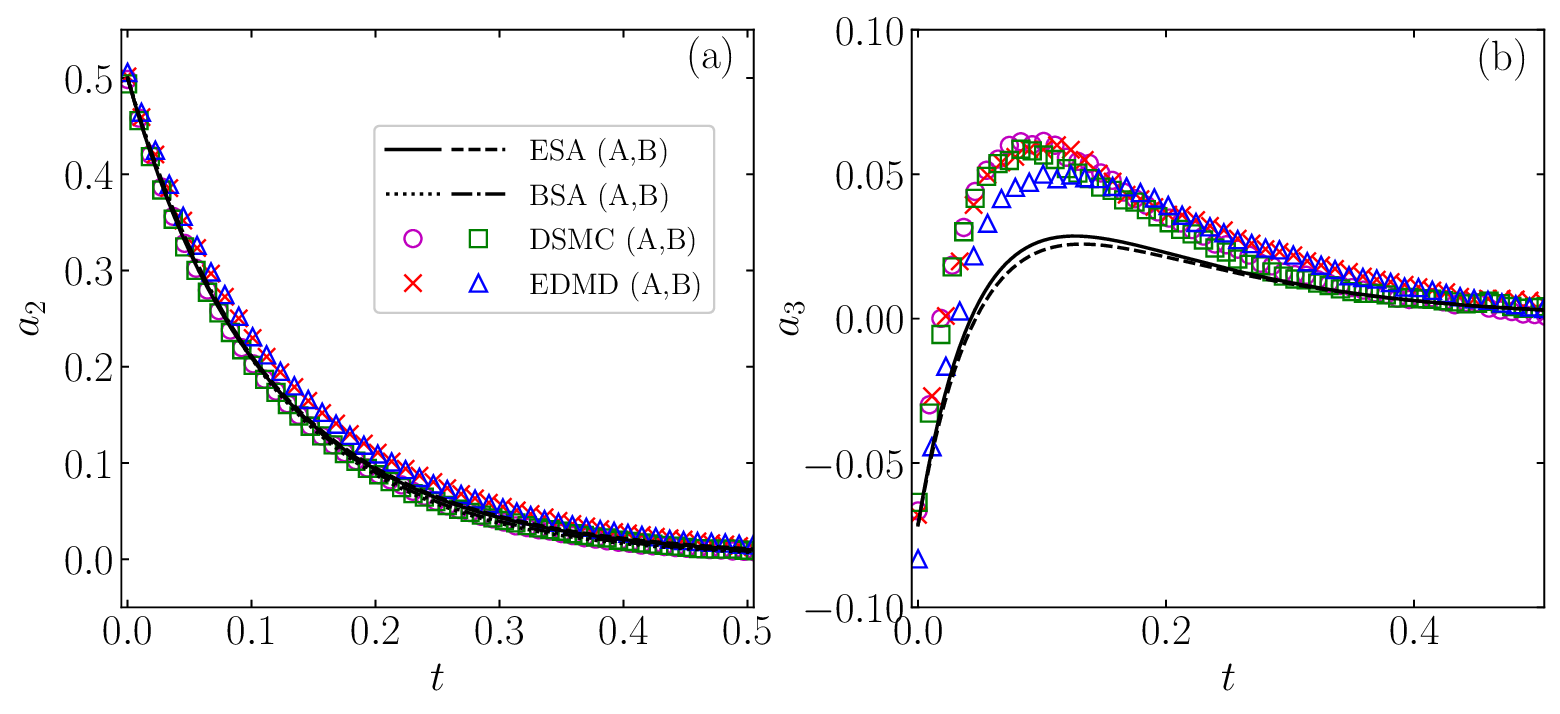}\\
    \includegraphics[width=\columnwidth]{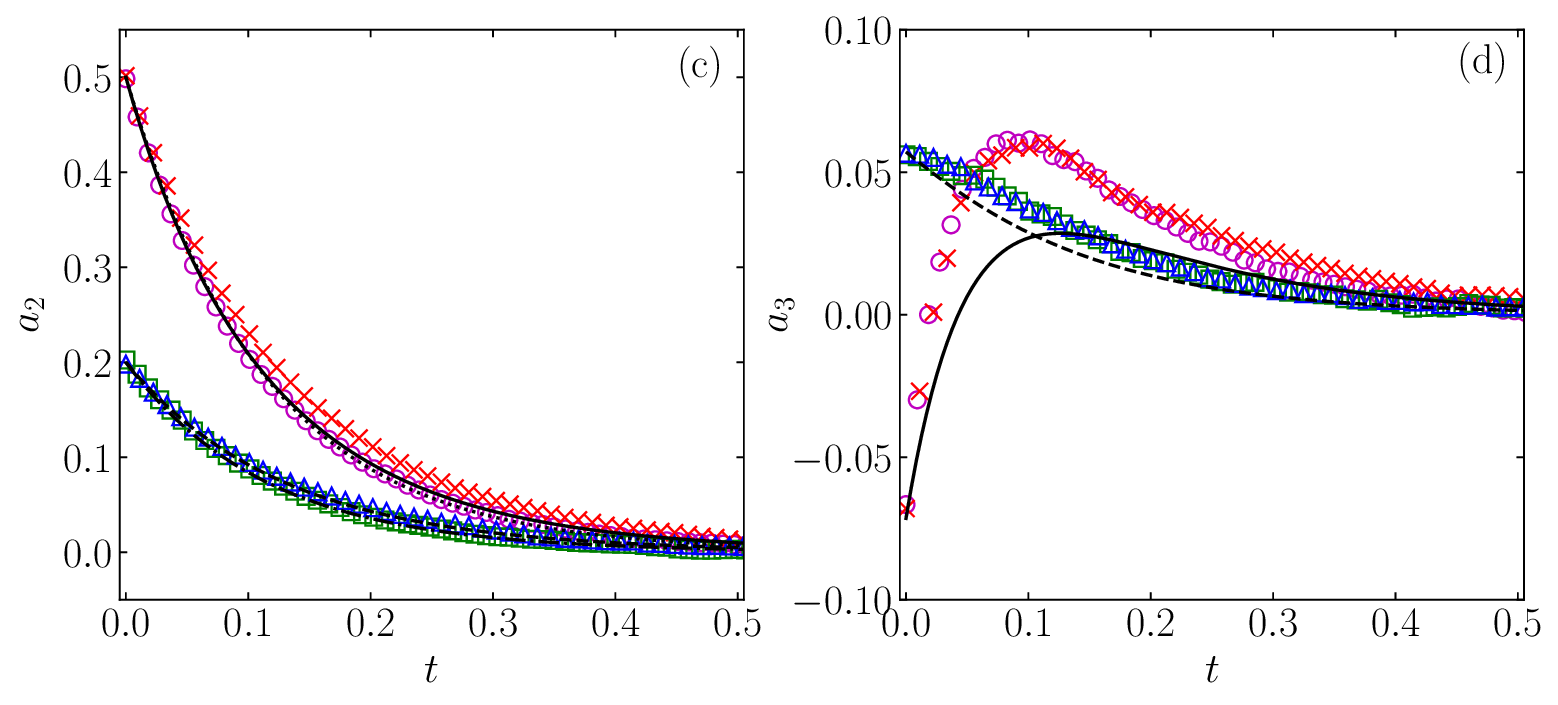}\\
    \includegraphics[width=\columnwidth]{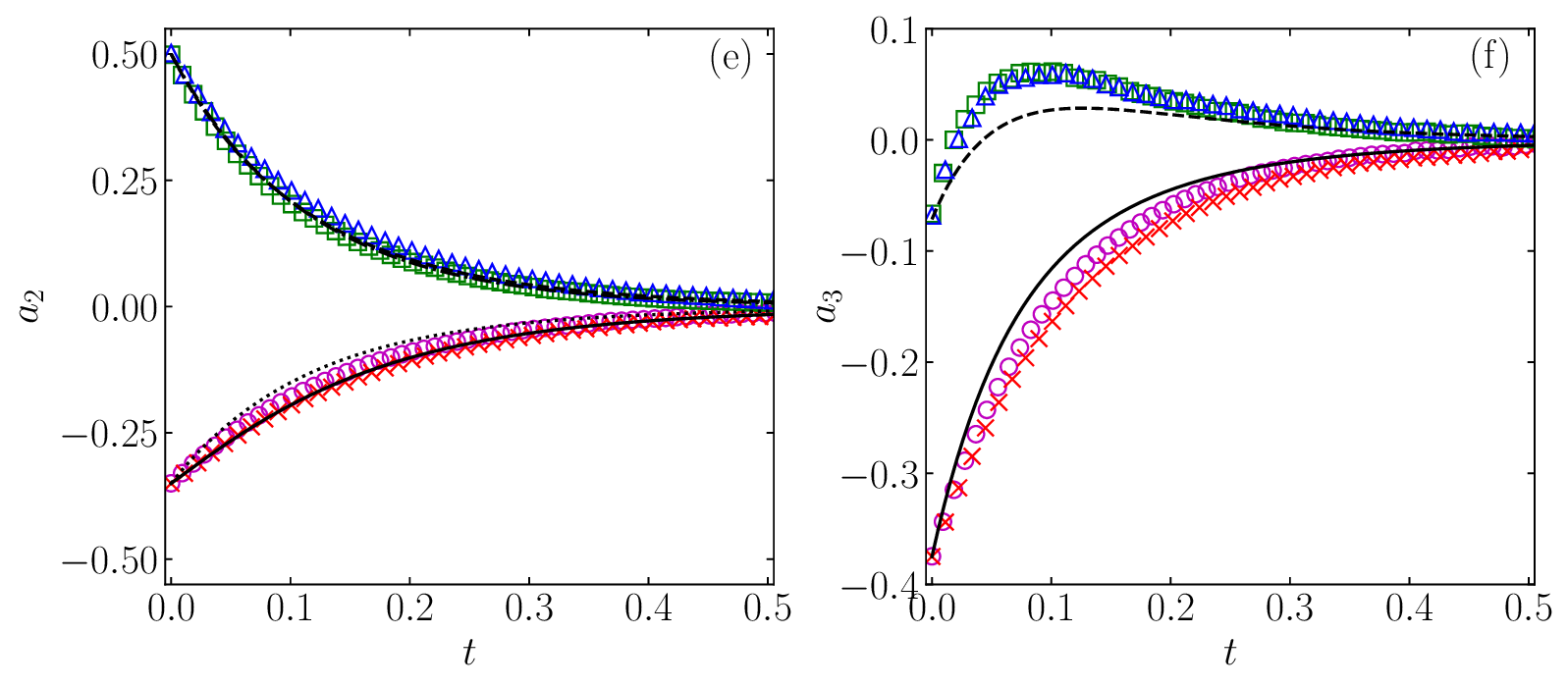}
    \caption{
    Same as in Fig.~\ref{fig:OME_I}, except that the quantities plotted are $a_2$ and $a_3$. }
    \label{fig:OME_I_cum}
\end{figure}

\begin{figure}[H]
    \includegraphics[width=\columnwidth]{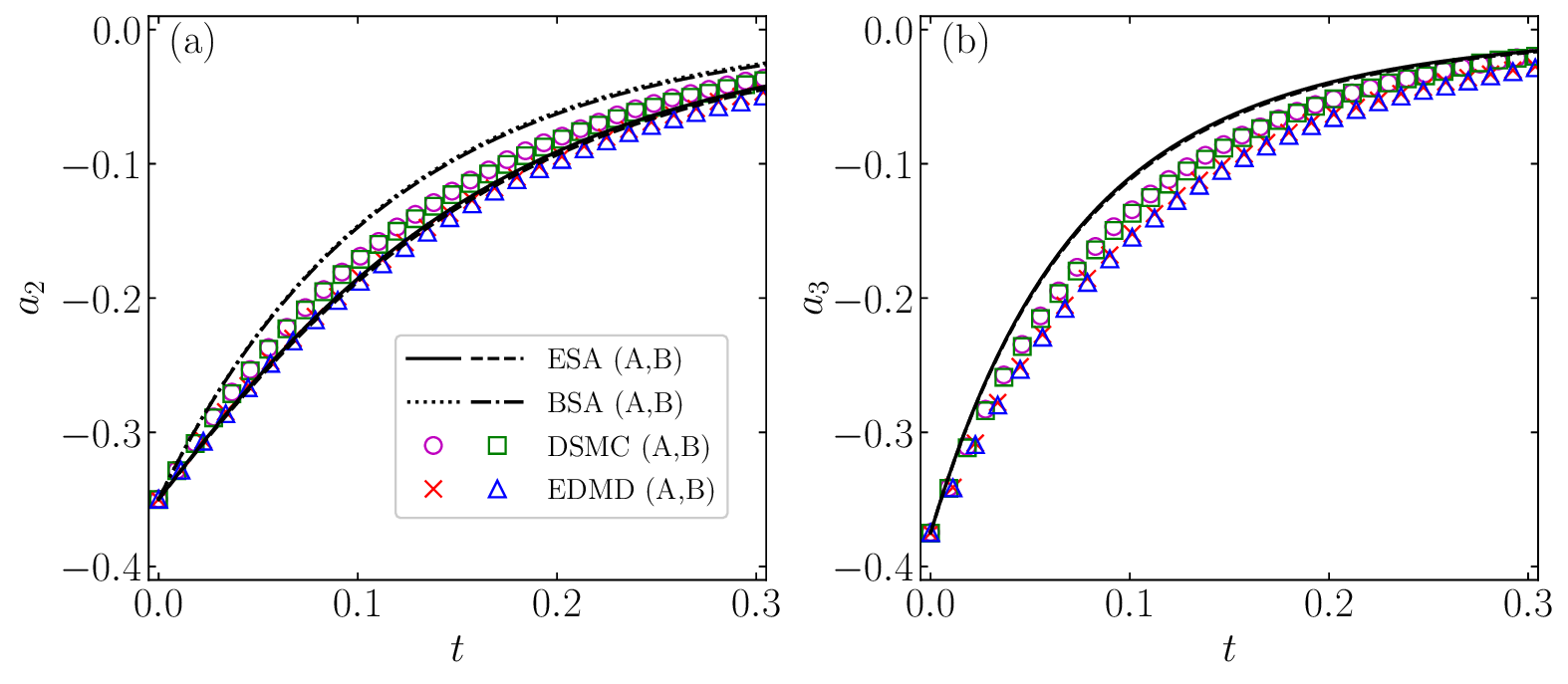}\\
    \includegraphics[width=\columnwidth]{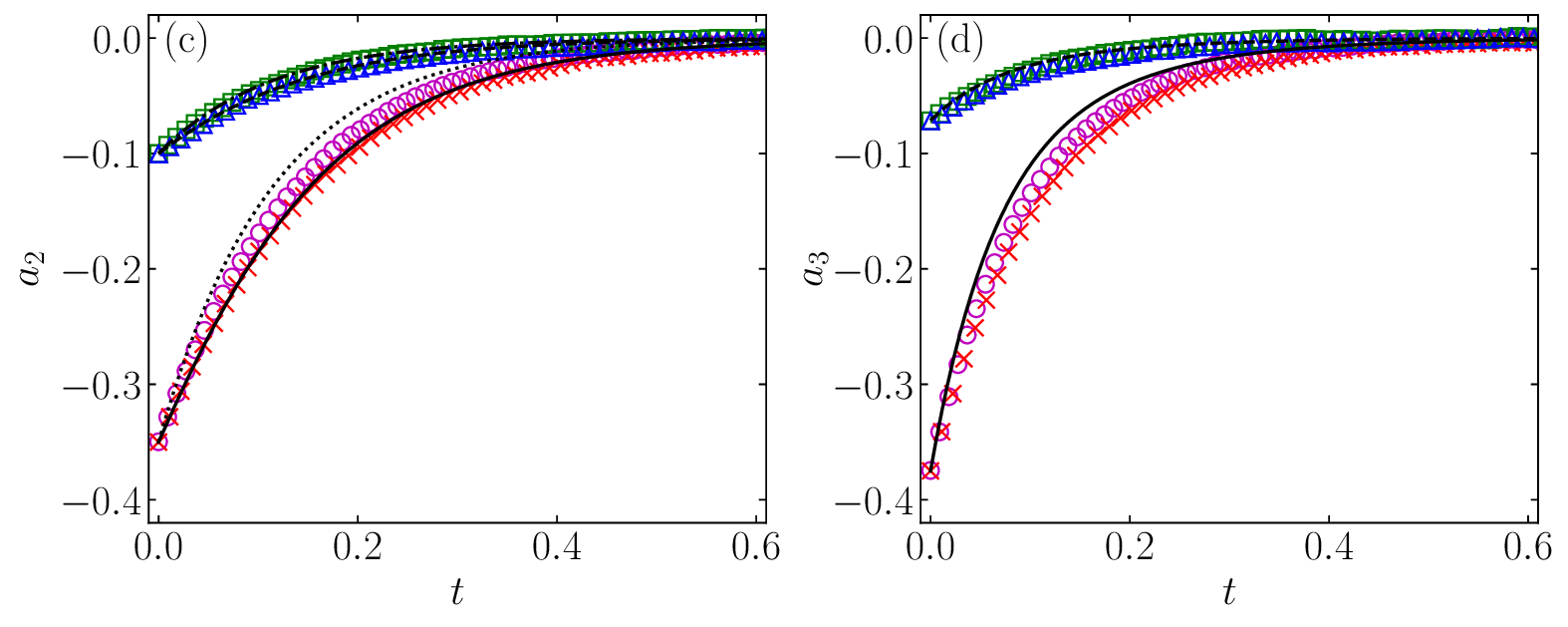}\\
    \includegraphics[width=\columnwidth]{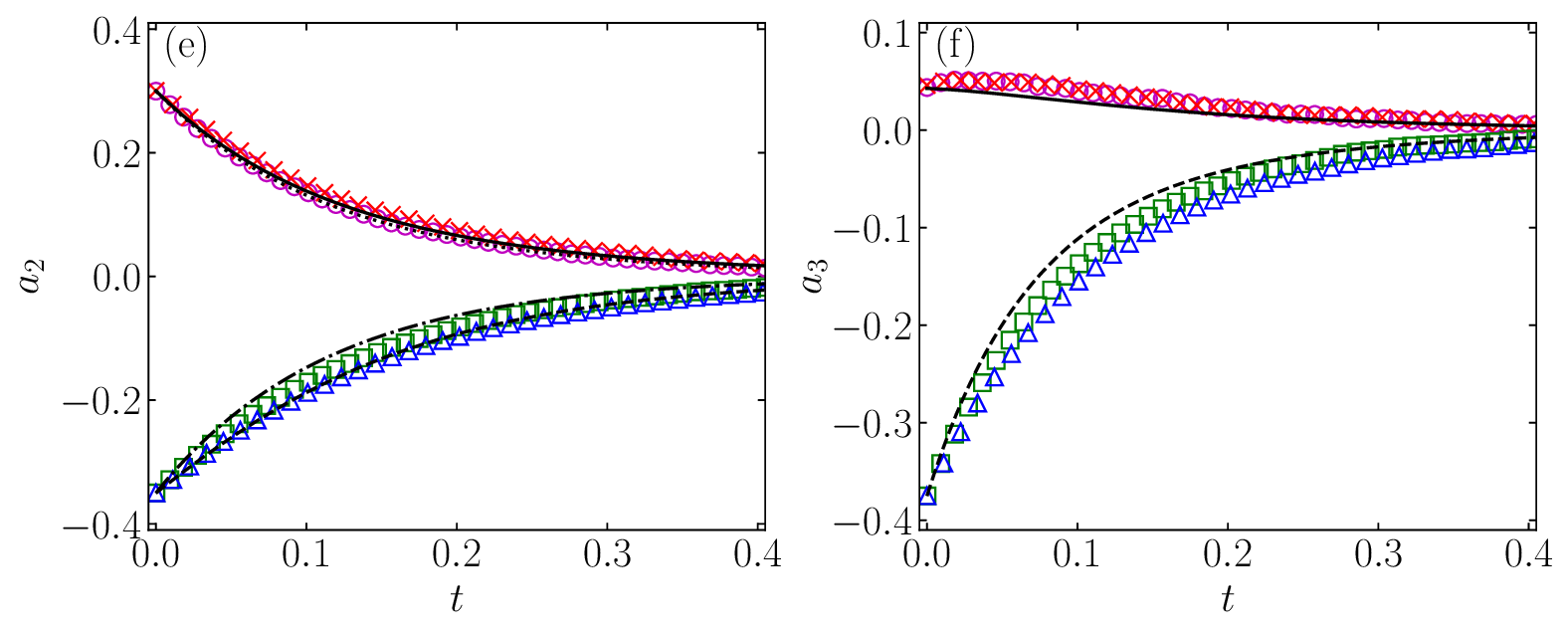}
    \caption{
    Same as in Fig.~\ref{fig:OME_II}, except that the quantities plotted are $a_2$ and $a_3$.}
    \label{fig:OME_II_cum}
\end{figure}

In this Appendix we present a comparison between the simulation results for the cumulants $a_2$ and $a_3$, and the theoretical predictions BSA (for $a_2$ only) and ESA. The results are displayed in Figs.~\ref{fig:ET_TE_12_cum}--\ref{fig:OME_II_cum}.

\begin{thebibliography}{108}
\bibitem [{not({\natexlab{a}})}]{note_22_01_1}%
  \BibitemOpen
  \href@noop {} {}  \bibinfo {note} {{S}trictly speaking, the
  KLD does not have the properties of a distance; e.g., it does not satisfy the
  triangle inequality.}\BibitemShut {Stop}%
\bibitem [{not({\natexlab{b}})}]{note_22_01_2}%
  \BibitemOpen
  \href@noop {} {}  \bibinfo {note} {{N}ote that, because of
  the second equality in Eq.\ \eqref{eq:KLD^LE}, the entropic ``distance''
  $\KLD$ from $f$ to $f^\eq$ is exactly given by the distance from $f$ to
  $f^\LE$ plus the distance from $f^\LE$ to $f^\eq$. The decomposition
  \eqref{eq:KLD_decomposition} can be schematically depicted as
  $$\mathop{\overbrace{\mathop{\underbrace{f(\mathbf{v},t)\longrightarrow
  f^\LE(\mathbf{v};}}_{\KLD^{\text{kin}}(t)}\mathop{\underbrace{\!T(t))\longrightarrow
  f^\eq(\mathbf{v})}}_{\KLD^\LE(T(t))}}}^{\KLD(t)}.$$}
\bibitem [{not({\natexlab{c}})}]{note_22_02_1}%
  \BibitemOpen
  \href@noop {} {}  \bibinfo {note} {{T}he sketch in
  Fig.~\ref{fig:sketch} was actually obtained from the solution of the BSA (see
  Sec.~\ref{sec:Sonine}) and the application of Eqs.~\eqref{eq:KLD^LE} and
  \eqref{eq:deltaKLDgamma} for the initial condition $\theta^0=2$,
  $a_2^0=0.5$.}\BibitemShut {Stop}%
\bibitem [{not({\natexlab{d}})}]{note_22_01_3}%
  \BibitemOpen
  \href@noop {} {}  \bibinfo {note} {{I}n that case,
  $\tau_\bb\to\infty$ and one has to change the definition of dimensionless
  time to $\zeta_0 t$.}\BibitemShut {Stop}%
\bibitem [{not({\natexlab{e}})}]{note_22_01_4}%
  \BibitemOpen
  \href@noop {} {}  \bibinfo {note} {{T}his was the approach
  carried out in a previous work to analyze the TME~\cite{SP20}}\BibitemShut
  {NoStop}%
\bibitem [{not({\natexlab{f}})}]{note_22_01_6}%
  \BibitemOpen
  \href@noop {} {}  \bibinfo {note} {{T}his approximation has
  been employed in Ref.~\cite{PSP21} to study the response of the nonlinear
  fluid to a low-temperature quench}\BibitemShut {NoStop}%
\bibitem [{not({\natexlab{g}})}]{note_22_01_5}%
  \BibitemOpen
  \href@noop {} {}  \bibinfo {note} {{F}or the EME, there is
  not a clear cut distinction between the direct and inverse EME (since the
  relative entropy $\KLD$ is positive definite), unless the initial
  temperatures are known}\BibitemShut {NoStop}%
\bibitem [{not({\natexlab{h}})}]{note_22_05_1}%
  \BibitemOpen
  \href@noop {} {}  \bibinfo {note} {{A} similar distance to
  a local equilibrium distribution has been recently introduced to analyze the
  relaxation of an Ising ferromagnet in the weak coupling
  limit~\cite{TYR22}.}\BibitemShut {Stop}%
\end{thebibliography}

%

\end{document}